\documentclass[fleqn,10pt]{wlscirep}

\usepackage[utf8]{inputenc}
\usepackage[T1]{fontenc}
\usepackage{bm}
\usepackage{cleveref}
\usepackage{amsthm}
\usepackage{bbold}
\usepackage{hyperref}
\usepackage{zref}
\usepackage{comment}
\usepackage{color,soul}
\usepackage{amsmath}
\usepackage{subcaption}
\usepackage{xcolor}

\usepackage{tikz}
\usetikzlibrary{arrows.meta,positioning,fit,calc,backgrounds}
\tikzset{
  var/.style      = {circle, draw, minimum size=18pt, inner sep=0pt},
  panel/.style    = {rounded corners, draw, inner sep=6pt},
  >={Stealth[length=2.2mm,width=2.2mm]}
}

\DeclareUnicodeCharacter{2212}{-}

\hypersetup{
    pdftitle={From Network Inequality to Network Fairness: A Perspective on Responsible Decision-Making},
    pdfauthor={Espin-Noboa et al.},
    pdfsubject={Nature Perspective},
    pdfkeywords={algorithmic fairness, social capital, procedural justice, network effects}
}

\newcommand{\para}[1]{\smallskip\noindent\textbf{#1}}

\newcommand{\blueit}[1]{\textcolor{blue}{#1}}

\newcommand{\titlebottom}{A Perspective on Responsible Decision-Making}

\title{From Network Inequality to Network Fairness:\\\titlebottom}

\author[1,$\dagger$]{Lisette Espín-Noboa}
\author[2,3]{Tina Eliassi-Rad}
\author[4]{Pak-Hang Wong}
\author[5]{Erich Prem}
\author[1]{Meike Zehlike}
\author[6,7]{Ricardo Baeza-Yates}
\author[8]{Suresh Venkatasubramanian}
\author[1,9,$\dagger$]{Fariba Karimi}
\affil[1]{Complexity Science Hub Vienna, Vienna, Austria}
\affil[2]{Northeastern University, Boston, USA}
\affil[3]{Santa Fe Institute, Santa Fe, USA}
\affil[4]{Hong Kong Baptist University, Kowloon Tong, Hong Kong}
\affil[5]{Vienna University of Technology, Vienna, Austria}
\affil[6]{KTH Royal Institute of Technology, Stockholm, Sweden}
\affil[7]{Universitat Pompeu Fabra, Barcelona, Spain}
\affil[8]{Brown University, Providence, USA}
\affil[9]{Graz University of Technology, Graz, Austria}
\affil[$\dagger$]{Corresponding authors: espin@csh.ac.at, karimi@tugraz.at}

\begin{abstract}
Social networks shape how individuals make decisions and how opportunities are distributed. However, the mechanisms that generate these networks often reflect pre-existing inequalities, and technologies that rely on network-derived signals risk further amplifying such disparities. Algorithmic fairness research largely treats networks as a fixed background, grounding analysis almost exclusively in distributive justice and overlooking how network structures systematically bias decision-making. In this Perspective, we identify ten network effects and trace how they create structural biases in the relationship between what we intend to measure and what we observe. Using academic hiring as an example, we show that network biases are not inherently harmful or beneficial. Determining their legitimacy requires examining the entire decision-making process through the lenses of both  distributive and procedural justice while engaging all affected stakeholders. We therefore call for a holistic, networked approach to fairness that moves beyond static group categories and recognizes the dynamic, relational, and structural nature of inequality.

\end{abstract}

\begin{document}

\flushbottom
\maketitle

\noindent \textbf{Key points:} 
\begin{itemize}
    \item Personal networks\footnote{The term ``personal network'' is used in this paper to refer to the egocentric network (or ego network) of an individual: the focal actor (ego), their direct contacts (alters), and the ties among those alters.~\cite{mccarty2011personal} A ``social network,'' in contrast, refers to the complete (sociocentric) network encompassing the personal networks of all individuals. Neither term should be confused with ``social network sites'' or ``social media platforms,'' which commonly refer to online platforms that enable people to connect and share information.~\cite{boyd_Ellison_2007}} shape our fate for better or worse. %
    \item Building and maintaining rich personal networks depends not merely on individual effort but also on structural conditions that affect people unequally.
    \item Decisions influenced by personal networks have long raised ethical concerns (e.g., nepotism, in-group favoritism, friend-of-a-friend advantage, guilt-by-association).
    \item Fairness in social networks---both socially and when algorithmically produced---must account for individuals, groups, and their connections.
    \item Distributive justice can only be one part of a broader socio-technical fairness framework.
\end{itemize}

\section*{Introduction}
We live in a highly interconnected world where social relations are essential to our well-being, coexistence, and economy.~\cite{jordan2008welfare, chetty2022social1}
Sociologists and network scientists have long studied these relations to understand how and why people connect.~\cite{mcpherson2001birds, newman2003social}
The popular adage ``it is not what you know, but who you know'' reflects the significance of social connections.
In venture capital, for instance, founding teams with ties to investors are significantly more likely to secure funding than equally promising teams without such connections.~\cite{shane2002network}
In criminal justice, risk assessment tools use an individual's ``criminal associates'' as a scoring factor, flagging people as higher risk based on the records of those they have been connected to.~\cite{jacobs2021measurement,nieto2023examining}
Connections can therefore generate both advantages and penalties.
Knowing the ``right'' people %
can provide access to jobs and other opportunities,~\cite{boyd2014networked,dies2025forecasting} while 
association with the ``wrong'' people can lead decision-makers to discount an individual’s own merits through 
``guilt-by-association.''~\cite{hussinger2019guilt}
Similar mechanisms operate in algorithmic systems.
Recommendation algorithms often learn our preferences based on the preferences of our connections.~\cite{gupta2013wtf} Since we tend to connect with similar others,~\cite{mcpherson2001birds} these algorithms end up amplifying those associations.~\cite{santos2021link}
These dynamics may extend further as large language models and AI agents are increasingly considered for social modeling and decision-making. 
When such models simulate social networks, they reproduce many structural patterns observed in real ones,~\cite{chang2025llms, gkartzios2025modeling, mallick2026llmssocialnetworkmodeling} 
yet tend to overemphasize political and demographic biases.~\cite{mehdizadeh2025homophily} 
When used to recommend people, their suggestions tend to cluster within existing collaborative circles, concentrating attention on small, tightly connected groups rather than surfacing equally qualified individuals beyond them.~\cite{barolo2025whose}

\textit{Network effects} pose a range of social and ethical challenges for those invested in equity because social ties are often treated symbolically as signals of individual merit, trust, or identity.~\cite{Bourdieu_2002}
Not everyone has equal access to the ``right'' connections, and marginalized groups frequently face structural barriers to building ties with powerful actors due to unequal opportunities shaped by place of birth, historical bias, stereotypes, and segregation.~\cite{braddock1987minorities, petersen2000offering, sampson2008neighborhood, loury2021anatomy} 
These disparities tend to be reinforced within social worlds where there are few incentives to form connections across existing social divisions (e.g., class, race, caste, etc.).~\cite{ibarra1992homophily, small2009unanticipated} Often, advantages accumulate around those who are already well connected. 
Thus, decisions that prioritize connections can systematically favor individuals with influential networks over equally capable others. 

The use of networks as informational signals can produce individual-level harms that are not always acknowledged.~\cite{dimaggio2012network,blumenthal2019potential,kopar2021critical,li2023bright}
However, algorithmic systems often pick up on these signals and replicate the harms, treating ``who one knows'' as a proxy for ``who one is.''~\cite{boyd2014s}  Moreover, even in the absence of deliberate self-disclosure, algorithms often infer sensitive attributes from publicly visible connections, exposing traits such as personality or behavior that individuals may prefer not to reveal.\cite{zheleva2009join,isaak2018user}

The current notions of fairness suggested by the scientific community~\cite{mehrabi2021survey} and enforced by certain governments~\cite{madiega2021artificial, canada2021equal, boyd2014networked} mainly concentrate on demographics (e.g., gender, ethnicity), disregarding the fact that discrimination can stem from personal connections and ultimately lead to disparities and unfairness.
In today's interconnected, algorithmically infused societies,~\cite{wagner2021measuring} networks often make the difference in outcomes, blurring the boundary between fairness, merit, and choice.
As technology studies scholars have noted, ``Networks are at the base of data analytics, yet our social and legal models focus on the individual.''~\cite{boyd2014networked}
To address this gap, we need new notions of fairness that take into account not only the personal characteristics of individuals but also their connections and position in networks.

We offer \textit{network fairness} as a new avenue for research in the fields of responsible AI and algorithmic fairness. 
We do not prescribe what a fair network should look like, nor do we set a normative standard for a network-fair process. 
Instead, we document and classify the mechanisms through which personal networks shape and reinforce inequality and inequity in decision-making processes.
We focus on ten~~\textit{network effects} that drive these dynamics (\Cref{tbl:effects}), 
shifting attention from individual to relational structures that influence opportunities and outcomes.
We organize these effects into a taxonomy based on their structural scope: \textit{node-level effects} driven by individual position or attributes, \textit{neighborhood-level effects} arising from local ties, and \textit{structural effects} rooted in broader topological patterns.
In practice, these effects rarely operate uniformly, because edge weights and signs (e.g., tie strength, trust, or antagonism) modulate their magnitude and direction. The same structural mechanism can therefore produce very different outcomes depending on the intensity and sign of the underlying relationships.
We develop this framework by first revisiting sociological work on social capital and network inequality. We then examine---and discuss the limitations of---existing efforts to conceptualize fairness in studies of social networks. %
Building on these foundations, we show how distributive concerns intersect with procedural justice and the perspectives of multiple stakeholders.
We conclude with an open call to researchers in philosophy, ethics, social science, computational social science, and computer science to study fairness with network effects in mind, and to treat fairness not only as a property of algorithms but as a shared social responsibility embedded in the systems and institutions that govern social decision-making.

\begin{table}[t!]
\centering
\caption{\textbf{Taxonomy of network effects.} Ten effects that drive inequality in network-mediated decisions, organized by structural scope: node-level, neighborhood-level, and structural-level.}
\label{tbl:effects}
\begin{tabular}
{p{0.003\textwidth}p{0.17\textwidth}p{0.17\textwidth}p{0.16\textwidth}p{0.40\textwidth}}
\toprule
\# & Network effect & Mechanism & Structure involved & Example \\ \midrule
\multicolumn{5}{l}{\textbf{Node-level effects}} \\ \addlinespace
1 & Rich-get-richer effect & preferential attachment & node degree & When eminent scientists get disproportionately more credit for their contributions to science while relatively unknown scientists tend to get disproportionately less credit for comparable contributions.~\cite{merton1968matthew}  \\
2 & First-mover advantage & non-Markovian temporal process & node degree and age & When the first papers in a field, essentially regardless of content, receive citations at a higher rate than papers published later.~\cite{newman2009first,kong2022influence} \\
3 & In-group favoritism & homophily & node attributes & When male workers who have male supervisors have higher chances of promotion than their female counterparts, or vice versa.~\cite{tajfel1979integrative}\\
\addlinespace \midrule
\multicolumn{5}{l}{\textbf{Neighborhood-level effects}} \\ \addlinespace
4 & Guilt-by-association & homophily & 1-hop neighborhood & In P2P lending, when the credit score of a person is defined by (the credit score of) her friends regardless of her own credit history.~\cite{de2019does, li2020fairness}  \\
5 & Chaperone effect & homophily and preferential attachment & 1-hop neighborhood, node degree and attributes & When someone (e.g., employee, student, or researcher) gets preferential treatment (e.g., gets a job, an internship, or a paper acceptance) because of her relation with someone (or something) very important.~\cite{sekara2018chaperone, brainard2022reviewers,crane1965scientists}  \\ 
6 & Reciprocity effect & reciprocity & 1-hop neighborhood & When a scientist cites an article by someone who cited her first to ``return'' the favor.~\cite{li2019reciprocity,wang2020early} \\ 
7 & Perception bias (e.g., majority illusion, friendship paradox, pluralistic ignorance) & homophily & 1-hop neighborhood, group size, node degree and attributes & 
When a few highly connected individuals frequently drink alcohol, their peers might incorrectly perceive heavy drinking as the norm, leading to increased alcohol consumption in the broader student network, despite most students drinking minimally.\cite{baer1991biases,lerman2016majority}
 \\
\addlinespace \midrule
\multicolumn{5}{l}{\textbf{Structural-level effects}} \\ \addlinespace
8 & Friend-of-a-friend advantage & brokerage (closed triad) & 1-hop and 2-hop neighborhood & When job applicants rank higher (or get the job) thanks to an insider referral.~\cite{shwed2014referrals} \\
9 & Broker effect & brokerage (open triad) & structural holes & Leaders who act as brokers within teams can cause a bottleneck in information flow that can decrease productivity.\cite{cummings2003structural, long2013bridges} \\
10 & Anti-transitivity effect & structural balance (triad) & signed edges, 1-hop and 2-hop neighborhood & When relational evaluations propagate transitively through social ties, such that ``the enemy of a friend is treated as an enemy,'' while ``the enemy of an enemy is treated as an ally.''\cite{lee1994friend,meger2023iterative} \\
\bottomrule
\end{tabular}
\end{table}

\section*{Social capital and network inequality}
\begin{figure}[ht]
\centering
\includegraphics[width=1\linewidth]{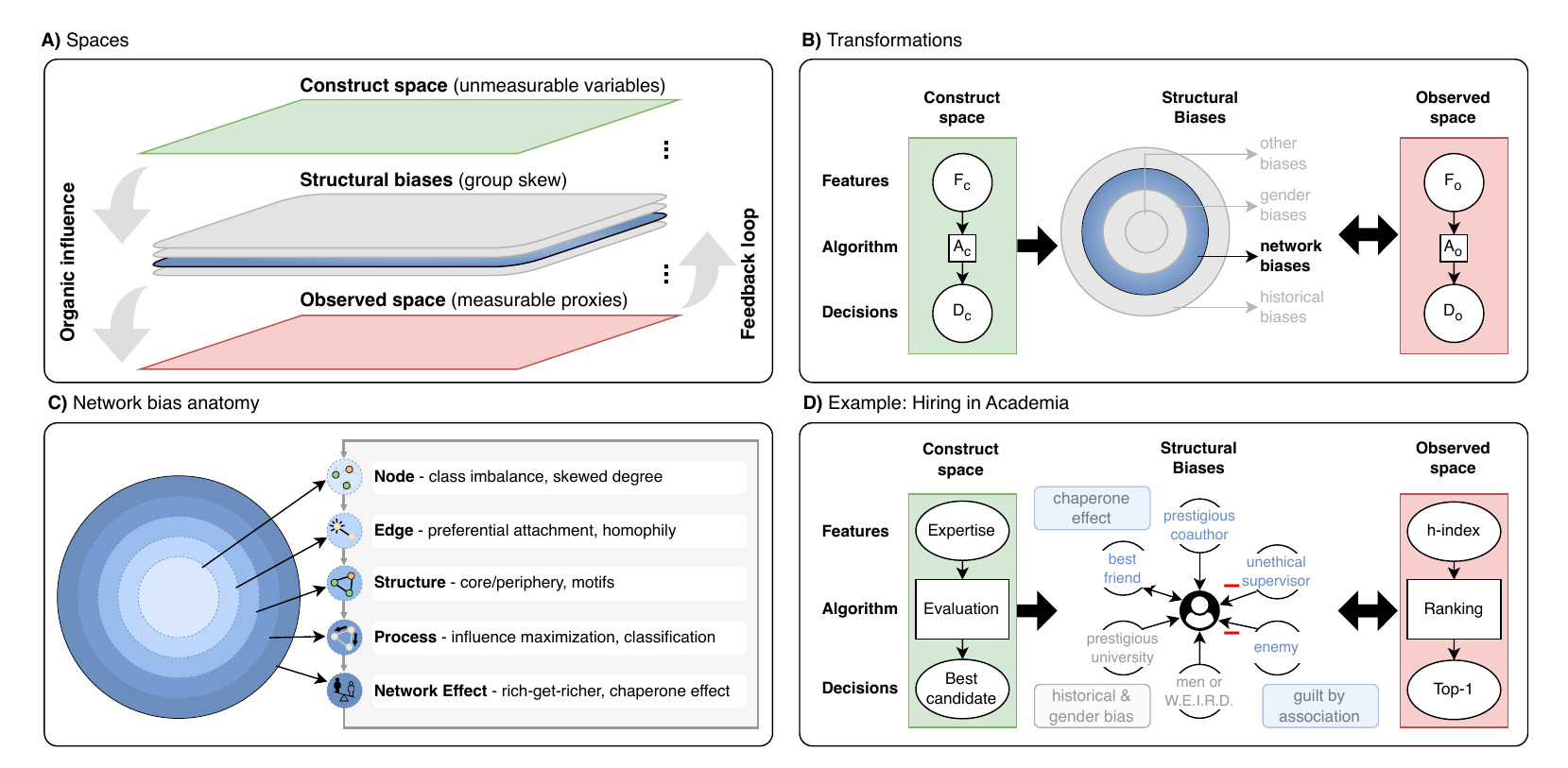}
\caption{\textbf{Biases in social decision-making processes}. 
\textbf{A)} {Spaces:} The \textit{construct} space encapsulates unmeasurable ideals, while the \textit{observed} space employs measurable proxies of the constructs, affected by underlying \textit{structural biases}. 
Acknowledging these biases requires adopting the ``we-are-all-equal'' (WAE) worldview, focusing on group fairness or non-discrimination.\cite{friedler2016possibility} 
We argue that the structural biases reflect a complex interplay between imbalanced group representations and biases rooted in \textit{personal and social networks}. 
Feedback loops can further amplify these biases in the observed space.
\textbf{B)} {Transformations:} 
A vertical transformation represents the decision-making process where input features $F_{\bullet}$ fuel an algorithm $A_{\bullet}$, yielding outputs or decisions $D_{\bullet}$ that reflect underlying rules or learned patterns. 
A horizontal transformation reflects the process of converting construct variables (${\bullet}_c$) into observed variables (${\bullet}_o$).~\cite{friedler2021possibility} 
We argue that personal and social networks significantly influence the entire decision-making pipeline ($F_{\bullet} \rightarrow D_{\bullet}$ and ${\bullet}_c \rightarrow {\bullet}_o$), \textit{impacting both individual and group fairness}.
\textbf{C)} Network bias anatomy: 
Nodes, edges, and network structure act as indicators of individual importance, influencing decision-making processes and contributing to social inequalities. 
\textbf{D)} Application to academic hiring. Construct-space variables (Expertise, Evaluation, Best candidate) are transformed through structural biases into observed proxies (h-index, Ranking, Top-1). These biases include not only those traditionally studied in fairness (e.g., historical and gender bias) but also network-based biases, which can be advantageous (e.g., the chaperone effect from a prestigious coauthor) or detrimental (e.g., guilt-by-association from being connected to an unethical supervisor).
While prior research mainly examines individual and group fairness, we advocate for a holistic and comprehensive approach that also considers network effects.
}
\label{fig:biases}
\end{figure}

The study of personal and social networks
has long been an effort to gain insights into their formation, evolution, and structure.~\cite{pescosolido2021personal} 
These insights have contributed to our understanding of \textit{social capital}---a valuable resource obtained from nurturing social connections.
Although definitions and interpretations of social capital vary,~\cite{adler2002social} those rooted in the theories of Putnam, Coleman, and Bourdieu emphasize the importance of relationships, interactions, and shared values as key components.~\cite{field2016social, robison2002social,lin2002social}
Because social capital is a latent construct that cannot be directly observed or measured (\Cref{fig:biases}A), sociologists have developed different strategies to approximate it.%
~\cite{borgatta1980level, hirsch1999umbrella,field2016social} 
In hiring contexts, for example, social capital is often operationalized through referrals, while skills are approximated using expertise measures such as academic transcripts (\Cref{fig:biases}D). 
However, these proxies embody contestable value judgments, introduce ambiguity through imperfect measurement, and risk encoding systematic biases into what appear to be objective assessments.\cite{adcock2001measurement,knox2022testing}

Many of these measurements reflect the entanglement of social and cultural capital.~\cite{Bourdieu_2002} To avoid this conflation, evaluating social capital requires careful attention to context, including cultural norms, political structures, and societal objectives.\cite{graeff2009social, baycan2022dark} 
Directing attention toward the benefits of social capital has been critical for recognizing the power of relationships, but the benefits associated with social capital are not universally applicable. 
While social capital can undoubtedly elevate an individual's quality of life, it is crucial to acknowledge how relational structures can also promote or exacerbate disparities within a society.
This is because individuals with strong social capital are not only likely to achieve better outcomes, but they are also better positioned to sustain and attract further advantageous connections. %
Consequently, social capital operates as a cumulative process, where initial advantages are reinforced over time. 
When institutions and decision-making processes systematically reward existing social capital, this dynamic can produce unequal distributions of opportunity and contribute to the persistence of social inequalities.\cite{ayios2014social,livan2019don}

\subsection*{Social capital as structural advantage}

Connections facilitate access to employment, economic mobility, and healthier lives,~\cite{lin2000inequality,chetty2022social1} while enabling individuals to reciprocate by sharing information, providing support, and creating opportunities for others. 
However, social capital often depends less on the number of connections and more on their strength, structural position, and quality. 
The \textit{strength} of ties is shaped by factors such as time, emotional closeness, intimacy, and reciprocity.\cite{granovetter1973strength}
Strong ties reflect connections that provide emotional support and trust; they tend to form between people who are similar in terms of their values and beliefs---but also their social networks. 
Weak ties, on the other hand, are more likely to connect people to \textit{different} social networks. 
Weak ties are important for accessing new information and opportunities.\cite{krackhardt2003strength,brown1987social,granovetter1983strength} For instance, job mobility in digitally oriented sectors often relies on weak ties, while other sectors depend more heavily on strong connections.\cite{rajkumar2022causal} 
Additionally, established practices of developing ties, such as through \textit{brokerage} or the bridging of \textit{structural holes},~\cite{burt2000network} expand access to diverse resources and control, broadening opportunities for career, education, and social mobility.
Similar structural advantages are evident in job referrals, which translate network position into value not only for individuals but also for organizations by improving recruitment efficiency and productivity.~\cite{shwed2014referrals} 

The \textit{quality} of a social tie is also a crucial aspect of an individual's social capital and is often reflected in the social capital of the person at the other end of the tie. 
In academia, for example, distinguished authors are more likely to get their papers published and cited, benefiting both them and their coauthors through a phenomenon known as the \textit{chaperone effect}.\cite{sekara2018chaperone,huber2022nobel,li2022untangling}
Similarly, having friends in higher socio-economic classes is associated with greater economic mobility.\cite{chetty2022social1}
These patterns reflect recursive mechanisms in which advantage propagates through social ties, a principle that also underlies network-based ranking algorithms such as PageRank.\cite{page1999pagerank,gleich2015pagerank} 
Social capital can therefore be inherited through structural position,~\cite{cordelli2015distributive} cultivated through social learning and contagion,~\cite{mas2009peers} or reinforced by similarity-driven connection patterns.~\cite{kandel1978homophily} %

Social capital, however, is not always the result of personal effort or merit. Network positionality matters.
Individuals can obtain---and be unable to obtain---social capital due to circumstances beyond their control, such as family background or place of birth, raising questions about whether success solely reflects individual achievement or unearned advantage concentrated among social elites.
A prominent example is the rise of so-called ``nepo babies,'' who inherit visibility and networks from their parents.\cite{foster2025nepo}
Conversely, limited or even negative social capital may result from harmful associations~\cite{christian2006social} or systemic inequalities~\cite{atkinson2015inequality} that constrain network formation and access to opportunity. 
Together, these dynamics show that social capital reflects a complex interplay between effort, luck, privilege, and structural constraint, highlighting the need to examine its sources and its effects in responsible social decision-making.

\subsection*{Discrimination on the basis of personal networks}

Regardless of where we stand within our networks, it is a common tendency for people to judge us partly based on our personal connections. The old proverb, ``show me your friends, and I will tell you who you are,'' affirms this inclination, emphasizing the idea that individuals tend to associate with those who share similar characteristics, values, or behaviors. 
This association is referred to as \textit{homophily}, and
it often reflects the preferences and deliberate choices of individuals in making connections with similar others. \cite{mcpherson2001birds} While homophily fosters cohesion, it paradoxically contributes to segregation and inequalities across groups. 
Excessive homophily creates a structure referred to as ``echo chambers,''\cite{currarini2009economic} facilitating the expansion of inequalities and misinformation by trapping information, opportunities, and trust within exclusive communities, such as ``old-boy networks'' or ``rich clubs,'' while weakening the bridging ties that might otherwise connect outsiders to new opportunities.~\cite{mcdonald2011s,jackson2021inequality,levy2019echo,ibarra1992homophily} 
In the labor market, for example, in-group preference can lead to more support in job searches and increased employment opportunities for individuals from the same gender or socio-economic class, while ``outsiders'' may not receive such benefits, regardless of their abilities.\cite{mcdonald2011s,coffman2018gender,jackson2019human}
Similarly, homophily can lead to glass ceilings, making it extremely challenging for minority groups to reach the highest ranks.\cite{avin2015homophily, karimi2018homophily, espin2022inequality, ferrara2022link, neuhauser2023improving} 

Homophily forms through a combination of individual choice and structural conditions.
While we may feel free to choose our connections, structural biases influence the available options,\cite{lindquist2015entrepreneurial} ultimately limiting the scope of our ``free'' choices.\cite{kossinets2009origins}
This implies that individuals facing pre-existing disadvantages may struggle more to access advantageous social capital, since those they (can) interact with are more likely to share similar disadvantages.\cite{jackson2021inequality} 
This situation can be intensified by the Matthew effect,~\cite{merton1968matthew} where the \textit{rich get richer} and the \textit{poor get poorer}.\footnote{Here, being rich (poor) means having a good (bad) quality of connections or high (low) levels of social capital. In network science jargon, this tendency to connect to \textit{popular}, often high-degree nodes is known as preferential attachment.\cite{barabasi1999emergence}}
One might argue, however, that individuals with limited social capital should strive to gain access to these influential circles. 
While this approach may yield results for a few, it typically demands significant and often futile efforts.
This is because the enduring consequences of years of disadvantage can sustain inequalities, causing people from under-represented groups to remain significantly behind unless interventions are implemented to alleviate the issue.~\cite{Horowitz2019}
Even when striving for greater social capital, people %
may face \textit{guilt-by-association} penalties for the actions of others in their networks. %
For example, individuals who collaborated with scientists involved in misconduct, even if they were not directly implicated, experience an 8–9\% reduction in citations.\cite{hussinger2019guilt}  Similar patterns have been observed in other scenarios, including the marginalization of families with incarcerated members,\cite{codd1998prisoners} 
and the transfer of perceived trustworthiness based on one's peers.\cite{de2019does}
These ``spillover'' %
effects, %
rooted in assumed homophily between connected individuals, become even more pronounced when they interact with other link formation mechanisms, 
such as \textit{reciprocity}, \textit{transitivity}, and \textit{consolidation}.~\cite{garip2021network}
When these mechanisms intertwine with homophily, they can exacerbate discrimination against minorities or outsiders~\cite{chiang2011network,laniado2016gender,grund2015ethnic} while preventing the spread of behaviors, resources, or ideas across sub-populations.\cite{centola2015social,zhao2021network}

\subsection*{Network position and structure}

An individual's position in a network affects %
both how they are perceived and how they perceive others.
This dynamic shapes worldviews, reinforces biases~\cite{bian2018evidence} and prior beliefs,\cite{nickerson1998confirmation,coffman2018gender} 
and influences perceptions of inequality.~\cite{schulz2022network} 
Some of these effects stem from the \textit{friendship paradox},\cite{alipourfard2020friendship} where, on average, individuals' friends tend to have more friends than the individuals themselves do.
Under certain conditions, such as high homophily or imbalanced group sizes, this paradox can create a \textit{majority illusion},\cite{grandi2023identifying,lerman2016majority} making a globally rare attribute appear common within a local network. 
Conversely, it can lead to underestimations of minority group sizes,\cite{lee2019homophily} potentially resulting in incorrect and unfair decisions.\cite{pronin2007perception}

Position is not only a matter of where one stands in a network, but also of when one enters it.
Timing can create structural disadvantages through \textit{first-mover advantages}, where those entering networks earlier typically accumulate more connections than later arrivals.~\cite{lieberman1988first,kong2022influence} 
Historical exclusion illustrates how past barriers create persistent networking disadvantages. 
For instance, in academia, both gender and racial disparities can be traced in part to historical restrictions that limited women's and ethnic minorities' access to education, professional positions, and influential networks.~\cite{schuck1974sexism,menges1983barriers,winegarden1972barriers,wilson2013men,zick2008ethnic,zbarauskaite2015minority}
These temporal networking effects can persist even after formal barriers are removed, perpetuating disadvantage through group-based stereotypes rather than individual merit.~\cite{coffman2018gender,shepherd2021inequality}

Once networks are established, their structure determines how influence spreads. 
Individuals' decisions are shaped by the actions and signals of others in their social environment.\cite{dimaggio2012network} 
This influence grows as more individuals in our network adopt a practice, share valuable information, or provide positive reinforcement, a phenomenon known as \textit{peer effects}.\cite{eckles2016estimating,centola2021change} 
For example, peer influence alone raises the odds of purchasing a service by over 60\% when a friend has adopted it,~\cite{bapna2015your} illustrating how personal networks drive everyday choices.
Similar peer effects are observed in %
educational performance,\cite{hoxby2000peer} academic productivity and prominence,\cite{li2022untangling} voting decisions,\cite{bond201261} smoking and obesity trends,\cite{christakis2008collective,christakis2007spread} and income mobility.\cite{chetty2022social1}
This influence extends beyond direct contacts, reaching even friends of friends.\cite{pinheiro2014origin}
But not all (in)direct ties carry positive associations.
Networks can also exhibit \textit{anti-transitivity} driven by balance-seeking behavior rather than assumed similarity.~\cite{heider1946attitudes,rezapour2024structural}
For instance, the principle that ``the friend of my enemy is my enemy'' leads actors to avoid ties with their adversaries' allies, creating polarized configurations where groups align into opposing sides to maintain structural balance.~\cite{lee1994friend,szell2010multirelational}

These perceptual and behavioral effects are further shaped by the overall structure of the network, in particular its edge density.
Large social networks are typically sparse, because individuals can maintain only a limited number of meaningful relationships (Dunbar's number).~\cite{dunbar2010many}
As a result, social decision-making\footnote{Decisions made by or for humans considering interpersonal relationships, group dynamics, and cultural norms.} often relies on indirect ties.
When no direct connection exists, intermediaries act as both connectors and sources of credibility, guiding whom to trust and why. 
This reliance on intermediaries (e.g., referrals or brokers) provides an \textit{efficient} heuristic for navigating large networks or bridging structural holes, but it can introduce systemic bias.\cite{beugnot2020gender}
As Borondo et al.~demonstrate, sparsity %
concentrates opportunities among well-positioned individuals, producing ``topocratic'' rather than merit-based allocations.\cite{borondo2014each} 
Paradoxically, efforts to promote meritocracy by imitating top performers can deepen this effect.  %
Strategic tie formation reinforces early advantages and further entrenches those already central.\cite{livan2019don,dwork2024equilibria,zhang2024network}
Even among actors with similar capabilities, connection patterns shaped by position, timing, and luck compound into cumulative advantage, yielding highly unequal and ultimately non-meritocratic outcomes.\cite{dwork2024equilibria}

\section*{Pitfalls of existing notions of fairness in social networks}

Existing notions of fairness focus primarily on individual or group fairness---framed in network terms as nodes and node attributes---often giving limited and indirect consideration to the relational components of the network itself.~\cite{zhang2022fairness}
Recent approaches~\cite{choudhary2022survey,mehrotra2022revisiting,liu2023group,zhang2024network} have started to bridge this gap by examining the edges between nodes and proposing strategies to diversify the links a node forms.\cite{neuhauser2023improving,barnes2025edge,beilinson2020fairness,wang2022information} 
These advances improve our understanding of fairness, particularly by addressing homophily and group size imbalances. 
However, they still overlook how network effects systematically distort decision outcomes, failing to address how network structure and social capital themselves can generate and perpetuate unfairness.

\subsection*{Node-attributed fairness in networks}
Fairness research in machine learning has mostly centered on tabular data, where observations are assumed to be independent and identically distributed (i.i.d.). Network data depart from this assumption because the attributes of a node may depend on the attributes of its connections. This interdependence complicates the direct use of traditional fairness approaches and highlights the need for methods tailored to network structures.\cite{choudhary2022survey,zhang2022fairness}
A common approach has been to adapt existing techniques by representing networks in tabular form through feature vectors or node embeddings.\cite{perozzi2014deepwalk} In a fairness context, embeddings are expected to preserve the graph’s structural properties while remaining independent of sensitive attributes.\cite{dong2022fairness} Balancing these two requirements is difficult when structural patterns correlate with protected characteristics. To address this tension, researchers have explored different strategies: pre-processing the graph to reduce bias,\cite{dong2022edits} modifying the objective function to include fairness constraints,\cite{khajehnejad2022crosswalk} or post-processing embeddings after training.\cite{choudhary2022survey,palowitch2019monet}
Beyond node representations, fairness concerns also arise in the relations between the nodes.\cite{li2021dyadic,yang2022obtaining} 
So-called \textit{dyadic fairness} adapts classic group fairness metrics---originally developed for non-network data---to evaluate whether connections across groups occur equitably.
Metrics such as disparate impact,\cite{laclau2021all} statistical parity,\cite{rahman2019fairwalk} equal opportunity,\cite{hardt2016equality} and acceptance rate parity~\cite{rahman2019fairwalk} have therefore been applied in dyadic settings, where edges are categorized according to the groups of the nodes they connect. While this provides a systematic way to assess fairness in network models,\cite{choudhary2022survey} it may fail to capture disparities that emerge from skewed or homophilic structures.
Mehrotra et al.\cite{mehrotra2022revisiting} highlight this limitation by showing that applying group fairness metrics at the dyadic level can produce misleading assessments, either by failing to capture imbalances in influence and reach or by overlooking biases in interactions between groups. 
To address this, they introduce two notions of group fairness that explicitly account for network structure and capture complementary forms of bias. 
Intra-network fairness evaluates how access to others in a network is distributed across groups, accounting for differences in structural advantage. 
Inter-group fairness, by contrast, assesses how equitably groups distribute their interactions across other groups. 
For example, in a hiring process, outcomes may appear fair across genders while interviewers disproportionately favor candidates from their own group. 
In such cases, intra-network fairness may be preserved, whereas inter-group fairness is undermined by homophily-driven bias.
Similar patterns are observed in algorithms for node classification, node ranking, link prediction, and influence maximization. These methods often amplify structural imbalances created by the interplay between homophily and degree inequality.\cite{stoica2018algorithmic,fabbri2020effect,stoica2020seeding,wang2022information,espin2021explaining,espin2022inequality,ferrara2022link} 

While this body of work has advanced our understanding of how networks shape algorithmic outcomes, its reliance on group-level metrics tied to the joint distribution of sensitive attributes may obscure individual-level disparities and does not fully resolve the breakdown of the i.i.d. assumption.\cite{choudhary2022survey,zhang2022fairness,liu2023group,saxena2024fairsna}

\subsection*{Relation-based fairness in networks}

The interdependence between individual outcomes and social structure has motivated a growing body of work on relation-based fairness in networks. 
Farnadi et al.~\cite{farnadi2018fairness, farnadi2019declarative} show that when outcomes are influenced by statistical dependencies between connected individuals---such as friends influencing each other’s decisions or attributes---traditional group fairness metrics may fail to detect discrimination. They propose ``FairPSL,'' a framework using probabilistic soft logic that can incorporate fairness constraints directly into relational models, allowing systems to reason about how bias may propagate through networked ties. 
Yang et al.~\cite{yang2024your} address such dependencies causally through \textit{interference-aware fairness}, distinguishing between \textit{self-fairness}, which captures discrimination arising from an individual's own sensitive attributes, and \textit{peer-fairness}, which captures discrimination induced by the sensitive attributes of their connections. %
Complementing this, Sium et al.~\cite{sium2024individual} extend the principle of individual fairness---that similar individuals should be treated similarly---to graph settings by defining similarity not just through node features, but also through local and global structural properties of the network, such as a person’s position or neighborhood in the graph. Their model ensures that individuals with equivalent roles or social positions are evaluated comparably, even when they are not feature-identical. 

Other lines of research emphasize fairness as a property of social relations themselves. Fish et al.~\cite{fish2022s} argue that existing fairness metrics in machine learning fail to capture harms that emerge not from unequal outcomes, but from the structure of social relationships---such as subordination, marginalization, or lack of recognition. They propose a relational theory of fairness that shifts the focus from distributions to social standing, and they illustrate how algorithmic systems can reinforce unjust social hierarchies even when parity metrics are satisfied. In a similar spirit, Birhane~\cite{birhane2021algorithmic} calls for a relational ethics approach that centers interdependence, context, and the lived experiences of marginalized communities, arguing that fairness must be grounded in the quality of relationships rather than technical abstractions. Bengtson et al.~\cite{bengtson2023relational} build on this view by applying theories of relational justice to networked settings. They frame fairness as ensuring that everyone enjoys either equal-quality relationships (egalitarianism) or at least a minimally acceptable level of connectedness and recognition (sufficientarianism), and demonstrate how these ideas translate into measurable properties of social graphs.

More recent work has investigated when and how networks might help reduce or amplify unfairness. Zhang et al.~\cite{zhang2024network} propose the concept of ``network fairness ambivalence,'' showing that the same network structures---such as tightly-knit communities or brokered connections---can either mitigate disadvantage by offering access to support, or reinforce inequality by isolating or excluding less connected individuals. Their analysis identifies structural conditions under which social capital improves or worsens group disparities. 
Arnaiz-Rodriguez et al.~\cite{arnaiz2025structural} define \textit{structural group unfairness} as disparities between groups in the social capital provided by their network position. They operationalize this social capital through access to information across the network, capturing differences in groups' isolation, reach, and control, and propose network interventions to reduce these disparities. Similarly, Balepur et al.~\cite{balepur2024intervening} examine how network structure can be actively reshaped to improve fairness, developing interventions that increase trust while accounting for community structure and showing how these interventions can improve downstream outcomes such as resource allocation or influence propagation.

This body of research collectively signals a shift toward placing networks---and the relationships that structure them---at the center of fairness analysis. Whether through social capital, trust, homophily, or structural position, the ties between individuals exert powerful influence on decision-making processes, algorithmic or not. While the concept of social capital has received extensive attention in sociology, economics, and political science, its implications for fairness---especially in computational settings---remain under-explored. There is a pressing need for more research that treats social structure not just as context but as a core driver of inequality. Advancing fairness in networked environments requires addressing both outcomes and the procedures that generate them, recognizing that unjust processes can persist even when statistical distributions appear balanced.
These limitations motivate a broader notion of \textit{network fairness} that accounts for social relationships and structural context.

\section*{Network fairness}

Our discussion of fairness in networks begins with the three main components of any decision-making process, whether algorithmic or human: the \textit{features} that describe individuals, the \textit{algorithm} or process that evaluates these features, and the \textit{decision} that follows (\Cref{fig:biases}B). 
In networks, the entities under consideration are the \textbf{\textit{nodes}} (e.g., individuals, organizations). Each node is described by two broad classes of features or attributes. The first are \textbf{\textit{individual features}}, which include non-sensitive characteristics such as education or skills, and sensitive ones such as race or ethnicity, gender, or socio-economic background. The second are \textbf{\textit{network features}}, which capture how each node is embedded in a broader structure. 
These may reflect the number of connections an individual has, the diversity of their contacts, their centrality or visibility in a community, the role they play as a broker between otherwise disconnected groups, or the extent to which they belong to tightly-knit clusters.
The \textbf{\textit{algorithm}} or decision-making process transforms features into \textbf{\textit{outcomes}}, which are then used to inform future \textbf{\textit{decisions}}.

Assessing whether a decision is fair is not straightforward. \textbf{\textit{Fairness}} lacks a universal definition~\cite{mehrabi2021survey} and is inherently context-dependent, shaped by moral, cultural, societal, and individual perspectives. Like social capital, its interpretation varies across populations and worldviews.\cite{friedler2016possibility,friedler2021possibility} The choice of fairness principle often reflects how one sees the world, a perspective influenced by individual values, social environments, and personal networks. This leads us to discuss two widely used worldviews. %

\subsection*{Worldviews}

Fairness is often framed through two main worldviews: the what-you-see-is-what-you-get (WYSIWYG) view and the we-are-all-equal (WAE) view of structural bias. %
WAE is a property of the construct space, whereas WYSIWYG describes the relation between the construct and the observed space (see~\Cref{fig:biases}A-B).~\cite{friedler2016possibility}

The \textbf{WYSIWYG} view assumes that observable outcomes directly reflect effort, talent, or merit.~\cite{friedler2016possibility,friedler2021possibility} Under this worldview, achievements are accepted without scrutiny, with no attention to unequal starting points or hidden structural factors. Fairness is assessed primarily at the \textit{individual} level and requires that similar individuals receive similar decisions. In practice, this view legitimizes existing differences by treating them as natural consequences of personal actions.

By contrast, the \textbf{WAE} view assumes that individuals are fundamentally similar with respect to the target outcome, and that observed disparities arise from unjust structural or contextual influences.~\cite{friedler2016possibility,friedler2021possibility} Fairness in this worldview requires correcting for such influences. It has inspired definitions that seek to neutralize the effect of \textit{group} membership on algorithmic decisions. Metrics such as statistical parity~\cite{dwork2012fairness} and disparate impact~\cite{feldman2015certifying} embody this principle by requiring comparable outcomes across socially salient groups, regardless of observed non-sensitive features.~\cite{zemel2013learning}

When decisions are influenced by networks, the distinction between these two views becomes blurred, because the same network feature can be earned through \textit{individual} effort or inherited through \textit{group} position.
We argue that, unlike sensitive attributes, which are typically considered only within the WAE view, \textbf{\textit{network features should play a role in both WYSIWYG and WAE perspectives}}, and that the principle separating the two roles is \textit{luck egalitarianism}:~\cite{cohen1989currency} inequalities that arise from informed choices are permissible, whereas those that arise from brute luck call for correction.
In network terms, this is the distinction between \textit{inherited social capital} (e.g., parental wealth, place of birth, the community one is born into) and \textit{earned social capital} (e.g., effortful networking, sustained collaboration), which echoes Rawls' \textit{Veil of Ignorance} and his principle of \textit{Fair Equality of Opportunity}.~\cite{rawls2009theory}
From this view, individual fairness should rely on network features only when they reflect earned capital,~\cite{sium2024individual} while group fairness can draw on both inherited and earned capital to define groups and assess disparities.
The two worldviews thus differ not in whether network features matter, but in which of them count as the individual's own.

\subsection*{Relational justice} %

Relational concerns are increasingly prominent in ethics, including feminist ethics~\cite{koggel2022feminist} and relational ethics,~\cite{Gunkel2025} which view a person's social standing as shaped through relationships rather than as an intrinsic property.
In social networks, this view has direct political implications.
Deciding which network features to measure, which ties count as valuable, and whose networks are worth more is not a neutral technical choice, but a political decision about which relationships are recognized and rewarded.
Those who make such decisions may also favor the kinds of networks to which they themselves belong, raising some ethical concerns.~\cite{mcpherson2001birds,li2019reciprocity,mcdonald2011s}
Determining which notion of fairness should guide a system therefore calls for democratic deliberation that weighs competing views of fairness and their underlying moral commitments.~\cite{wong2020democratizing}

In this context, such deliberation can draw on three principles that place the quality and structure of social relations at the center of justice.
\textit{Relational egalitarianism}~\cite{nath2020relational} seeks to replace social hierarchies with relations of equal standing, in which individuals treat one another with equal respect and regard.
In networks, this means preventing unequal access to connections from becoming a source of social domination.~\cite{anderson1999point,fish2022s}
\textit{Prioritarianism}~\cite{adler2022prioritarianism} instead gives greater weight to improving the position of those who are worst off.
Applied to networks, it directs attention to those with the least social capital, even if the same resources would yield greater gains for better-connected individuals.
\textit{Relational sufficientarianism}~\cite{bengtson2023relational} asks whether everyone reaches a minimum level of social and moral standing.
In networked settings, this requires that no one falls below the threshold of access and recognition needed to benefit from the opportunities the network provides.

These principles rarely operate in isolation.
They may conflict with {efficiency}, which prioritizes overall benefit regardless of its distribution, and their application depends on the social, cultural, and institutional context in which fairness is judged.~\cite{konow2001fair}
Whichever principle is adopted, assessing fairness in networks requires treating each decision as part of an interconnected system, in which relationships shape both who benefits and whether the decision-making process itself is regarded as legitimate.
Relational justice therefore extends the assessment of fairness beyond distributive outcomes to the procedures and social relations through which those outcomes are produced.

\subsection*{Network fair processes}

Procedural justice asks whether the process that produced an outcome was legitimate.~\cite{lind1992procedural}
It requires that decisions rest on accurate information, apply consistent rules, suppress bias, remain open to correction, and represent those affected.~\cite{leventhal1980should}
In algorithmic settings, these criteria must hold at every stage of the decision pipeline (\Cref{fig:biases}B).~\cite{suresh2021framework}
Networks complicate each of them, because network features describe not only the person being evaluated but also the people they are connected to, including the decision maker and other stakeholders, and carry whatever biases those connections encode (\Cref{fig:biases}C-D).
A network-fair process therefore raises questions that individual and group fairness do not capture.
Accuracy demands asking whether social capital is a valid proxy for ability or trustworthiness, or merely a marker of privileged access.~\cite{lin2002social,jacobs2021measurement}
Bias suppression demands asking whether network position is legitimate to use at all,~\cite{grgic2018beyond} or should be treated as a protected attribute, and if so, when.~\cite{mcdonald2011s,liu2023group}
It also demands asking whether the individual features are themselves free of network bias,~\cite{merton1968matthew} and if not, whether they should be corrected, as WAE would require,~\cite{friedler2016possibility} or at least disclosed.
Correction is harder to guarantee than for individual features, because network-derived evidence rests on the behavior of others, which the person evaluated cannot change.~\cite{ustun2019actionable,koggel2022feminist}
Representation is at stake already at data collection, since the sampling design and the network structure itself jointly determine who is visible in the data.~\cite{wagner2017sampling,espin2021explaining,lee2019homophily}
It is at stake in the decision itself, since network data may expose connections whose information shapes the outcome without their consent.~\cite{zheleva2009join}
And it is at stake after the decision, which feeds back into the network and determines whose position improves or worsens.~\cite{ferrara2022link,mansoury2020feedback}

These questions progressively widen the circle of those with a stake in the process.
\textbf{Stakeholders} therefore include those who shape the decision, those directly affected, those whose connections were used as evidence, and those indirectly reached through feedback loops.~\cite{ferrara2022link}
People care about how a decision was made, not only whether it favored them.
Having a voice and being treated with respect are among the most consistent drivers of that judgment.~\cite{lind1992procedural,colquitt2001justice}
Because procedures signal whom a system values, those who judge them fair are more likely to accept outcomes and keep cooperating,~\cite{tyler1992relational} which in networks sustains the very connections on which future decisions rely.
Fairness must also be assessed across sequences of decisions rather than single instances, since small disparities accumulate.~\cite{d2020fairness,si2022enabling}
And fairness is only one of the goals stakeholders hold, alongside well-being, autonomy, and other moral claims.~\cite{mepham2000framework,oneil2020near}
Because these goals interact, and a decision that serves one stakeholder may harm another, network fairness must be approached as a \textbf{multi-stakeholder}, \textbf{multi-goal} problem in which competing interests are weighed rather than optimized one at a time.~\cite{kaya2025mapping,farnadi2018fairness,farnadi2019declarative}

\subsection*{Topological structure of fair networks}
The framework introduced in this Perspective raises a question that precedes both algorithms and outcomes: what would a \textit{fair network} look like?
Rather than correcting decisions after they are made or adjusting the algorithms that produce them, one can ask whether the social structures that feed into those decisions could themselves be designed, or reshaped, to be fair.
This shifts the focus from fairness of decisions to fairness of the relational substrate on which decisions depend.

Answering this question is far from straightforward, because there is no single structural property that makes a network fair.
One intuitive assumption is that fairness is equivalent to \textit{balance}: if no demographic group is disproportionately central, well-connected, or large relative to another, then the network should not encode a structural advantage that downstream processes can inherit.
Under this assumption, a fair network would be one in which groups are similarly represented in size, connectivity, and influence.
Much of the existing literature on network structure and bias operates within this logic, identifying imbalances and proposing ways to correct them: redistributing connections from over-connected groups to under-connected ones,~\cite{neuhauser2023improving,wang2020early,balepur2024intervening} increasing the representation of minorities through quotas or targeted recruitment,~\cite{bastarrica2018affirmative,gomes2019class} or promoting cross-group ties that bridge segregated communities.~\cite{livan2019don,barnes2025edge,chetty2022social2}

Another assumption is that fairness is equivalent to \textit{randomness}: if a network were wired entirely at random, no group would enjoy a systematic structural advantage.
Random graphs such as Erd\H{o}s–R\'{e}nyi models~\cite{erdHos1960evolution} are often invoked as benchmarks of neutrality, but randomness does not translate to fairness.~\cite{bower2022random}
A randomly wired network erases the meaningful structure (mentorship, trust, collaboration) that makes social capital valuable in the first place.
It would be ``neutral'' only in the trivial sense that nobody benefits, which is not the same as everyone benefiting equitably.
A perfectly balanced or neutral network does not guarantee meritocratic outcomes either,~\cite{borondo2014each} because equal starting conditions do not ensure equitable equilibria once agents act strategically.~\cite{dwork2024equilibria}

If neither balance nor randomness suffices, what does?
From a normative perspective, a fair network would be one intentionally designed to promote egalitarian principles: mitigating inherited socio-economic inequalities, ensuring that every member has meaningful access to information, endorsement, and opportunity, and fostering conditions under which effort and talent, rather than positional privilege, determine outcomes.~\cite{borondo2014each,cordelli2015distributive,balepur2024intervening}
Designing such a network is challenging: it entails costs and trade-offs for both those seeking to build social capital and those already well connected, and strategic link formation can produce unintended consequences even under well-intentioned designs.~\cite{dwork2024equilibria}
When organic change is difficult to achieve, algorithmic tools can assist by recommending missing links, suggesting connections that would improve structural equity, or simulating the downstream effects of proposed rewiring.~\cite{tsioutsiouliklis2022link,current2022fairegm,becker2023improving}
In this sense, deliberate network design can serve as a pre-processing approach to fairness, complementing in-processing and post-processing strategies,~\cite{mehrabi2021survey} and embedding relational equity into the very substrate from which decisions are drawn.

\begin{figure}
    \centering
    \includegraphics[width=1.0\linewidth]{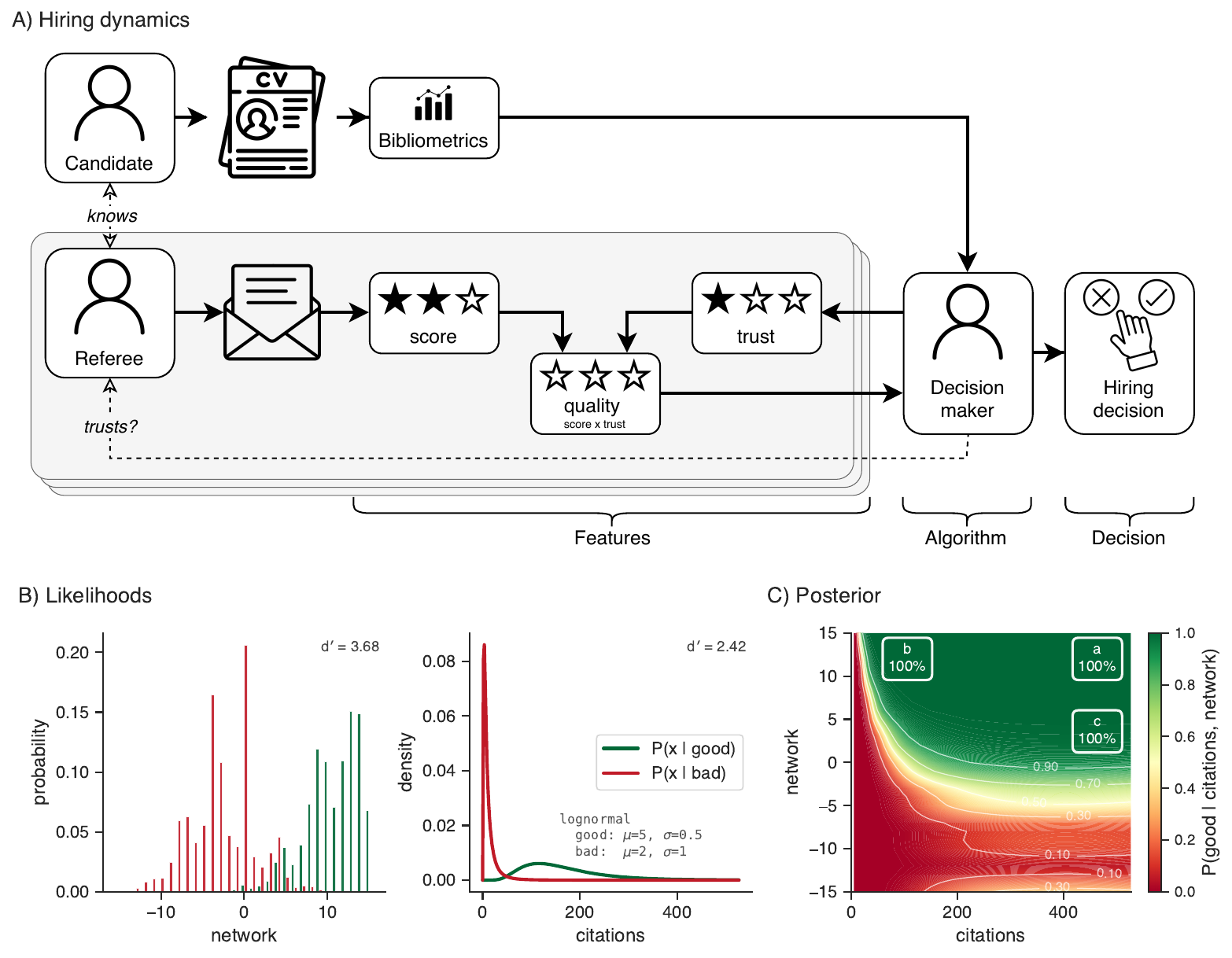}
    \caption{
    \textbf{Network signals can override bibliometrics in academic hiring decisions.}
    \textbf{A)} Schematic of the academic hiring process. 
    Candidates are evaluated through bibliometric indicators (e.g., citations) and recommendation letters, where each letter is weighted by the hiring committee's trust in the referee. 
    Together, these observable features serve as proxies for latent candidate fit and are combined by a decision algorithm to produce a hiring outcome.
    \textbf{B)} Likelihood distributions of network and citation signals for good and bad candidates (derived from a Bayesian model;~\Cref{app:sec:hiring}). Network signals provide stronger discriminability between candidate types ($d' = 3.68$) than citation counts ($d' = 2.42$), suggesting that social capital is a more informative cue than bibliometric output in this model.
    \textbf{C)} Posterior probability of being a good candidate as a function of both signals. 
    Contour lines indicate constant posterior levels. 
    Highlighted regions correspond to candidates achieving similarly high posterior probabilities through different signal combinations: 
    (a) strong network and bibliometric signals, 
    (b) strong network but weak bibliometric signal, and 
    (c) strong bibliometric but moderate network signal. 
    Region (b) illustrates that network signals can override weak bibliometric performance, while region (c) shows the opposite. 
    Overall, these regions demonstrate that a high probability of candidate fit can arise from trade-offs across signal dimensions, not only from strength in both.
    }
    \label{fig:hiring_case}
\end{figure}

\section*{Use case: Hiring in Academia}
\label{sec:hiring}

\emph{Network fairness} asks whether decisions are fair when they rely on network-derived signals that are unequally distributed. 
This question is particularly salient in hiring (\Cref{fig:hiring_case}A), where committees evaluate candidates under uncertainty and often treat referrals as signals of competence or fit. 
Referral quality can be predictive, but access to strong endorsements reflects network position, creating ``friend-of-a-friend'' or ``chaperone'' advantages.~\cite{shwed2014referrals}
Hiring decisions also rely on non-network metrics such as h-index, publications, or citations, commonly interpreted as merit~\cite{reymert2021bibliometrics} but shaped by cumulative ``rich-get-richer'' dynamics.~\cite{merton1968matthew,kong2022influence}

\para{Operationalizing referrals.} %
We model hiring as a classification problem under uncertainty. 
Each candidate has a \emph{latent} type in construct space---good or bad fit---that the committee cannot directly observe.
What they observe are two noisy proxies: network and non-network signals (\Cref{fig:hiring_case}A). 
The network signal aggregates recommendation scores weighted by the trust placed in each referee. 
The non-network signal is a bibliometric indicator drawn from the CV, such as the number of received citations. 
Good candidates tend to attract stronger endorsements and higher citation counts (\Cref{fig:hiring_case}B), but for each signal the distributions of good and bad candidates may overlap
(see~\Cref{app:sec:hiring} for model details and additional non-network signals). %
The question, then, is whether candidates with the same latent fit, get similar opportunities to be considered for hiring?

\para{The top-right trap.}
Hiring committees face two errors: selecting a poor fit (false positive) or passing over an excellent candidate (false negative). 
Which matters more depends on context---sometimes a bad hire is costlier;~\cite{laurano2015true} sometimes missing talent is the greater loss.~\cite{acharya2020rational} 
In practice, committees often mitigate this uncertainty by requiring candidates to be strong on \emph{both} dimensions: the ``top-right'' corner of the referral--competence plane, a pattern broadly echoed in our focus groups (\Cref{app:sec:focus}), reflecting the rich-get-richer effect in hiring.~\cite{conroy2021rethinking}
The posterior probability of being a good fit, however, tells a different story (\Cref{fig:hiring_case}C). 
Candidates who score high on both dimensions (region~\textbf{a}) are, as expected, very likely a good fit. 
Yet candidates with strong referrals and weak citations (region~\textbf{b}), or strong citations and moderate referrals (region~\textbf{c}), also reach the same high posterior probability. 
Consequently, restricting attention to the top-right corner does not only filters out bad candidates but also excludes many who would have been successful hires.
This raises a fairness tension.
With respect to the observed signals, the rule treats similar profiles similarly, consistent with individual fairness. With respect to the posterior---the best available estimate of latent fit---candidates with identical posteriors receive different outcomes depending on which signal drives their score. Candidates equally likely to be good fits are thus not offered the same opportunities.

\para{How many do we miss?}
To quantify the extent of that exclusion, we simulate $1{,}000$~candidates per signal pair. 
We draw each candidate's latent type (good or bad fit) with a prior of $35\%$~good hires, then generate network and bibliometric signals from the likelihoods in~\Cref{fig:hiring_case}B. 
We then compare two screening rules: the posterior boundary ($p \geq 0.5$), which defines the pool of candidates the model considers more likely good than bad (\Cref{fig:hiring_case}C), and a naive top-right rule that requires both signals to exceed a high threshold simultaneously (\Cref{fig:hiring_case}C, region (a)).
The posterior boundary admits $350$ candidates into the pool---consistent with the prior. 
The top-right rule admits only~$4$.
Across all three bibliometric specifications (citations, publications, h-index), $\approx 99\%$ of candidates whom the combined evidence identifies as likely good hires are \emph{excluded} by the top-right heuristic (\Cref{app:tbl:counts}).
Of course, only one candidate is typically hired in the end. 
But the boundary determines who gets the \emph{opportunity} to be considered---the pool from which the final selection is made. 
Two candidates with the same latent fit (same posterior probability) may find themselves on opposite sides of a top-right threshold simply because their observable strengths lie on different dimensions. 
The top-right rule thus creates disparities in opportunity among equally qualified candidates, driven not by their actual fit but by the particular combination of signals they happen to exhibit (see~\Cref{app:sec:simulation} for more details).

\para{Implications.}
The posterior contours (\Cref{fig:hiring_case}C) suggest a straightforward improvement: allow strength on one dimension to compensate for the other, rather than only requiring both to be high. 
Whether we allow such trade-offs is itself a fairness question.
Women and racial minorities often have less access to the professional networks that generate strong endorsements,~\cite{shwed2014referrals,ibarra1992homophily} even when their non-network credentials are strong. %
Similarly, highly capable autistic individuals may excel on measurable skills yet lack the social ties that produce strong referrals.~\cite{bolick2008takes, crespi2016autism} 
A rule that demands both dimensions be high penalizes precisely those candidates, even when they would be good fits.
The same logic applies beyond hiring. 
Whenever outcomes depend on a mix of network and non-network signals---whom we cite,~\cite{merton1968matthew} whom we review favorably,~\cite{brainard2022reviewers} how talent is scouted in ballet~\cite{herrera2023quantifying}---rigid thresholds on each axis risk compounding existing inequalities. 
In some domains the asymmetry is more extreme: in politics, for instance, network position can override observable merit almost entirely,~\cite{lee1994friend,robertson1999corruption} leaving little room for non-network signals to compensate. 
Across all these settings, posterior probabilities can reveal which candidates are unfairly excluded under notions of individual \textit{network} fairness,~\cite{sium2024individual} but they do not tell decision-makers how much those mistakes actually cost. 
To answer that question, one must weigh the payoffs of correct and incorrect decisions, allowing institutions to calibrate the trade-off between the risk of a bad hire and the cost of missed talent---an approach we illustrate in~\Cref{app:sec:payoff}.

\subsection*{Ethical conundrum}
The hiring example exposes a broader challenge: once we acknowledge that networks shape who would be hired, several difficult questions follow. In our focus groups (\Cref{app:sec:focus}), the same candidate profile elicited strikingly different judgments---some evaluators valued effort as a sign of perseverance, while others read it as a lack of  talent; some weighted referrals heavily, others dismissed them as noise. These disagreements are not idiosyncratic; they reflect genuine ethical tensions that a network fairness perspective makes explicit. We outline three paradoxes and the normative tensions they create.

\para{The paradox of social capital.}
Social capital is often treated as either an asset or a bias, but its dual role creates a persistent tension. Individuals with strong networks may benefit from trust, visibility, and support---but are they successful because of their networks, or do they attract strong networks because they are good and successful? This chicken-and-egg dynamic makes it difficult to determine whether outcomes reflect individual merit or inherited structural advantage. Network-based decisions can lead to both false positives---less competent individuals who benefit from strong connections---and false negatives---highly qualified individuals excluded due to limited access or visibility. %

\para{The paradox of effort.}
  In many decision-making processes, effort is assumed to signal moral deservingness. Yet not all effort results in reward, and not all rewards reflect effort. Social networks can amplify or suppress the returns on individual effort depending on one's position or embeddedness. Research shows that people are more likely to value effort when they see it being consistently rewarded,~\cite{inzlicht2018effort, lin2024effort} and that perceived fairness in compensation can influence motivation.~\cite{cohn2015fair} However, when effort is disconnected from outcomes---either because of structural exclusion or biased reward systems---people may disengage. Moreover, effort can sometimes be perceived negatively. For example, when two individuals achieve the same result, the one who had to exert more effort may be seen as less naturally capable, raising doubts about future performance.~\cite{nicholls1976effort} In networked settings, this becomes more complex, as individuals with similar effort may receive very different outcomes depending on who supports or recognizes them. 
  This disparity is compounded by the difficulty of measuring effort itself, which raises both ethical and methodological challenges. 
  Signals of effort can already be shaped by social capital. In academic settings, for instance, effort might be proxied by the number of extracurricular activities such as internships, which themselves depend on access to mentorship, collaborations, and visibility within influential networks. As a result, what looks like effort may instead reflect underlying relational advantage.

\para{The paradox of meritocracy.}
Merit is often treated as an objective basis for fair evaluation, but it is socially constructed and context-dependent. Measures of merit such as job fit, productivity, or credentials may also embed bias when they are influenced by social capital. A candidate may appear more ``qualified'' not because of ability alone, but because of accumulated advantages such as mentorship, elite networks, or access to informal endorsements.~\cite{freyer2022inherited} Guaranteeing a meritocratic system depends on whether opportunities to demonstrate merit are equally accessible.~\cite{borondo2014each} Otherwise, merit becomes a reflection of prior privilege or connectedness rather than personal achievement.

\para{Normative dilemmas.}
These paradoxes give rise to difficult normative questions. Should systems reward effort, even if it does not always result in high performance? How can effort be measured, given that outcomes are often shaped by luck, access, and context? Are network signals always %
unfair? 
While Rawls argues for rewarding individuals based on effort rather than inherited advantages or social circumstances,~\cite{alexander1985fair} empirical evidence complicates this view as the boundary between them is not always clear and can itself be contested.
For instance, in some regions ``education'' is considered more a result of personal decisions and, hence, ``earned,'' whereas in other regions it may be considered more a matter of social factors or destiny.~\cite{baker2005national,brown2016credentials}
Therefore, context plays a critical role in how fairness is interpreted. Research from economic sociology shows that the use of social networks in hiring is not inherently unfair, but its acceptability depends on institutional norms and sector-specific expectations.~\cite{chua2021economic} In sectors like public administration and health care, networks are viewed with suspicion due to meritocratic ideals. In contrast, in industries like construction or retail, referrals and connections are seen as a practical necessity. 
This variation reminds us that fairness is not only a technical property of algorithms, but also a %
value judgment shaped by social norms and institutional contexts.

\section*{Outlook}

This perspective argues that fairness in decision-making systems cannot be fully captured by definitions focused only on individual attributes or outcome distributions. 
By placing relationships, social capital, and structural position at the center of analysis, we have shown how seemingly neutral decisions can reinforce hidden inequalities. 
Moving forward, advancing network fairness requires coordinated efforts across conceptual, institutional, and methodological domains.

\para{Placing networks at the core of fairness.}
Personal networks shape who is visible, who is trusted, and ultimately who is selected for opportunities, with effects that depend on the number of connections, their strength, quality, and position. 
Ignoring these dynamics risks addressing symptoms while leaving structural sources of inequality intact. 
Fairness research must therefore treat networks and relations alongside individual attributes as first-class objects of analysis, recognizing that fairness depends on how individuals are embedded in social structures.

\para{A multi-stakeholder, multi-goal approach.}
Fairness evaluations should consider the perspectives of all stakeholders involved in the decision-making process---not only the individuals for whom decisions are made, but also decision-makers, institutions, and the broader communities affected by systemic outcomes. 
Perceptions of fairness often depend on who makes the decision and who bears its consequences,~\cite{messick1979fairness, nguyen2024definitions} and including multiple stakeholders helps surface conflicts and trade-offs that may remain invisible from a purely algorithmic perspective.~\cite{yurrita2022towards} 
Beyond stakeholders, future work must also account for the multiple moral and normative goals that societies strive to achieve. 
Fairness is rarely the only value at stake---autonomy, well-being, and the common good are equally critical considerations.~\cite{europe2019guidelines, bell2023possibility, hu2024achieving} 
For example, autonomy may involve giving individuals meaningful control over how they are represented in decision-making systems, while allowing decision-makers to account for context-specific constraints---goals that fairness alone does not necessarily address. 
Although satisfying multiple fairness criteria simultaneously is often challenging, and in some cases theoretically impossible,~\cite{kleinberg2016inherent} recent work shows that relaxing or approximating fairness guarantees can make it possible to satisfy multiple constraints in parallel.~\cite{bell2023possibility,zehlike2025beyond} 
This suggests that fairness should be understood not as the strict satisfaction of a single criterion, but as a balance across competing objectives, where stakeholders may achieve full satisfaction of some goals or partial gains across several. 
Importantly, such trade-offs can generate tensions: promoting autonomy for one stakeholder---such as a decision-maker exercising discretion---may reduce perceived fairness for another, such as an applicant disadvantaged by that discretion.
Addressing fairness in networks therefore requires not only identifying overlapping goals, but also designing mechanisms for managing moral trade-offs and value conflicts. 
Ultimately, responsible decision-making in networks is not only a technical challenge but a social and ethical one, and solving it demands tools and frameworks that reflect that complexity.

\para{From distributive to procedural justice.}
Much of the algorithmic fairness literature focuses on distributive justice, assessing whether outcomes such as hiring decisions, loan approvals, or recommendations are allocated equitably across groups or individuals. 
While essential, this framing %
alone is insufficient. 
Theories of \textit{procedural justice} emphasize that fairness also requires transparency and legitimacy in decision processes, not only in their outcomes.~\cite{lind1992procedural} 
In networked settings, decisions may appear distributively fair at a given time but conceal procedural failures, such as opaque criteria, biased feedback loops, or reliance on features that encode structural privilege rather than genuine qualification. 
Fairness is therefore inherently dynamic in systems where networks evolve and decisions accumulate, and evaluating it at a single point risks conflating luck with merit and overlooking systemic patterns that emerge within multi-stage decisions and across repeated interactions over time.~\cite{si2022enabling, d2020fairness}
Moving forward requires shifting from outcome-based assessments to a more holistic view of fairness as a property of the entire decision process. 
This entails scrutinizing the features and feedback mechanisms that shape decisions and ensuring transparency and equity at each stage, from who is visible to how decisions are ultimately made.

\subsection*{Interventions toward network fairness}

These conceptual shifts call for interventions that reshape how opportunities are created and allocated in networked systems at three levels: relational, institutional, and methodological.

\para{Relational interventions: Building social capital.}
Fairness interventions must go beyond redistributive measures such as quotas or affirmative action, which, while important, are often insufficient on their own.~\cite{corradi2025admission}
Without addressing how groups interact and gain access to influence, such measures may fail to reduce power imbalances or increase representation in positions of authority. 
Evidence shows that combining redistribution with behavioral changes---such as forming ties to influential actors---can improve the visibility of under-represented groups in top-ranking positions.~\cite{neuhauser2023improving}
Effective interventions therefore require \textit{policy cocktails} that combine redistributive aims with deliberate efforts to reshape the relational mechanisms that sustain advantage.~\cite{jackson2021inequality}
A central strategy is to actively build social capital before high-stakes decisions are made, by fostering ties of support, trust, and information that shape future opportunities.~\cite{jackson2021inequality, mishra2020social}
At the \textit{individual} level, social capital depends less on personality than on context: access to environments that enable meaningful interaction and openness to new connections.~\cite{ferrazzi2014never, small2019role}
At the \textit{organizational} level, institutions and influential actors can act as brokers by funding exchanges, supporting mentorship, and creating opportunities for interaction across otherwise disconnected groups.~\cite{small2009unanticipated, helliwell2006well,cordelli2015distributive, gilani2020creating,bachmann2026cumulative}

\para{Institutional interventions: Data and accountability.}
A key step is to collect data systematically and responsibly. 
Institutions should implement structured decision-making pipelines and record each stage of their processes. 
This enables audits that evaluate procedures and identify structural biases, including unconscious biases~\cite{greenwald1995implicit} and the network effects discussed in this Perspective, which often shape decisions but remain unobserved.~\cite{dies2025forecasting}
Once such data are available, they can be used to audit past decisions. 
For instance, historical hiring data can be analyzed using models such as the one presented in~\Cref{app:sec:hiring} to assess whether past decisions were fair and to track false positives over time, allowing institutions to update payoffs and adjust decision boundaries accordingly.

\para{Methodological interventions: Formalizing network effects and fairness.}
The central challenge is to formalize how network effects enter fairness assessments in a principled and measurable way. 
This requires defining decision boundaries that integrate network and non-network signals, quantifying how these signals shape error and bias, and updating these boundaries as new evidence becomes available. 
A first step is to ground these models in social science, network science, and complex systems by incorporating structural network properties such as tie strength, position, and community structure into the modeling of opportunity and access.
Building on this foundation, methodological frameworks are needed to reason under competing objectives and uncertainty. 
Multi-criteria decision analysis provides a basis for evaluating trade-offs across stakeholders and goals,~\cite{greco2016multiple, wu2022multi} while multidimensional Bayesian approaches enable principled updating under uncertainty.~\cite{goldner2025multidimensional}
Complementary perspectives from evolutionary game theory capture how cooperative or discriminatory norms emerge and stabilize among interacting agents,~\cite{samson2018multi,hilbe2014cooperation,salahshour2025perceptual} and fuzzy logic extends these approaches to settings where criteria are inherently ambiguous and cannot be expressed through sharp thresholds.~\cite{loor2017refocusing, tapia2016fusion, aplak2013fuzzy}
Together, these approaches support a shift from single-objective optimization toward multi-objective, multi-stakeholder reasoning that reflects the relational nature of decision-making.
The goal is not only to adjust outcomes, but to make explicit how network structures shape visibility, evaluation, and selection, and to provide tools to redesign these processes in a transparent and accountable way.

In summary, human judgment, classical machine learning, and today's generative AI all draw on relational signals, and all are therefore exposed to the same network biases, albeit with increasing scale and decreasing transparency. 
This perspective has shown that fairness cannot be reduced to observable attributes because the proxies on which decisions are built are themselves products of network structure, whichever worldview one adopts. 
We have organized this influence into ten network effects spanning individual nodes, their neighborhoods, and the wider structure. 
We call on researchers and practitioners to identify the domains where these effects are most consequential, to quantify their impact on decision outcomes, and to develop actionable interventions at the relational, institutional, and methodological levels.

\section*{Acknowledgments}
The authors thank danah boyd, Christian Hilbe, Cynthia Dwork, Matthew O. Jackson, Cris Moore, Hungtang Ko, Pedro Márquez-Zacarías, Aastha Nath, János Kertész, Elisa Omodei, Gerardo Iñiguez, Vito Servedio, Kjartan van Driel, Ana M. Jaramillo, Rafael Prieto-Curiel, Reinhard Munz, the ANETI Lab in Budapest, the participants of the Ethics in Complex Systems Workshop at the University of Zürich (2025), the participants of the Computer Science Talks \#15 at the Graz University of Technology (2025), and the participants of the AI-GAP workshop held in L'Aquila (2023) for their fruitful feedback and discussions.

\section*{Funding}
L.E.N. acknowledges support from the Vienna Science and Technology Fund (WWTF) under project no. ICT20-07 and from the Austrian Research Promotion Agency (FFG) under project no. 873927 (ESSENCSE).

\section*{Author contributions}
L.E.-N. and F.K. conceptualized the work. L.E.-N. conducted the simulations and focus groups and led the writing of the manuscript. F.K. supervised the work. T.E.-R., R.B.-Y., and S.V. contributed to the development of the network fairness perspective, P.-H.W. and E.P. to the ethical and procedural justice components, and M.Z. to the algorithmic fairness components. All authors contributed to the interpretation and development of the arguments and reviewed and edited the manuscript.

\section*{Declaration of generative AI use}
During the preparation of this manuscript, the authors used large language models---Claude Opus 5 and Claude Fable 5.1 (Anthropic) and GPT-5.5 (OpenAI)---to proofread the final draft for grammar, spelling and clarity. These tools were not used to generate scientific content, analyze data, produce figures or draft substantive text, and no AI tool is listed as an author. The authors reviewed and edited all suggested changes and take full responsibility for the content of this publication.

\section*{Competing interests}
The authors declare no competing interests. 

\section*{Publisher's note}
Springer Nature remains neutral with regard to jurisdictional claims in published maps and institutional affiliations.

\newpage
\section*{Supplementary information}

{\raggedright
  {\large\bfseries From Network Inequality to Network Fairness:~A Perspective on Responsible Decision-Making\par}
  \vspace{0.4em}
  {\small Lisette Espín-Noboa, Tina Eliassi-Rad, Pak-Hang Wong, Erich Prem, Meike Zehlike, Ricardo Baeza-Yates, Suresh Venkatasubramanian, and Fariba Karimi\par}
}
\vspace{1.5em}

\appendix
\numberwithin{figure}{section}
\numberwithin{table}{section}

\renewcommand{\thesubfigure}{\Alph{subfigure}}
\captionsetup[subfigure]{labelformat=brace, labelfont=bf}

\section{Hiring scenario}
\label{app:sec:hiring}

In academic hiring, candidates are evaluated on two types of observable 
features: \emph{network} signals (e.g., referrals) and \emph{non-network} 
signals (e.g., citation or publication counts). A natural but flawed response is to require both to be high simultaneously---the ``top-right trap.'' 
Because both signals are noisy proxies of latent candidate fit, this rule is doubly unreliable: it violates individual fairness by giving equally qualified candidates different outcomes depending on which signal happens to be stronger, and it carries an institutional cost by overlooking many strong candidates who excel on one dimension but not both. 
Alternative decision rules---grounded in a Bayesian posterior over candidate fit---allow one signal to compensate for the other without compromising the quality of the hire. 
We derive these boundaries analytically and evaluate them 
on a simulated candidate pool, providing the formal setup underlying the main text findings.

\subsection{Observable features}

We distinguish two types of observables. 
\emph{Network} features arise from who is connected to whom or who vouches for whom.
\emph{Non-network} features are measurable quantities that do not depend on network structure. 
Both can be informative about the latent quality of the candidate (good vs. bad fit), which is unobservable at the time of the decision.

\para{Network:}
In academic hiring, reference letters are the primary source of network-derived information. 
Each letter conveys two signals: what the referee says about the candidate (\emph{reference score}) and how much the decision-maker trusts the referee (\emph{referee trust}). 
We define the network signal as \emph{referral quality}: the product of reference score and referee trust, aggregated across three letters. 
The multiplicative form reflects a natural asymmetry: a glowing recommendation carries weight only if the referee is known and credible to the evaluator; the same letter from an unknown referee is far less informative.

\para{Non-network:}
We use three bibliometrics as non-network signals: h-index, publication count, and citation count. 
These metrics are widely used in academic hiring and are right-skewed or count-like; we model them with distributional forms that match empirical and theoretical evidence (power-law, Poisson, log-normal) as detailed below.

\subsection{Bayesian model}
\label{app:sec:model}

We model the hiring decision as Bayesian inference over a latent binary 
state---good or bad candidate fit---given two observable signals. 
We assume that, given candidate fit, the network and non-network signals are conditionally independent: once latent quality is known, a candidate's 
referral quality carries no additional information about their bibliometric 
record. 
This allows all combinations to occur---a candidate with few citations may have strong or weak referrals, and so may a highly cited one---which is 
precisely what enables signal compensation in the model. 
We adopt this assumption to keep the posterior closed-form; the framework can accommodate dependence with a joint likelihood.

Given the observables $\mathrm{net}$ (the network signal, i.e., weighted referral) and $\mathrm{non}$ (the non-network signal, i.e., a bibliometric), the posterior probability that a candidate is a good fit is given by Bayes' rule (prior $\times$ likelihood $/$ evidence):
\begin{equation}
\label{eq:bayes-rule}
P(\text{good} \mid {\mathrm{net}}, {\mathrm{non}})
= \frac{ P(\text{good})\, P({\mathrm{net}}, {\mathrm{non}} \mid \text{good}) }{ P({\mathrm{net}}, {\mathrm{non}}) },
\end{equation}
where 
\begin{equation}
\label{eq:evidence}
P({\mathrm{net}}, {\mathrm{non}}) = P(\text{good})\, P({\mathrm{net}}, {\mathrm{non}} \mid \text{good}) + P(\text{bad})\, P({\mathrm{net}}, {\mathrm{non}} \mid \text{bad}).
\end{equation}
Under conditional independence, the joint likelihoods factorize and the 
posterior simplifies to
\begin{equation}
\label{eq:posterior-independent}
P(\text{good} \mid {\mathrm{net}}, {\mathrm{non}})
= \frac{ P(\text{good})\, P({\mathrm{net}} \mid \text{good})\, P({\mathrm{non}} \mid \text{good}) }
       { P(\text{good})\, P({\mathrm{net}} \mid \text{good})\, P({\mathrm{non}} \mid \text{good}) 
       + P(\text{bad})\, P({\mathrm{net}} \mid \text{bad})\, P({\mathrm{non}} \mid \text{bad}) }.
\end{equation}
The \emph{prior} $P(\text{good})$ encodes the baseline probability that a 
randomly drawn candidate is a good fit before observing any signals.
The \emph{likelihoods} describe how probable the observed signal values are given candidate fit. 
The natural decision rule derived from this posterior is to hire whenever a candidate is more likely good than bad---that is, whenever $P(\text{good} \mid \mathrm{net}, \mathrm{non}) \geq 0.5$. 
Next, we describe the likelihoods and parameter choices before examining when and why this threshold should be adjusted.

\begin{table}[t]
\centering
\caption{
\textbf{Simulation parameters for the Bayesian hiring model.}
Parameter values used throughout all simulations. The prior $P(good) = 0.35$ reflects an assumption that good candidates are a minority of the applicant pool. Payoffs are defined asymmetrically: missing a good candidate (reject good: $−100$) is penalized twice as heavily as hiring a bad one (hire bad: $−50$), while correct decisions carry no cost. %
 The network signal is computed as the sum of $score \times trust$ across three recommendation letters, 
 where reference scores and trust weights follow stylized discrete distributions that differ markedly between good and bad candidates.  
 Bibliometric signals are modeled using standard count distributions: citation counts follow log-normal distributions (capturing the heavy-tailed nature of citation data), publication counts follow Poisson distributions, and h-index values follow truncated power laws---all with parameters chosen to produce realistic but clearly separated distributions between good and bad candidates.
}
\label{tbl:si-params}
\small
\begin{tabular}{@{}lll@{}}
  \toprule
  \textbf{Component} & \textbf{Parameter} & \textbf{Value} \\
  \midrule
  Prior & $P(\text{good})$ & 0.35 \\
  \midrule
  Payoffs & hire good or reject bad & $0$ \\
   &  hire bad  & $-50$ \\
   &  reject good & $-100$ \\
  \midrule
  Network: referral quality & aggregation & sum over 3 letters \\
  & reference score (numeric) & excellent $5$, good $4$, average $3$, bad $2$, very\_bad $1$ \\
  & referee trust (numeric) & trustworthy $1$, unknown $0$, untrustworthy $-1$ \\
  & $P(\text{ref.\ score} \mid \text{good})$  & excellent 0.48, good 0.35, average 0.10, bad 0.05, very\_bad 0.02 \\
  & $P(\text{ref.\ score} \mid \text{bad})$  & excellent 0.10, good 0.47, average 0.30, bad 0.10, very\_bad 0.03 \\
  & $P(\text{trust} \mid \text{good})$  & trustworthy 0.85, unknown 0.14, untrustworthy 0.01 \\
  & $P(\text{trust} \mid \text{bad})$  & trustworthy 0.10, unknown 0.55, untrustworthy 0.35 \\
  \midrule
  Non-network: h-index & power law (good) & $\alpha=1.8$, $x_{\min}=1$, $x_{\max}=150$ \\
  & power law (bad) & $\alpha=2.8$, $x_{\min}=1$, $x_{\max}=150$ \\
  \midrule
  Non-network: n\_publications & Poisson (good) & $\lambda=20$ \\
  & Poisson (bad) & $\lambda=5$ \\
  \midrule
  Non-network: n\_citations & log-normal (good) & $\mu=5.0$, $\sigma=0.5$ \\
  & log-normal (bad) & $\mu=2.0$, $\sigma=1.0$ \\
  \bottomrule
\end{tabular}
\end{table}

\begin{figure}[t]
    \centering
    \includegraphics[width=1.0\linewidth]{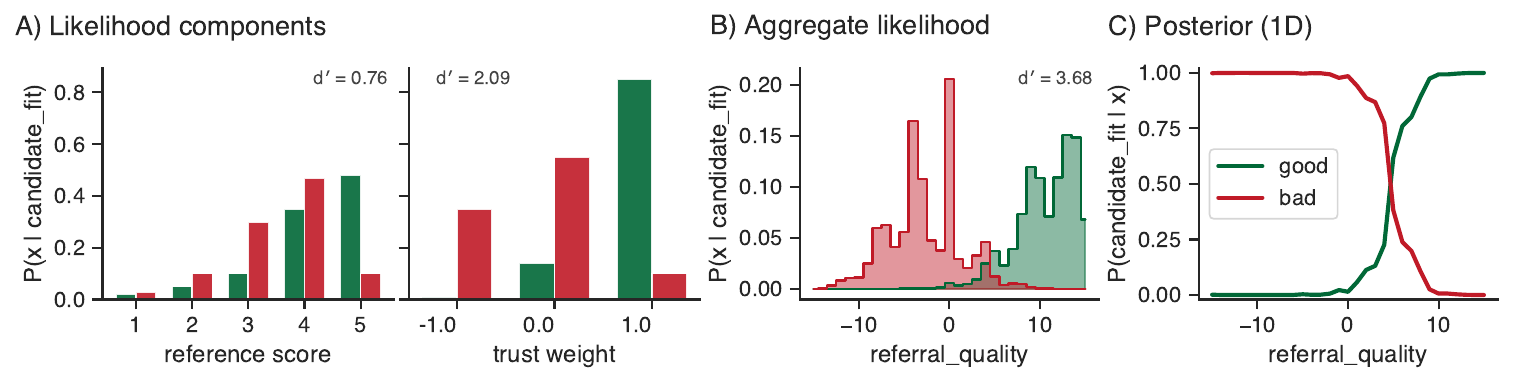}
    \caption{
    \textbf{Generative model of the network signal and its discriminability between good and bad candidates.}
    \textbf{A)}~Likelihood distributions of the two components of the network signal, conditioned on candidate fit. Reference score (left) ranges from very bad (1) to excellent (5); trust weight (right) ranges from untrustworthy (−1) to trustworthy (1). Trust weight is a more discriminative signal than reference score alone ($d' = 2.09$ vs. $0.76$, respectively).
    \textbf{B)}~Likelihood of the aggregate network signal, computed as $\sum_i \text{score}_i \times \text{trust}_i$ across three recommendation letters. Combining both components substantially improves discriminability ($d' = 3.68$). 
    \textbf{C)}~Posterior probability $P(\text{good} \mid x)$ given the aggregate network signal alone, using prior $P(\text{good})=0.35$. 
    The steep sigmoidal curves confirm that the aggregate signal is a highly informative cue for candidate quality, and forms the basis of the network signal used throughout the paper.
    }
    \label{app:fig:network_signals}
\end{figure}

\begin{figure}[t]
    \centering
    \includegraphics[width=1.0\linewidth]{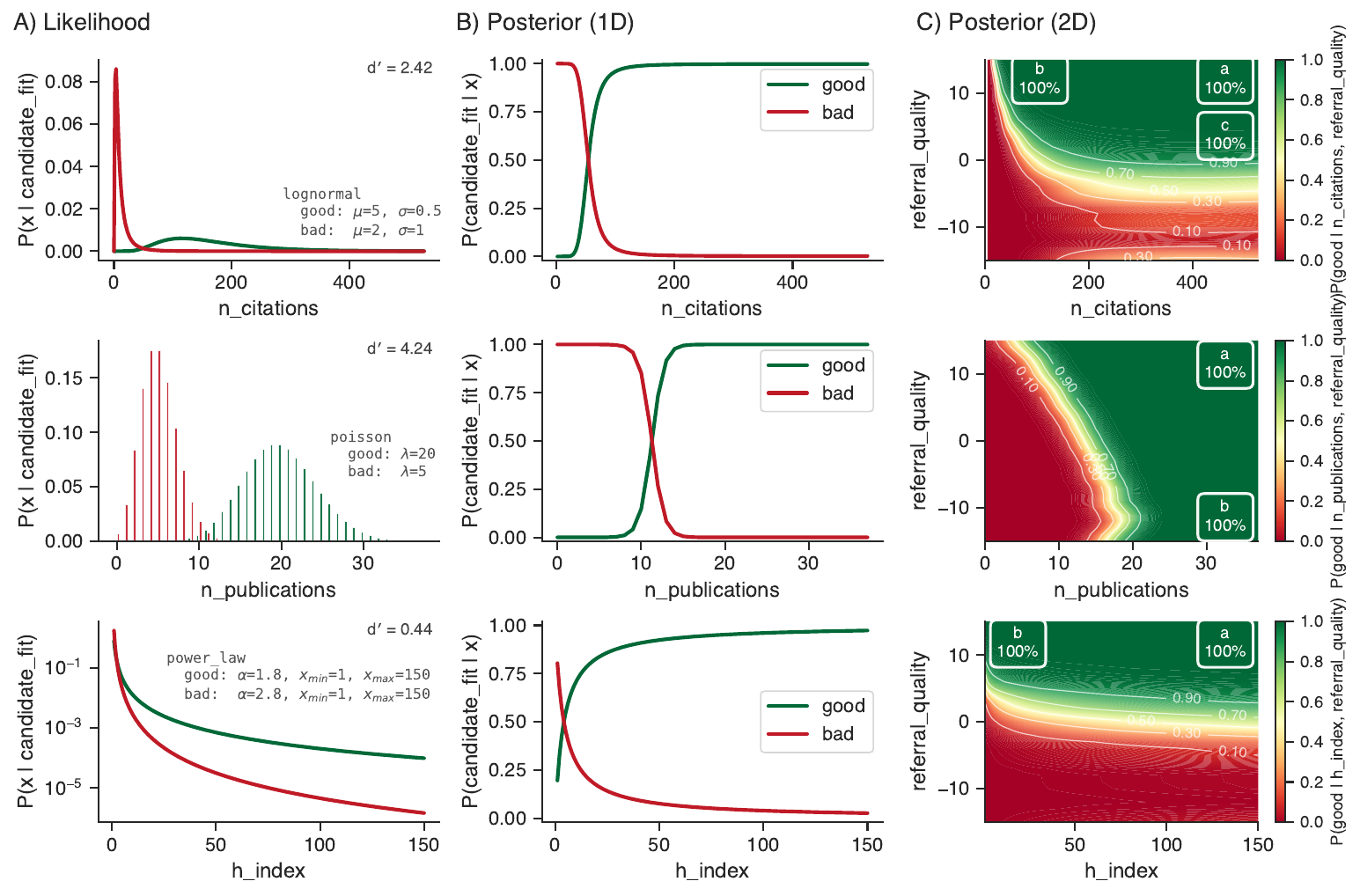}
    \caption{
    \textbf{Generative model of the non-network signal and posterior distributions.}
    Each row corresponds to a different bibliometric: number of citations (log-normal, top---used in the main text),
    number of publications (Poisson, middle), 
    and h-index (power law, bottom).
    \textbf{A)}~Likelihood $P(\text{x} \mid \text{good/bad})$; good candidates tend toward higher values in all three metrics.
    \textbf{B)}~One-dimensional posterior $P(\text{good} \mid \text{x})$ using only the non-network signal, with prior $P(\text{good})=0.35$.
    \textbf{C)}~Two-dimensional posterior $P(\text{good} \mid \text{network}, \text{non-network})$ combining both signals. Contours show how a strong network signal can compensate for a weaker bibliometric record, and vice versa.
    The network signal model is the same as in~\Cref{app:fig:network_signals}.
    }
    \label{app:fig:nonnetwork_likelihoods_posteriors}
\end{figure}

\subsubsection{Likelihoods, distributions, and parameter choices}
\label{app:sec:distributions}

\Cref{tbl:si-params} summarizes all parameters. 
The values are chosen to illustrate the mechanism and keep the exposition transparent; they are not fitted to real hiring data. 
In applied settings, the prior and likelihoods can be estimated from historical data linking observables to outcomes, and payoffs can be set from institutional priorities.

\para{Prior.}
We assume that $35\%$\ of candidates would turn out to be good hires. 
This reflects a selective setting where the proportion of strong fits is modest; the qualitative conclusions hold for other proportions.
 
\para{Network likelihood: referral score, trust, and quality.}
Each candidate receives three reference letters, each with 
a reference score (excellent to very bad) and a referee trust level 
(trustworthy, unknown, untrustworthy). 
Referral quality per letter is $\mathrm{score} \times \mathrm{trust}$; the aggregate network signal is the sum of quality scores over the three letters. 
Good candidates are more likely to receive higher scores from trustworthy referees; bad candidates tend to receive good or average scores from unknown or untrustworthy referees (\Cref{app:fig:network_signals}A). 
Trust is the strongest source of separation. 
Without trust, score distributions overlap considerably ($d'_{\text{score}} = 0.76$ vs.\ $d'_{\text{trust}} = 2.09$), and the aggregate achieves strong separation ($d'_{\text{quality}} = 3.68$), driven largely by trust amplifying differences in reference scores (\Cref{app:fig:network_signals}B).
Throughout, $d'$ denotes the standardized separation
$\lvert \mu_g - \mu_b \rvert / \sqrt{0.5\,(\sigma_g^2 + \sigma_b^2)}$.~\cite{das2021method}

\para{Non-network likelihood: h-index, publications, and citations.}
We use one non-network feature per simulation run---citation count, publication count, or h-index---modeled as log-normal, Poisson, and power-law distributions respectively, with separate parameters for good and bad candidates such that higher values are more probable under good candidates (\Cref{app:fig:nonnetwork_likelihoods_posteriors}A). 
These distributional families are well supported in the literature: log-normal or heavy-tailed distributions for citations,~\cite{radicchi2008universality} Poisson for publication counts,~\cite{xie2020predicting} and Pareto-type scaling for impact measures.~\cite{spearman2010survey}

\subsubsection{Posterior distributions}
\label{app:sec:posterior}

\Cref{app:fig:nonnetwork_likelihoods_posteriors} shows the likelihood (A), 1D posterior (B), and 2D posterior (C) for all three bibliometric metrics. 
The key quantity is the standardized separation index $d'$ (column~A): it determines which signal dominates the 2D posterior. 
For citations ($d' = 2.42$) and h-index ($d' = 0.44$), the non-network signal is less discriminative than the network signal ($d'_{\text{quality}} = 3.68$, see~\Cref{app:fig:network_signals}), which consequently dominates the 2D posterior---the contours in column~C are nearly horizontal, indicating that referral quality drives the hiring decision regardless of bibliometric value. 
For publications ($d' = 4.24$), the situation reverses: the non-network signal is more discriminative than the network signal, the contours become nearly vertical, and bibliometrics drive the posterior. 
This contrast is deliberate---our three specifications span a range from network-dominated to non-network-dominated regimes, allowing us to examine how decision boundaries and the top-right trap behave across settings.

\subsubsection{Payoffs and the payoff-optimal threshold}
\label{app:sec:payoff}

The posterior-based rule ($p \geq 0.5$) treats both types of error symmetrically: it selects a candidate whenever the candidate is more likely good than bad. 
In practice, however, the two errors rarely carry equal weight---missing a strong candidate may be far costlier than hiring a weak one. 
We make this explicit by assigning numerical payoffs to each of the four outcomes. 
Correct decisions carry no cost: $u_{\mathrm{hire,good}} = 0$ and $u_{\mathrm{reject,bad}} = 0$. 
Errors are penalized asymmetrically: $u_{\mathrm{hire,bad}} = -50$ and $u_{\mathrm{reject,good}} = -100$; thus, missing a good candidate is twice as costly as hiring a bad one (\Cref{tbl:si-params}).
 
Let $p = P(\text{good} \mid \mathrm{net}, \mathrm{non})$. 
The expected payoff of hiring is
\begin{equation}
E[\text{payoff} \mid \text{hire}] = p \cdot 0 + (1-p)(-50) = -50 + 50p,
\end{equation}
and of rejecting is
\begin{equation}
E[\text{payoff} \mid \text{reject}] = p(-100) + (1-p) \cdot 0 = -100p.
\end{equation}
Hiring is preferred whenever $E[\text{payoff} \mid \text{hire}] \geq E[\text{payoff} \mid \text{reject}]$, which yields the payoff-optimal threshold:
\begin{equation}
p^* = \frac{u_{\mathrm{reject,bad}} - u_{\mathrm{hire,bad}}}{u_{\mathrm{hire,good}} - u_{\mathrm{hire,bad}} - u_{\mathrm{reject,good}} + u_{\mathrm{reject,bad}}} = \frac{0-(-50)}{0-(-50)-(-100)+0} = \frac{1}{3} \approx 0.33.
\end{equation}
Because $p^* = 0.33 < 0.5$, the payoff-optimal rule is strictly more lenient than the posterior-based rule: for candidates with $p \in [0.33, 0.5)$, the payoff rule hires but the posterior rule rejects. 
The posterior rule is preferable when a symmetric treatment of errors is acceptable; the payoff rule when the costs of false positives and false negatives differ and can be specified.

\subsubsection{Top-right corner heuristic}
\label{app:sec:topright}

Both model-based rules are compensatory, in that a very strong network signal can offset a modest bibliometric record and vice versa, allowing a candidate to cross the hiring threshold on the strength of either signal. 
A third rule removes this compensation and hires only candidates who score highly on \emph{both} signals. 
This top-right heuristic is a conjunctive, non-compensatory rule that requires no probabilistic model and no posterior computation, only a threshold on each signal. 
The rule is intuitively appealing because a candidate who excels on every dimension appears to be the safest choice. 
Requiring both thresholds to be met at once, however, excludes most candidates, including many with high posterior probability. 
A candidate whose strong network signal would be sufficient to offset a moderate bibliometric record under the posterior rule is simply excluded. 
The top-right heuristic is therefore not a conservative approximation of the posterior rule; it is a categorically different criterion that privileges joint strength over probabilistic fitness. 
The simulations below quantify this cost.

\subsection{Comparing decision rules}
\label{app:sec:simulation}
 
The three rules---posterior-based ($p \geq 0.5$), payoff-optimal ($p \geq p^* = 0.33$), and top-right---can be compared analytically in terms of their decision boundaries, but quantifying their performance requires a candidate pool with known true states.

\para{Simulated population.}
For each of $1{,}000$\ candidates, we draw the latent state (good with probability $P(\text{good}) = 0.35$), then draw observables from the corresponding likelihoods.
The network signal is the sum of referral-quality scores across three reference letters, each drawn from $P(\text{ref.\ score} \mid \text{good/bad})$ and $P(\text{trust} \mid \text{good/bad})$.
The non-network signal is drawn from $P(\text{bibliometric} \mid \text{good/bad})$.
We compute the posterior for every candidate and apply each decision rule.

\begin{figure}[ht]
  \centering
  \includegraphics[width=1.\linewidth]{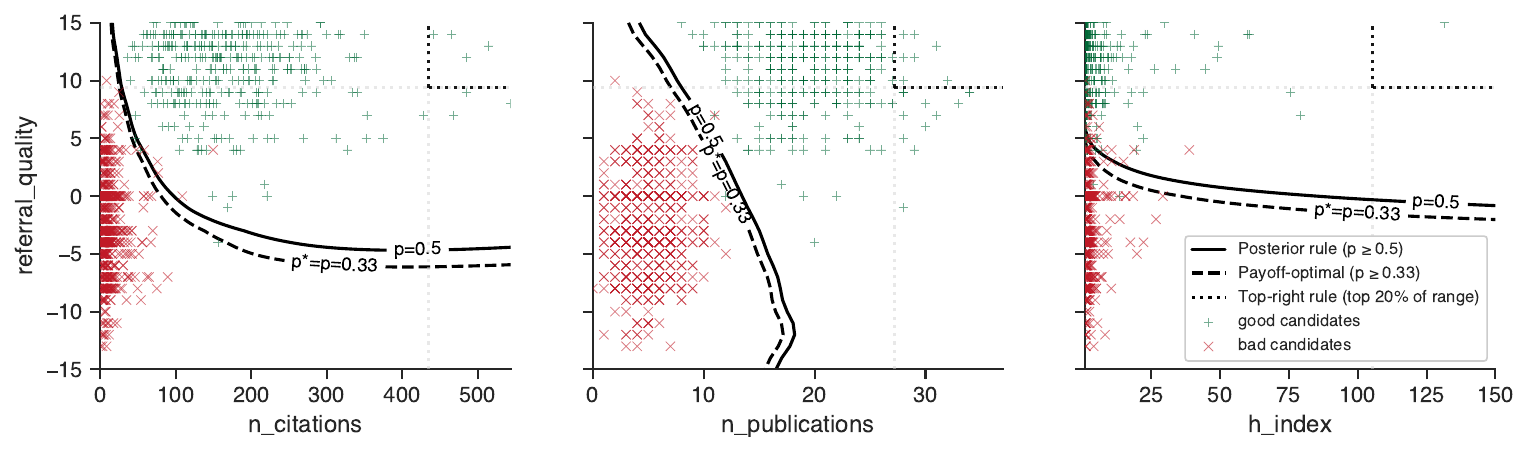}
  \caption{
  \textbf{Simulated candidates and decision boundaries.}
  Each panel shows simulated good (green~$+$) and bad (red~$\times$) candidates as a function of referral quality (y-axis) and a bibliometric signal (x-axis): citations (left), publications (middle), and h-index (right).
  The solid curve marks the posterior-based boundary ($p \geq 0.5$); the dashed curve the payoff-optimal boundary ($p \geq 0.33$), derived under an asymmetric payoff structure in which rejecting a good candidate is twice as costly as hiring a bad one (\Cref{app:sec:payoff}).
  The gap between the two curves identifies candidates the payoff rule would advance but the posterior rule would not.
  The dotted rectangle marks the top-right heuristic, which requires both signals to fall within the top 20\% of their respective value ranges.
  }
  \label{app:fig:posterior-contours}
\end{figure}

\begin{table}[ht]
\caption{
\textbf{Number of candidates classified as good or bad under three decision rules.}
Each row combines referral quality (network) with one bibliometric measure.
Out of $1{,}000$\ simulated candidates, the posterior- and payoff-based rules classify roughly $350$ as good, consistent with the prior $P(\text{good}) = 0.35$.
The top-right rule (top 20\%) classifies nearly all candidates as bad, reflecting its strict requirement for high values on both signal dimensions simultaneously.
}
\label{app:tbl:counts}
\centering
\begin{tabular}{lccc}
\toprule
Observed signals & Posterior-based & Payoff-based & Top-right rule \\
(net., non-net.) & good / bad & good / bad & good / bad \\
\midrule
referral\_quality, n\_citations & 350 / 650 & 352 / 648 & 4 / 996 \\
referral\_quality, n\_publications & 349 / 651 & 349 / 651 & 7 / 993 \\
referral\_quality, h\_index & 351 / 649 & 364 / 636 & 1 / 999 \\
\bottomrule
\end{tabular}
\end{table}

\begin{table}[ht!]
\caption{
\textbf{Classification performance of three decision rules across signal pairs.}
Results are shown as posterior (post.) / payoff-optimal (pay.) / top-20\% of the value range (topk) for $1{,}000$~simulated candidates.
Bold denotes the best scores per decision rule and signal pair.
The 1D accuracy ratio compares network-only to non-network-only accuracy; values $>1$ (\blueit{blue}) indicate that the network signal is more informative in isolation.
Realized payoff is computed under the payoff-optimal rule using the payoffs in~\Cref{tbl:si-params}.
}
\label{tbl:si-eval}
\centering
\begin{tabular}{lccccc}
\toprule
{Observed signals} & {Accuracy$\uparrow$} & {Precision$\uparrow$} & {Recall$\uparrow$} & {1D Accuracy ratio} & {Realized$\uparrow$} \\
(net., non-net.) & post.\ / pay.\ / topk & post.\ / pay.\ / topk & post.\ / pay.\ / topk & post.\ / pay.\ / topk & payoff \\
\midrule
referral\_quality, n\_citations & \textbf{1.00} / 0.99 / \textbf{0.66} & 0.99 / 0.99 / \textbf{1.00} & 0.99 / 0.99 / 0.01 & 0.99 / 0.99 / \blueit{1.35} & $-$450 \\
referral\_quality, n\_publications & \textbf{1.00} / \textbf{1.00} / \textbf{0.66} & \textbf{1.00} / \textbf{1.00} / \textbf{1.00} & \textbf{1.00} / \textbf{1.00} / \textbf{0.02} & 0.97 / 0.97 / \blueit{1.33} & \textbf{$-$150} \\
referral\_quality, h\_index & 0.96 / 0.96 / 0.65 & 0.93 / 0.92 / \textbf{1.00} & 0.94 / 0.96 / 0.00 & \blueit{1.39} / \blueit{1.52} / \blueit{1.37} & $-$2,700 \\
\bottomrule
\end{tabular}
\end{table}

\para{Decision boundaries.}
\Cref{app:fig:posterior-contours} overlays the three decision boundaries on the simulated population.
The posterior- and payoff-based boundaries are smooth curves along which the two signals trade off: a deficit on one axis can be offset by strength on the other.
The payoff-based boundary lies below the posterior-based one; the gap between them identifies candidates the payoff rule would advance but the posterior rule would not.
Both model-based rules identify roughly $350$ candidates as strong fits based on the observable signals (\Cref{app:tbl:counts})---not all need to be hired, but all are equally qualified given the evidence and merit equal consideration.
The top-right rule, by contrast, occupies only the extreme corner of the signal space, advancing at most $7$ of $1{,}000$\ candidates (\Cref{app:tbl:counts}) and thereby eliminating over 99\% of equally qualified candidates before they can be considered at all.

\para{Performance.}
\Cref{tbl:si-eval} evaluates the three rules against true (simulated) candidate quality.
Under our asymmetric payoff structure, in which false negatives are penalized twice as heavily as false positives, recall is the most decision-relevant metric.
The posterior- and payoff-based rules achieve near-perfect accuracy, precision, and recall across all signal combinations, whereas the top-right rule achieves high precision but near-zero recall: it rarely errs on whom it advances, but advances almost no one.
The payoff-optimal rule recovers slightly more truly good candidates (higher recall) at a marginal cost to precision---a favorable trade-off given the asymmetric penalties.
The 1D accuracy ratio reveals the relative informativeness of each signal in isolation.
By design (\Cref{app:sec:distributions}), the network signal is more discriminative than citations and h-index, while publications is the most discriminative signal overall.
This is reflected in the 1D ratio, which is substantially above $1$ only for h-index ($1.39$ / $1.52$), and in the realized payoff, which ranges from $-150$ (publications) to $−2{,}700$ (h-index).
Under the top-right rule, the 1D ratio exceeds $1$ for all three signal pairs, indicating that when candidates are filtered to the extreme corner of the signal space, the network signal consistently overrides the non-network signal in determining accuracy.

\section{Focus group}
\label{app:sec:focus}

So far we have treated the decision boundary as fixed, but in practice the relative weight placed on network versus non-network signals may vary with the hiring context. 
A department recruiting a PhD student might tolerate a weaker publication record if the candidate comes with strong referrals, whereas a senior professorship may demand demonstrated strength on both dimensions. 
To explore whether people intuitively adjust these thresholds by role---and whether the top-right corner retains its appeal when context is made explicit---we ran three focus groups.

Participants were asked to reason about hiring when candidates are described along two dimensions: \emph{social capital} (network-based signals such as referrals and professional connections) and \emph{skills} (non-network signals such as bibliometrics and measurable performance). Participants were drawn from three seminar talks delivered by the first author in Vienna, Budapest, and Graz. They included individuals from undergraduate to faculty levels and represented a demographically diverse audience. The tasks were designed to capture (i)~whether people instinctively favor the ``top-right'' candidate when presented with a simple quadrant layout, and (ii)~whether the minimum levels of social capital and skills they require vary by hiring context (e.g., hiring a PhD student vs.\ a professor).

\begin{figure*}[t]
  \centering
  \begin{subfigure}[t]{0.5\textwidth}
    \centering
    \caption{Whom would you hire?}
    \includegraphics[height=2.5in]{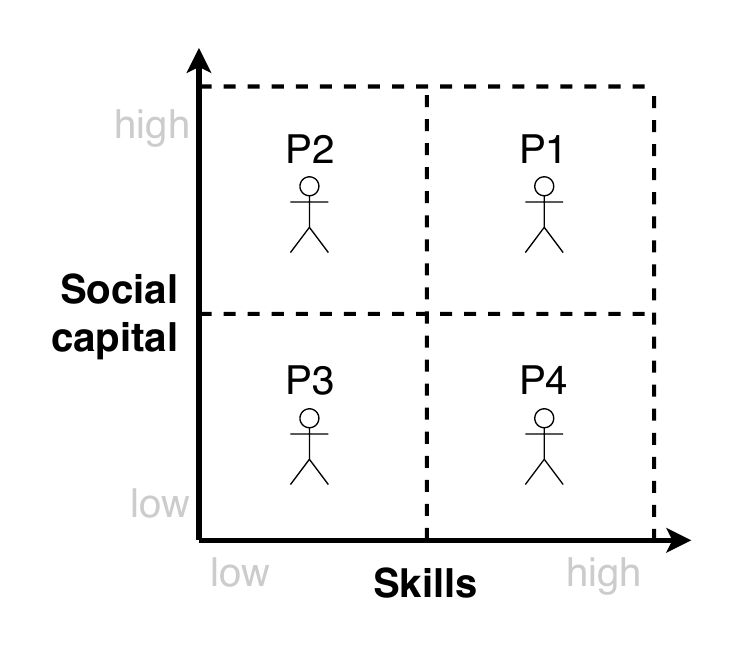}
    \label{app:fig:focus-quadrant:who}
  \end{subfigure}%
  \begin{subfigure}[t]{0.5\textwidth}
    \centering
    \caption{Minimum levels required}
    \includegraphics[height=2.5in]{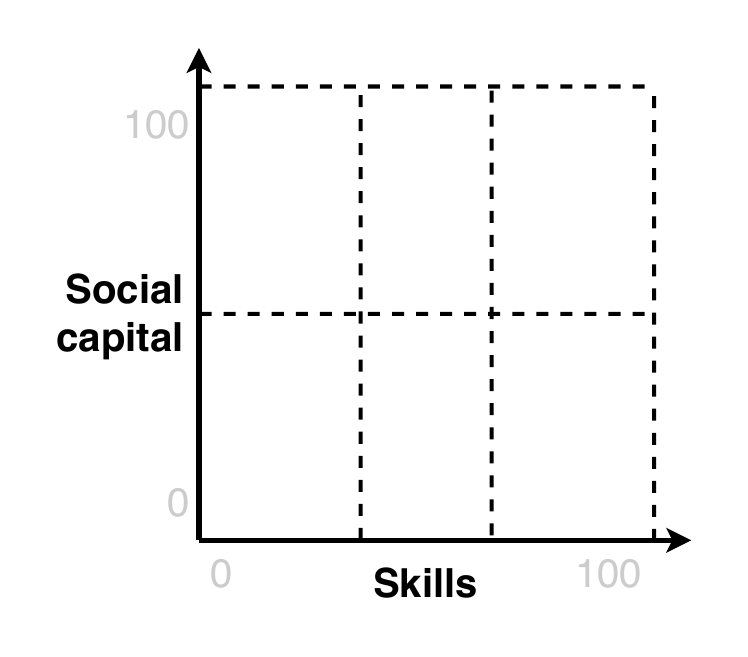}
    \label{app:fig:focus-quadrant:rank}
  \end{subfigure}
  \caption{\textbf{Two focus-group tasks in the social capital--skills space.}
  Axes: ``Social capital'' (network-based signals) and ``Skills'' (non-network metrics).
  \textbf{A) Whom would you hire?} Four candidate profiles (P1--P4), one per quadrant. Participants ranked them for hiring. %
  \textbf{B) Minimum levels required.} Empty grid shown to participants who had to place the minimum-required profile anywhere in the space for each role (PhD student, postdoc, professor). %
  }
  \label{app:fig:focus-quadrant}
\end{figure*}

\para{Ranking task.}
The Vienna group was asked: ``You need to hire one of these candidates. In order of preference, how would you rank them, from best to worst?.'' We showed four candidate profiles, each placed in one quadrant of a two-dimensional space with axes ``Social capital'' and ``Skills'' (\Cref{app:fig:focus-quadrant:who}). 
Of 29 participants, 26 completed the full ranking (three ranked only their first choice, all P1, and were excluded). Participants clearly preferred P1 > P4 > P2 > P3 (mean ranks 1.35, 2.31, 2.65, 3.69).~\footnote{P1 was ranked first by 22/26 participants (85\%), P4 second by 17/26 (65\%), P2 third by 17/26 (65\%), and P3 last by 20/26 (77\%). First choices were heavily concentrated on P1: P2 and P4 received 2/26 (8\%) each, and no participant ranked P3 first.}
P1, the top-right candidate (high social capital, high skills), was the dominant first choice. When P1 was set aside, participants favored P4 (high skills, low social capital) over P2 (high social capital, low skills), suggesting that they weighted skills more than social capital. P3, low on both dimensions, was clearly the least desirable. This strong preference for the top-right corner illustrates the intuitive appeal of requiring both dimensions to be high---the same instinct that, in our simulations, leads to the top-right trap when applied as a strict decision rule.

\begin{figure}[t]
  \centering
  \includegraphics[width=0.97\linewidth]{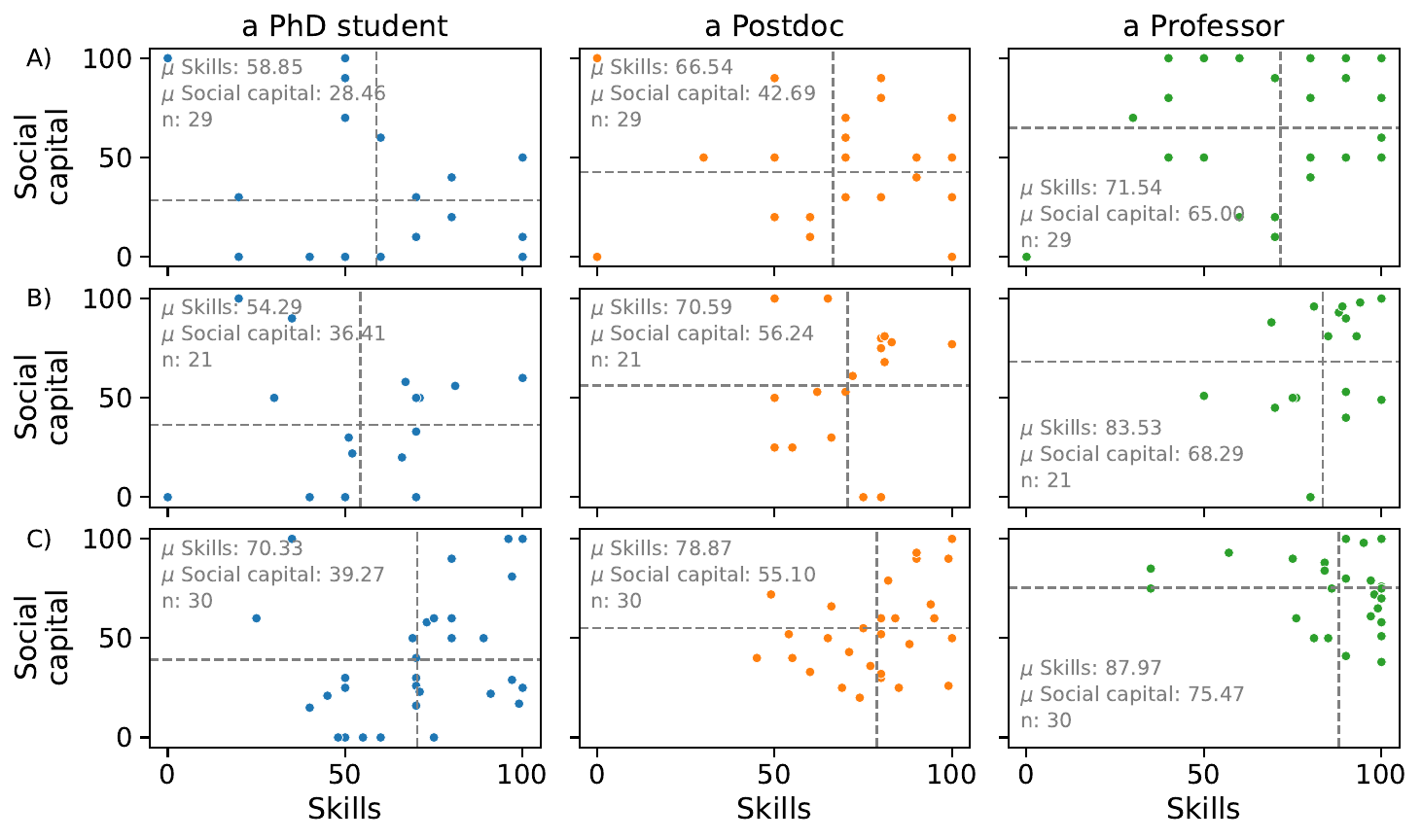}
  \caption{\textbf{Minimum required skills and social capital by role and focus group.}
  Each column corresponds to a role (PhD student, postdoc, professor); each row (A-C) to one of the three focus groups (Vienna, Budapest, and Graz, respectively) with sample size $n$ as indicated. Points are individual participants' stated minimum thresholds on the two axes (0--100); dashed lines show the mean skills and mean social capital per group and role. Across all three groups, mean requirements increase with seniority: participants demand higher minimum skills and higher minimum social capital for a professor than for a postdoc, and higher for a postdoc than for a PhD student, illustrating context-dependent relaxation of the ``top-right'' bar.}
  \label{app:fig:focus-requirements}
\end{figure}

\para{Minimum requirements by context.}
In practice, candidates are not confined to four discrete quadrants but lie anywhere in the social capital--skills space, and the bar for hiring depends on the role (e.g., PhD student vs.\ professor). To capture how participants set that bar, we asked in all three focus groups: ``What do you think are the minimum `skills' and `social capital' required for a good candidate when hiring \emph{[\_\_\_]}?'' The placeholder was filled with different roles: PhD student, postdoc, and professor. Participants were shown the same quadrant space (social capital vs. skills) and could place the minimum-required profile anywhere in it, choosing a level on each axis independently (\Cref{app:fig:focus-quadrant:rank}). \Cref{app:fig:focus-requirements} shows individual responses and means by role and focus group. Results show a clear pattern: required social capital and required skills both increased with seniority. Participants demanded higher minimum social capital and higher minimum skills for a professor than for a postdoc, and higher for a postdoc than for a PhD student. The stated thresholds are not symmetric across dimensions: for a given role, the minimum skill level required often differs from the minimum social capital required.
So the effective bar is not a single ``top-right'' corner but two potentially different cutoffs---the ``top'' on each axis can differ. 
Participants required more skills than social capital, and most responses fell in the right-hand quadrants containing P1 and P4 (\Cref{app:fig:focus-quadrant:who}), corresponding to a skills requirement above 50\% (\Cref{app:fig:focus-requirements}).
This aligns with the ranking results, where P1 and P4 were the top two choices. It contrasts with a strict top-right rule, which treats only the corner (both axes high) as acceptable. In other words, people do not apply a single fixed top-right bar. They relax the implied thresholds when the role is less senior and raise them when the role is more senior. That suggests that, in practice, people are willing to let context (here, the level of the position) shape how much they require on each axis. Such flexibility hints at the kind of boundary relaxation that our model recommends---e.g., payoff-based or posterior-based rules that allow one signal to compensate for the other---rather than a rigid top-right rule.

\para{Interpretation and implications.}
First, the ranking task confirms that the top-right candidate has strong intuitive appeal: when presented with a simple quadrant, a large majority of participants chose the candidate who was high on both dimensions. That preference mirrors the type of rule our simulations warn against---one that requires both referral quality and bibliometrics to be high and in doing so excludes many candidates whom a combined evaluation would classify as strong. Second, the minimum-requirements task shows that the same participants, when asked to set thresholds by role, do not apply a single rigid bar: they require more of both dimensions for senior roles and less for junior ones. So in practice, people already relax the ``top-right'' requirement depending on context; the question is whether hiring institutions do so in a principled way (e.g., by using a combined posterior or payoff-optimal rule) or by ad hoc role-dependent thresholds. From a network-fairness perspective, the risk of the top-right trap is that it systematically disadvantages candidates who are strong on one dimension but not both---often those with less access to referrals or less conventional credentials. The focus groups suggest that people can recognize context-dependent standards when prompted; the challenge is to embed that intuition in transparent, consistent decision rules that allow one signal to compensate for the other rather than requiring both to clear a high bar.


\begin{thebibliography}{100}
\urlstyle{rm}
\expandafter\ifx\csname url\endcsname\relax
  \def\url#1{\texttt{#1}}\fi
\expandafter\ifx\csname urlprefix\endcsname\relax\def\urlprefix{URL }\fi
\expandafter\ifx\csname doiprefix\endcsname\relax\def\doiprefix{DOI: }\fi
\providecommand{\bibinfo}[2]{#2}
\providecommand{\eprint}[2][]{\url{#2}}

\bibitem{mccarty2011personal}
\bibinfo{author}{McCarty, C.}, \bibinfo{author}{Gamper, M.},
  \bibinfo{author}{Lubbers, M.} \& \bibinfo{author}{Molina, J.~L.}
\newblock \bibinfo{title}{Personal network analysis}.
\newblock In \emph{\bibinfo{booktitle}{Encyclopedia of Social Networks}},
  vol.~\bibinfo{volume}{2}, \bibinfo{pages}{702--703},
  \doiprefix\url{10.4135/9781412994170.n286} (\bibinfo{publisher}{SAGE
  Publications, Inc.}, \bibinfo{year}{2011}).

\bibitem{boyd_Ellison_2007}
\bibinfo{author}{boyd, d.} \& \bibinfo{author}{Ellison, N.}
\newblock \bibinfo{journal}{\bibinfo{title}{Social network sites: Definition,
  history, and scholarship}}.
\newblock {\emph{\JournalTitle{Journal of Computer-Mediated Communication}}}
  \textbf{\bibinfo{volume}{13}}, \bibinfo{pages}{210–230},
  \doiprefix\url{10.1111/j.1083-6101.2007.00393.x} (\bibinfo{year}{2007}).

\bibitem{jordan2008welfare}
\bibinfo{author}{Jordan, B.}
\newblock \emph{\bibinfo{title}{Welfare and well-being: {S}ocial value in
  public policy}} (\bibinfo{publisher}{Policy Press}, \bibinfo{year}{2008}).

\bibitem{chetty2022social1}
\bibinfo{author}{Chetty, R.} \emph{et~al.}
\newblock \bibinfo{journal}{\bibinfo{title}{Social capital i: measurement and
  associations with economic mobility}}.
\newblock {\emph{\JournalTitle{Nature}}} \textbf{\bibinfo{volume}{608}},
  \bibinfo{pages}{108--121}, \doiprefix\url{10.1038/s41586-022-04996-4}
  (\bibinfo{year}{2022}).

\bibitem{mcpherson2001birds}
\bibinfo{author}{McPherson, M.}, \bibinfo{author}{Smith-Lovin, L.} \&
  \bibinfo{author}{Cook, J.~M.}
\newblock \bibinfo{journal}{\bibinfo{title}{Birds of a feather: {Homophily} in
  social networks}}.
\newblock {\emph{\JournalTitle{Annual review of sociology}}}
  \bibinfo{pages}{415--444} (\bibinfo{year}{2001}).

\bibitem{newman2003social}
\bibinfo{author}{Newman, M.~E.} \& \bibinfo{author}{Park, J.}
\newblock \bibinfo{journal}{\bibinfo{title}{Why social networks are different
  from other types of networks}}.
\newblock {\emph{\JournalTitle{Physical review E}}}
  \textbf{\bibinfo{volume}{68}}, \bibinfo{pages}{036122}
  (\bibinfo{year}{2003}).

\bibitem{shane2002network}
\bibinfo{author}{Shane, S.} \& \bibinfo{author}{Cable, D.}
\newblock \bibinfo{journal}{\bibinfo{title}{Network ties, reputation, and the
  financing of new ventures}}.
\newblock {\emph{\JournalTitle{Management science}}}
  \textbf{\bibinfo{volume}{48}}, \bibinfo{pages}{364--381}
  (\bibinfo{year}{2002}).

\bibitem{jacobs2021measurement}
\bibinfo{author}{Jacobs, A.~Z.} \& \bibinfo{author}{Wallach, H.}
\newblock \bibinfo{title}{Measurement and fairness}.
\newblock In \emph{\bibinfo{booktitle}{Proceedings of the 2021 ACM conference
  on fairness, accountability, and transparency}}, \bibinfo{pages}{375--385}
  (\bibinfo{year}{2021}).

\bibitem{nieto2023examining}
\bibinfo{author}{Nieto, A.}, \bibinfo{author}{Davies, T.} \&
  \bibinfo{author}{Borrion, H.}
\newblock \bibinfo{journal}{\bibinfo{title}{Examining the importance of
  existing relationships for co-offending: {A} temporal network analysis in
  {Bogot{\'a}}, {Colombia} (2005--2018)}}.
\newblock {\emph{\JournalTitle{Applied Network Science}}}
  \textbf{\bibinfo{volume}{8}}, \bibinfo{pages}{4} (\bibinfo{year}{2023}).

\bibitem{boyd2014networked}
\bibinfo{author}{boyd, d.}, \bibinfo{author}{Levy, K.} \&
  \bibinfo{author}{Marwick, A.}
\newblock \bibinfo{journal}{\bibinfo{title}{The networked nature of algorithmic
  discrimination}}.
\newblock {\emph{\JournalTitle{Data and Discrimination: Collected Essays. Open
  Technology Institute}}}  (\bibinfo{year}{2014}).

\bibitem{dies2025forecasting}
\bibinfo{author}{Dies, S.}, \bibinfo{author}{Liu, D.} \&
  \bibinfo{author}{Eliassi-Rad, T.}
\newblock \bibinfo{journal}{\bibinfo{title}{Forecasting faculty placement from
  patterns in co-authorship networks}}.
\newblock {\emph{\JournalTitle{arXiv preprint arXiv:2507.14696}}}
  (\bibinfo{year}{2025}).

\bibitem{hussinger2019guilt}
\bibinfo{author}{Hussinger, K.} \& \bibinfo{author}{Pellens, M.}
\newblock \bibinfo{journal}{\bibinfo{title}{Guilt by association: {How}
  scientific misconduct harms prior collaborators}}.
\newblock {\emph{\JournalTitle{Research policy}}}
  \textbf{\bibinfo{volume}{48}}, \bibinfo{pages}{516--530}
  (\bibinfo{year}{2019}).

\bibitem{gupta2013wtf}
\bibinfo{author}{Gupta, P.} \emph{et~al.}
\newblock \bibinfo{title}{{WTF}: {The} who to follow service at {Twitter}}.
\newblock In \emph{\bibinfo{booktitle}{Proceedings of the 22nd international
  conference on World Wide Web}}, \bibinfo{pages}{505--514}
  (\bibinfo{year}{2013}).

\bibitem{santos2021link}
\bibinfo{author}{Santos, F.~P.}, \bibinfo{author}{Lelkes, Y.} \&
  \bibinfo{author}{Levin, S.~A.}
\newblock \bibinfo{journal}{\bibinfo{title}{Link recommendation algorithms and
  dynamics of polarization in online social networks}}.
\newblock {\emph{\JournalTitle{Proceedings of the National Academy of
  Sciences}}} \textbf{\bibinfo{volume}{118}}, \bibinfo{pages}{e2102141118}
  (\bibinfo{year}{2021}).

\bibitem{chang2025llms}
\bibinfo{author}{Chang, S.} \emph{et~al.}
\newblock \bibinfo{title}{Llms generate structurally realistic social networks
  but overestimate political homophily}.
\newblock In \emph{\bibinfo{booktitle}{Proceedings of the International AAAI
  Conference on Web and Social Media}}, vol.~\bibinfo{volume}{19},
  \bibinfo{pages}{341--371} (\bibinfo{year}{2025}).

\bibitem{gkartzios2025modeling}
\bibinfo{author}{Gkartzios, C.}, \bibinfo{author}{Pitoura, E.} \&
  \bibinfo{author}{Tsaparas, P.}
\newblock \bibinfo{title}{Modeling network formation with llm agents: {The}
  role of demographics and personality}.
\newblock In \emph{\bibinfo{booktitle}{2025 IEEE International Conference on
  Data Mining Workshops (ICDMW)}}, \bibinfo{pages}{137--146}
  (\bibinfo{organization}{IEEE}, \bibinfo{year}{2025}).

\bibitem{mallick2026llmssocialnetworkmodeling}
\bibinfo{author}{Mallick, S.}, \bibinfo{author}{Thomo, A.} \&
  \bibinfo{author}{Saxena, A.}
\newblock \bibinfo{title}{Llms for social network modeling: {From} network
  generation to dynamic processes} (\bibinfo{year}{2026}).
\newblock \eprint{2609.08049}.

\bibitem{mehdizadeh2025homophily}
\bibinfo{author}{Mehdizadeh, A.} \& \bibinfo{author}{Hilbert, M.}
\newblock \bibinfo{journal}{\bibinfo{title}{Homophily-induced emergence of
  biased structures in llm-based multi-agent ai systems}}.
\newblock {\emph{\JournalTitle{Social Network Analysis and Mining}}}
  \textbf{\bibinfo{volume}{15}}, \bibinfo{pages}{108} (\bibinfo{year}{2025}).

\bibitem{barolo2025whose}
\bibinfo{author}{Barolo, D.} \emph{et~al.}
\newblock \bibinfo{journal}{\bibinfo{title}{Whose name comes up? auditing
  llm-based scholar recommendations}}.
\newblock {\emph{\JournalTitle{arXiv preprint arXiv:2506.00074}}}
  (\bibinfo{year}{2025}).

\bibitem{Bourdieu_2002}
\bibinfo{author}{Bourdieu, P.}
\newblock \emph{\bibinfo{title}{Distinction: {A} social critique of the
  judgement of taste}} (\bibinfo{publisher}{Harvard Univ. Press},
  \bibinfo{address}{Cambridge, Mass}, \bibinfo{year}{2002}).

\bibitem{braddock1987minorities}
\bibinfo{author}{Braddock, J.~H.} \& \bibinfo{author}{McPartland, J.~M.}
\newblock \bibinfo{journal}{\bibinfo{title}{How minorities continue to be
  excluded from equal employment opportunities: {Research} on labor market and
  institutional barriers}}.
\newblock {\emph{\JournalTitle{Journal of Social Issues}}}
  \textbf{\bibinfo{volume}{43}}, \bibinfo{pages}{5--39} (\bibinfo{year}{1987}).

\bibitem{petersen2000offering}
\bibinfo{author}{Petersen, T.}, \bibinfo{author}{Saporta, I.} \&
  \bibinfo{author}{Seidel, M.-D.~L.}
\newblock \bibinfo{journal}{\bibinfo{title}{Offering a job: {Meritocracy} and
  social networks}}.
\newblock {\emph{\JournalTitle{American Journal of Sociology}}}
  \textbf{\bibinfo{volume}{106}}, \bibinfo{pages}{763--816}
  (\bibinfo{year}{2000}).

\bibitem{sampson2008neighborhood}
\bibinfo{author}{Sampson, R.~J.} \& \bibinfo{author}{Sharkey, P.}
\newblock \bibinfo{journal}{\bibinfo{title}{Neighborhood selection and the
  social reproduction of concentrated racial inequality}}.
\newblock {\emph{\JournalTitle{Demography}}} \textbf{\bibinfo{volume}{45}},
  \bibinfo{pages}{1--29} (\bibinfo{year}{2008}).

\bibitem{loury2021anatomy}
\bibinfo{author}{Loury, G.~C.}
\newblock \emph{\bibinfo{title}{The anatomy of racial inequality: {With} a new
  preface}} (\bibinfo{publisher}{Harvard University Press},
  \bibinfo{year}{2021}).

\bibitem{ibarra1992homophily}
\bibinfo{author}{Ibarra, H.}
\newblock \bibinfo{journal}{\bibinfo{title}{Homophily and differential returns:
  {S}ex differences in network structure and access in an advertising firm}}.
\newblock {\emph{\JournalTitle{Administrative science quarterly}}}
  \bibinfo{pages}{422--447} (\bibinfo{year}{1992}).

\bibitem{small2009unanticipated}
\bibinfo{author}{Small, M.~L.}
\newblock \emph{\bibinfo{title}{Unanticipated gains: {Origins} of network
  inequality in everyday life}} (\bibinfo{publisher}{Oxford University Press},
  \bibinfo{year}{2009}).

\bibitem{dimaggio2012network}
\bibinfo{author}{DiMaggio, P.} \& \bibinfo{author}{Garip, F.}
\newblock \bibinfo{journal}{\bibinfo{title}{Network effects and social
  inequality}}.
\newblock {\emph{\JournalTitle{Annual review of sociology}}}
  (\bibinfo{year}{2012}).

\bibitem{blumenthal2019potential}
\bibinfo{author}{Blumenthal-Barby, J.} \emph{et~al.}
\newblock \bibinfo{journal}{\bibinfo{title}{Potential unintended consequences
  of recent shared decision making policy initiatives}}.
\newblock {\emph{\JournalTitle{Health Affairs}}} \textbf{\bibinfo{volume}{38}},
  \bibinfo{pages}{1876--1881} (\bibinfo{year}{2019}).

\bibitem{kopar2021critical}
\bibinfo{author}{Kopar, P.~K.}, \bibinfo{author}{Kramer, J.~B.},
  \bibinfo{author}{Brown, D.~E.} \& \bibinfo{author}{Bochicchio, G.~V.}
\newblock \bibinfo{journal}{\bibinfo{title}{Critical ethics: {H}ow to balance
  patient autonomy with fairness when patients refuse coronavirus disease 2019
  testing}}.
\newblock {\emph{\JournalTitle{Critical Care Explorations}}}
  \textbf{\bibinfo{volume}{3}} (\bibinfo{year}{2021}).

\bibitem{li2023bright}
\bibinfo{author}{Li, X.}, \bibinfo{author}{Guo, X.} \& \bibinfo{author}{Shi,
  Z.}
\newblock \bibinfo{journal}{\bibinfo{title}{Bright sides and dark sides:
  {Unveiling} the double-edged sword effects of social networks}}.
\newblock {\emph{\JournalTitle{Social Science \& Medicine}}}
  \textbf{\bibinfo{volume}{329}}, \bibinfo{pages}{116035}
  (\bibinfo{year}{2023}).

\bibitem{boyd2014s}
\bibinfo{author}{boyd, d.}
\newblock \emph{\bibinfo{title}{It's complicated: {The} social lives of
  networked teens}} (\bibinfo{publisher}{Yale University Press},
  \bibinfo{year}{2014}).

\bibitem{zheleva2009join}
\bibinfo{author}{Zheleva, E.} \& \bibinfo{author}{Getoor, L.}
\newblock \bibinfo{title}{To join or not to join: {The} illusion of privacy in
  social networks with mixed public and private user profiles}.
\newblock In \emph{\bibinfo{booktitle}{Proceedings of the 18th international
  conference on World wide web}}, \bibinfo{pages}{531--540}
  (\bibinfo{year}{2009}).

\bibitem{isaak2018user}
\bibinfo{author}{Isaak, J.} \& \bibinfo{author}{Hanna, M.~J.}
\newblock \bibinfo{journal}{\bibinfo{title}{User data privacy: {F}acebook,
  {C}ambridge {A}nalytica, and privacy protection}}.
\newblock {\emph{\JournalTitle{Computer}}} \textbf{\bibinfo{volume}{51}},
  \bibinfo{pages}{56--59} (\bibinfo{year}{2018}).

\bibitem{mehrabi2021survey}
\bibinfo{author}{Mehrabi, N.}, \bibinfo{author}{Morstatter, F.},
  \bibinfo{author}{Saxena, N.}, \bibinfo{author}{Lerman, K.} \&
  \bibinfo{author}{Galstyan, A.}
\newblock \bibinfo{journal}{\bibinfo{title}{A survey on bias and fairness in
  machine learning}}.
\newblock {\emph{\JournalTitle{ACM Computing Surveys (CSUR)}}}
  \textbf{\bibinfo{volume}{54}}, \bibinfo{pages}{1--35} (\bibinfo{year}{2021}).

\bibitem{madiega2021artificial}
\bibinfo{author}{Madiega, T.~A.}
\newblock \bibinfo{journal}{\bibinfo{title}{Artificial intelligence act}}.
\newblock {\emph{\JournalTitle{{European Parliament: {E}uropean Parliamentary
  Research Service}}}}  (\bibinfo{year}{2021}).

\bibitem{canada2021equal}
\bibinfo{author}{{Government of Canada}}.
\newblock \bibinfo{title}{Equal pay for work of equal value}
  (\bibinfo{year}{2021}).
\newblock \bibinfo{note}{Accessed online: 2022-11-02},
  \eprint{https://www.canada.ca/en/treasury-board-secretariat/topics/pay/equitable-compensation.html}.

\bibitem{wagner2021measuring}
\bibinfo{author}{Wagner, C.} \emph{et~al.}
\newblock \bibinfo{journal}{\bibinfo{title}{Measuring algorithmically infused
  societies}}.
\newblock {\emph{\JournalTitle{Nature}}} \textbf{\bibinfo{volume}{595}},
  \bibinfo{pages}{197--204} (\bibinfo{year}{2021}).

\bibitem{merton1968matthew}
\bibinfo{author}{Merton, R.~K.}
\newblock \bibinfo{journal}{\bibinfo{title}{The matthew effect in science:
  {T}he reward and communication systems of science are considered.}}
\newblock {\emph{\JournalTitle{Science}}} \textbf{\bibinfo{volume}{159}},
  \bibinfo{pages}{56--63} (\bibinfo{year}{1968}).

\bibitem{newman2009first}
\bibinfo{author}{Newman, M.~E.}
\newblock \bibinfo{journal}{\bibinfo{title}{The first-mover advantage in
  scientific publication}}.
\newblock {\emph{\JournalTitle{EPL (Europhysics Letters)}}}
  \textbf{\bibinfo{volume}{86}}, \bibinfo{pages}{68001} (\bibinfo{year}{2009}).

\bibitem{kong2022influence}
\bibinfo{author}{Kong, H.}, \bibinfo{author}{Martin-Gutierrez, S.} \&
  \bibinfo{author}{Karimi, F.}
\newblock \bibinfo{journal}{\bibinfo{title}{Influence of the first-mover
  advantage on the gender disparities in physics citations}}.
\newblock {\emph{\JournalTitle{Communications Physics}}}
  \textbf{\bibinfo{volume}{5}}, \bibinfo{pages}{1--11} (\bibinfo{year}{2022}).

\bibitem{tajfel1979integrative}
\bibinfo{author}{Tajfel, H.} \& \bibinfo{author}{Turner, J.}
\newblock \bibinfo{title}{An integrative theory of intergroup conflict}.
\newblock In \bibinfo{editor}{Hatch, M.~J.}, \bibinfo{editor}{Schultz, M.},
  \bibinfo{editor}{Hatch, M.~J.} \& \bibinfo{editor}{Schultz, M.} (eds.)
  \emph{\bibinfo{booktitle}{Organizational Identity: A Reader}},
  \doiprefix\url{10.1093/oso/9780199269464.003.0005}
  (\bibinfo{publisher}{Oxford University Press}, \bibinfo{year}{2000}).
\newblock
  \eprint{https://academic.oup.com/book/0/chapter/421807417/chapter-pdf/52307450/isbn-9780199269464-book-part-5.pdf}.

\bibitem{de2019does}
\bibinfo{author}{De~Cnudde, S.} \emph{et~al.}
\newblock \bibinfo{journal}{\bibinfo{title}{What does your {Facebook} profile
  reveal about your creditworthiness? using alternative data for
  microfinance}}.
\newblock {\emph{\JournalTitle{Journal of the Operational Research Society}}}
  \textbf{\bibinfo{volume}{70}}, \bibinfo{pages}{353--363}
  (\bibinfo{year}{2019}).

\bibitem{li2020fairness}
\bibinfo{author}{Li, Y.}, \bibinfo{author}{Ning, Y.}, \bibinfo{author}{Liu,
  R.}, \bibinfo{author}{Wu, Y.} \& \bibinfo{author}{Hui~Wang, W.}
\newblock \bibinfo{title}{Fairness of classification using users’ social
  relationships in online peer-to-peer lending}.
\newblock In \emph{\bibinfo{booktitle}{Companion Proceedings of the Web
  Conference 2020}}, \bibinfo{pages}{733--742} (\bibinfo{year}{2020}).

\bibitem{sekara2018chaperone}
\bibinfo{author}{Sekara, V.} \emph{et~al.}
\newblock \bibinfo{journal}{\bibinfo{title}{The chaperone effect in scientific
  publishing}}.
\newblock {\emph{\JournalTitle{Proceedings of the National Academy of
  Sciences}}} \textbf{\bibinfo{volume}{115}}, \bibinfo{pages}{12603--12607},
  \doiprefix\url{10.1073/pnas.1800471115} (\bibinfo{year}{2018}).
\newblock \eprint{https://www.pnas.org/doi/pdf/10.1073/pnas.1800471115}.

\bibitem{brainard2022reviewers}
\bibinfo{author}{Brainard, J.}
\newblock \bibinfo{journal}{\bibinfo{title}{Reviewers award higher marks when a
  paper’s author is famous}}.
\newblock {\emph{\JournalTitle{Science (New York, NY)}}}
  \textbf{\bibinfo{volume}{377}}, \bibinfo{pages}{1251--1251}
  (\bibinfo{year}{2022}).

\bibitem{crane1965scientists}
\bibinfo{author}{Crane, D.}
\newblock \bibinfo{journal}{\bibinfo{title}{Scientists at major and minor
  universities: {A} study of productivity and recognition}}.
\newblock {\emph{\JournalTitle{American sociological review}}}
  \bibinfo{pages}{699--714} (\bibinfo{year}{1965}).

\bibitem{li2019reciprocity}
\bibinfo{author}{Li, W.}, \bibinfo{author}{Aste, T.},
  \bibinfo{author}{Caccioli, F.} \& \bibinfo{author}{Livan, G.}
\newblock \bibinfo{journal}{\bibinfo{title}{Reciprocity and impact in academic
  careers}}.
\newblock {\emph{\JournalTitle{EPJ Data Science}}}
  \textbf{\bibinfo{volume}{8}}, \bibinfo{pages}{20} (\bibinfo{year}{2019}).

\bibitem{wang2020early}
\bibinfo{author}{Wang, W.} \emph{et~al.}
\newblock \bibinfo{journal}{\bibinfo{title}{Early-stage reciprocity in
  sustainable scientific collaboration}}.
\newblock {\emph{\JournalTitle{Journal of Informetrics}}}
  \textbf{\bibinfo{volume}{14}}, \bibinfo{pages}{101041}
  (\bibinfo{year}{2020}).

\bibitem{baer1991biases}
\bibinfo{author}{Baer, J.~S.}, \bibinfo{author}{Stacy, A.} \&
  \bibinfo{author}{Larimer, M.}
\newblock \bibinfo{journal}{\bibinfo{title}{Biases in the perception of
  drinking norms among college students.}}
\newblock {\emph{\JournalTitle{Journal of studies on alcohol}}}
  \textbf{\bibinfo{volume}{52}}, \bibinfo{pages}{580--586}
  (\bibinfo{year}{1991}).

\bibitem{lerman2016majority}
\bibinfo{author}{Lerman, K.}, \bibinfo{author}{Yan, X.} \& \bibinfo{author}{Wu,
  X.-Z.}
\newblock \bibinfo{journal}{\bibinfo{title}{The ``majority illusion'' in social
  networks}}.
\newblock {\emph{\JournalTitle{PloS one}}} \textbf{\bibinfo{volume}{11}},
  \bibinfo{pages}{e0147617} (\bibinfo{year}{2016}).

\bibitem{shwed2014referrals}
\bibinfo{author}{Shwed, U.} \& \bibinfo{author}{Kalev, A.}
\newblock \bibinfo{journal}{\bibinfo{title}{Are referrals more productive or
  more likeable? social networks and the evaluation of merit}}.
\newblock {\emph{\JournalTitle{American Behavioral Scientist}}}
  \textbf{\bibinfo{volume}{58}}, \bibinfo{pages}{288--308}
  (\bibinfo{year}{2014}).

\bibitem{cummings2003structural}
\bibinfo{author}{Cummings, J.~N.} \& \bibinfo{author}{Cross, R.}
\newblock \bibinfo{journal}{\bibinfo{title}{Structural properties of work
  groups and their consequences for performance}}.
\newblock {\emph{\JournalTitle{Social networks}}}
  \textbf{\bibinfo{volume}{25}}, \bibinfo{pages}{197--210}
  (\bibinfo{year}{2003}).

\bibitem{long2013bridges}
\bibinfo{author}{Long, J.~C.}, \bibinfo{author}{Cunningham, F.~C.} \&
  \bibinfo{author}{Braithwaite, J.}
\newblock \bibinfo{journal}{\bibinfo{title}{Bridges, brokers and boundary
  spanners in collaborative networks: {A} systematic review}}.
\newblock {\emph{\JournalTitle{BMC health services research}}}
  \textbf{\bibinfo{volume}{13}}, \bibinfo{pages}{1--13} (\bibinfo{year}{2013}).

\bibitem{lee1994friend}
\bibinfo{author}{Lee, S.~C.}, \bibinfo{author}{Muncaster, R.~G.} \&
  \bibinfo{author}{Zinnes, D.~A.}
\newblock \bibinfo{journal}{\bibinfo{title}{`the friend of my enemy is my
  enemy': {Modeling} triadic internation relationships}}.
\newblock {\emph{\JournalTitle{Synthese}}} \textbf{\bibinfo{volume}{100}},
  \bibinfo{pages}{333--358} (\bibinfo{year}{1994}).

\bibitem{meger2023iterative}
\bibinfo{author}{Meger, E.} \& \bibinfo{author}{Raz, A.}
\newblock \bibinfo{journal}{\bibinfo{title}{The iterative independent model}}.
\newblock {\emph{\JournalTitle{Discrete Applied Mathematics}}}
  \textbf{\bibinfo{volume}{341}}, \bibinfo{pages}{242--256}
  (\bibinfo{year}{2023}).

\bibitem{pescosolido2021personal}
\bibinfo{author}{Pescosolido, B.} \& \bibinfo{author}{Smith, E.~B.}
\newblock \emph{\bibinfo{title}{Personal networks: {C}lassic readings and new
  directions in egocentric analysis}} (\bibinfo{publisher}{Cambridge University
  Press}, \bibinfo{year}{2021}).

\bibitem{adler2002social}
\bibinfo{author}{Adler, P.~S.} \& \bibinfo{author}{Kwon, S.-W.}
\newblock \bibinfo{journal}{\bibinfo{title}{Social capital: {P}rospects for a
  new concept}}.
\newblock {\emph{\JournalTitle{Academy of management review}}}
  \textbf{\bibinfo{volume}{27}}, \bibinfo{pages}{17--40}
  (\bibinfo{year}{2002}).

\bibitem{field2016social}
\bibinfo{author}{Field, J.}
\newblock \emph{\bibinfo{title}{Social capital}}
  (\bibinfo{publisher}{Routledge}, \bibinfo{year}{2016}).

\bibitem{robison2002social}
\bibinfo{author}{Robison, L.~J.}, \bibinfo{author}{Schmid, A.~A.} \&
  \bibinfo{author}{Siles, M.~E.}
\newblock \bibinfo{journal}{\bibinfo{title}{Is social capital really capital?}}
\newblock {\emph{\JournalTitle{Review of social economy}}}
  \textbf{\bibinfo{volume}{60}}, \bibinfo{pages}{1--21} (\bibinfo{year}{2002}).

\bibitem{lin2002social}
\bibinfo{author}{Lin, N.}
\newblock \emph{\bibinfo{title}{Social capital: {A} theory of social structure
  and action}}, vol.~\bibinfo{volume}{19} (\bibinfo{publisher}{Cambridge
  university press}, \bibinfo{year}{2002}).

\bibitem{borgatta1980level}
\bibinfo{author}{Borgatta, E.~F.} \& \bibinfo{author}{Bohrnstedt, G.~W.}
\newblock \bibinfo{journal}{\bibinfo{title}{Level of measurement: {O}nce over
  again}}.
\newblock {\emph{\JournalTitle{Sociological Methods \& Research}}}
  \textbf{\bibinfo{volume}{9}}, \bibinfo{pages}{147--160}
  (\bibinfo{year}{1980}).

\bibitem{hirsch1999umbrella}
\bibinfo{author}{Hirsch, P.~M.} \& \bibinfo{author}{Levin, D.~Z.}
\newblock \bibinfo{journal}{\bibinfo{title}{Umbrella advocates versus validity
  police: {A} life-cycle model}}.
\newblock {\emph{\JournalTitle{Organization Science}}}
  \textbf{\bibinfo{volume}{10}}, \bibinfo{pages}{199--212}
  (\bibinfo{year}{1999}).

\bibitem{adcock2001measurement}
\bibinfo{author}{Adcock, R.} \& \bibinfo{author}{Collier, D.}
\newblock \bibinfo{journal}{\bibinfo{title}{Measurement validity: {A} shared
  standard for qualitative and quantitative research}}.
\newblock {\emph{\JournalTitle{American political science review}}}
  \textbf{\bibinfo{volume}{95}}, \bibinfo{pages}{529--546}
  (\bibinfo{year}{2001}).

\bibitem{knox2022testing}
\bibinfo{author}{Knox, D.}, \bibinfo{author}{Lucas, C.} \&
  \bibinfo{author}{Cho, W. K.~T.}
\newblock \bibinfo{journal}{\bibinfo{title}{Testing causal theories with
  learned proxies}}.
\newblock {\emph{\JournalTitle{Annual Review of Political Science}}}
  \textbf{\bibinfo{volume}{25}}, \bibinfo{pages}{419--441}
  (\bibinfo{year}{2022}).

\bibitem{friedler2016possibility}
\bibinfo{author}{Friedler, S.~A.}, \bibinfo{author}{Scheidegger, C.} \&
  \bibinfo{author}{Venkatasubramanian, S.}
\newblock \bibinfo{journal}{\bibinfo{title}{On the (im)possibility of
  fairness}}.
\newblock {\emph{\JournalTitle{arXiv preprint arXiv:1609.07236}}}
  (\bibinfo{year}{2016}).

\bibitem{friedler2021possibility}
\bibinfo{author}{Friedler, S.~A.}, \bibinfo{author}{Scheidegger, C.} \&
  \bibinfo{author}{Venkatasubramanian, S.}
\newblock \bibinfo{journal}{\bibinfo{title}{The (im)possibility of fairness:
  {D}ifferent value systems require different mechanisms for fair decision
  making}}.
\newblock {\emph{\JournalTitle{Communications of the ACM}}}
  \textbf{\bibinfo{volume}{64}}, \bibinfo{pages}{136--143}
  (\bibinfo{year}{2021}).

\bibitem{graeff2009social}
\bibinfo{author}{Graeff, P.}
\newblock \bibinfo{journal}{\bibinfo{title}{Social capital: {T}he dark side}}.
\newblock {\emph{\JournalTitle{Handbook of social capital}}}
  \bibinfo{pages}{143--161} (\bibinfo{year}{2009}).

\bibitem{baycan2022dark}
\bibinfo{author}{Baycan, T.} \& \bibinfo{author}{{\"O}ner, {\"O}.}
\newblock \bibinfo{journal}{\bibinfo{title}{The dark side of social capital:
  {A} contextual perspective}}.
\newblock {\emph{\JournalTitle{The Annals of Regional Science}}}
  \bibinfo{pages}{1--20} (\bibinfo{year}{2022}).

\bibitem{ayios2014social}
\bibinfo{author}{Ayios, A.}, \bibinfo{author}{Jeurissen, R.},
  \bibinfo{author}{Manning, P.} \& \bibinfo{author}{Spence, L.~J.}
\newblock \bibinfo{journal}{\bibinfo{title}{Social capital: {A} review from an
  ethics perspective}}.
\newblock {\emph{\JournalTitle{Business Ethics: {A} European Review}}}
  \textbf{\bibinfo{volume}{23}}, \bibinfo{pages}{108--124}
  (\bibinfo{year}{2014}).

\bibitem{livan2019don}
\bibinfo{author}{Livan, G.}
\newblock \bibinfo{journal}{\bibinfo{title}{Don’t follow the leader: {H}ow
  ranking performance reduces meritocracy}}.
\newblock {\emph{\JournalTitle{Royal Society Open Science}}}
  \textbf{\bibinfo{volume}{6}}, \bibinfo{pages}{191255} (\bibinfo{year}{2019}).

\bibitem{lin2000inequality}
\bibinfo{author}{Lin, N.}
\newblock \bibinfo{journal}{\bibinfo{title}{Inequality in social capital}}.
\newblock {\emph{\JournalTitle{Contemporary sociology}}}
  \textbf{\bibinfo{volume}{29}}, \bibinfo{pages}{785--795}
  (\bibinfo{year}{2000}).

\bibitem{granovetter1973strength}
\bibinfo{author}{Granovetter, M.~S.}
\newblock \bibinfo{journal}{\bibinfo{title}{The strength of weak ties}}.
\newblock {\emph{\JournalTitle{American journal of sociology}}}
  \textbf{\bibinfo{volume}{78}}, \bibinfo{pages}{1360--1380}
  (\bibinfo{year}{1973}).

\bibitem{krackhardt2003strength}
\bibinfo{author}{Krackhardt, D.}
\newblock \bibinfo{journal}{\bibinfo{title}{The strength of strong ties}}.
\newblock {\emph{\JournalTitle{Networks in the knowledge economy}}}
  \textbf{\bibinfo{volume}{82}} (\bibinfo{year}{2003}).

\bibitem{brown1987social}
\bibinfo{author}{Brown, J.~J.} \& \bibinfo{author}{Reingen, P.~H.}
\newblock \bibinfo{journal}{\bibinfo{title}{Social ties and word-of-mouth
  referral behavior}}.
\newblock {\emph{\JournalTitle{Journal of Consumer research}}}
  \textbf{\bibinfo{volume}{14}}, \bibinfo{pages}{350--362}
  (\bibinfo{year}{1987}).

\bibitem{granovetter1983strength}
\bibinfo{author}{Granovetter, M.}
\newblock \bibinfo{journal}{\bibinfo{title}{The strength of weak ties: {A}
  network theory revisited}}.
\newblock {\emph{\JournalTitle{Sociological theory}}} \bibinfo{pages}{201--233}
  (\bibinfo{year}{1983}).

\bibitem{rajkumar2022causal}
\bibinfo{author}{Rajkumar, K.}, \bibinfo{author}{Saint-Jacques, G.},
  \bibinfo{author}{Bojinov, I.}, \bibinfo{author}{Brynjolfsson, E.} \&
  \bibinfo{author}{Aral, S.}
\newblock \bibinfo{journal}{\bibinfo{title}{A causal test of the strength of
  weak ties}}.
\newblock {\emph{\JournalTitle{Science}}} \textbf{\bibinfo{volume}{377}},
  \bibinfo{pages}{1304--1310} (\bibinfo{year}{2022}).

\bibitem{burt2000network}
\bibinfo{author}{Burt, R.~S.}
\newblock \bibinfo{journal}{\bibinfo{title}{The network structure of social
  capital}}.
\newblock {\emph{\JournalTitle{Research in organizational behavior}}}
  \textbf{\bibinfo{volume}{22}}, \bibinfo{pages}{345--423}
  (\bibinfo{year}{2000}).

\bibitem{huber2022nobel}
\bibinfo{author}{Huber, J.} \emph{et~al.}
\newblock \bibinfo{journal}{\bibinfo{title}{Nobel and novice: {A}uthor
  prominence affects peer review}}.
\newblock {\emph{\JournalTitle{Proceedings of the National Academy of
  Sciences}}} \textbf{\bibinfo{volume}{119}}, \bibinfo{pages}{e2205779119}
  (\bibinfo{year}{2022}).

\bibitem{li2022untangling}
\bibinfo{author}{Li, W.}, \bibinfo{author}{Zhang, S.}, \bibinfo{author}{Zheng,
  Z.}, \bibinfo{author}{Cranmer, S.~J.} \& \bibinfo{author}{Clauset, A.}
\newblock \bibinfo{journal}{\bibinfo{title}{Untangling the network effects of
  productivity and prominence among scientists}}.
\newblock {\emph{\JournalTitle{Nature communications}}}
  \textbf{\bibinfo{volume}{13}}, \bibinfo{pages}{4907} (\bibinfo{year}{2022}).

\bibitem{page1999pagerank}
\bibinfo{author}{Page, L.}, \bibinfo{author}{Brin, S.},
  \bibinfo{author}{Motwani, R.} \& \bibinfo{author}{Winograd, T.}
\newblock \bibinfo{title}{The {PageRank} citation ranking: {Bringing} order to
  the web.}
\newblock \bibinfo{type}{Tech. Rep.}, \bibinfo{institution}{Stanford InfoLab}
  (\bibinfo{year}{1999}).

\bibitem{gleich2015pagerank}
\bibinfo{author}{Gleich, D.~F.}
\newblock \bibinfo{journal}{\bibinfo{title}{{PageRank} beyond the {Web}}}.
\newblock {\emph{\JournalTitle{SIAM Review}}} \textbf{\bibinfo{volume}{57}},
  \bibinfo{pages}{321--363} (\bibinfo{year}{2015}).

\bibitem{cordelli2015distributive}
\bibinfo{author}{Cordelli, C.}
\newblock \bibinfo{journal}{\bibinfo{title}{Distributive justice and the
  problem of friendship}}.
\newblock {\emph{\JournalTitle{Political Studies}}}
  \textbf{\bibinfo{volume}{63}}, \bibinfo{pages}{679--695}
  (\bibinfo{year}{2015}).

\bibitem{mas2009peers}
\bibinfo{author}{Mas, A.} \& \bibinfo{author}{Moretti, E.}
\newblock \bibinfo{journal}{\bibinfo{title}{Peers at work}}.
\newblock {\emph{\JournalTitle{American Economic Review}}}
  \textbf{\bibinfo{volume}{99}}, \bibinfo{pages}{112--145}
  (\bibinfo{year}{2009}).

\bibitem{kandel1978homophily}
\bibinfo{author}{Kandel, D.~B.}
\newblock \bibinfo{journal}{\bibinfo{title}{Homophily, selection, and
  socialization in adolescent friendships}}.
\newblock {\emph{\JournalTitle{American journal of Sociology}}}
  \textbf{\bibinfo{volume}{84}}, \bibinfo{pages}{427--436}
  (\bibinfo{year}{1978}).

\bibitem{foster2025nepo}
\bibinfo{author}{Foster, J.} \& \bibinfo{author}{Maroto, M.}
\newblock \bibinfo{journal}{\bibinfo{title}{Nepo babies and the myth of
  meritocracy}}.
\newblock {\emph{\JournalTitle{The Sociological Quarterly}}}
  \textbf{\bibinfo{volume}{66}}, \bibinfo{pages}{215--237}
  (\bibinfo{year}{2025}).

\bibitem{christian2006social}
\bibinfo{author}{Christian, J.}, \bibinfo{author}{Mellow, J.} \&
  \bibinfo{author}{Thomas, S.}
\newblock \bibinfo{journal}{\bibinfo{title}{Social and economic implications of
  family connections to prisoners}}.
\newblock {\emph{\JournalTitle{Journal of Criminal justice}}}
  \textbf{\bibinfo{volume}{34}}, \bibinfo{pages}{443--452}
  (\bibinfo{year}{2006}).

\bibitem{atkinson2015inequality}
\bibinfo{author}{Atkinson, A.~B.}
\newblock \emph{\bibinfo{title}{Inequality: {What} can be done?}}
  (\bibinfo{publisher}{Harvard University Press}, \bibinfo{year}{2015}).

\bibitem{currarini2009economic}
\bibinfo{author}{Currarini, S.}, \bibinfo{author}{Jackson, M.~O.} \&
  \bibinfo{author}{Pin, P.}
\newblock \bibinfo{journal}{\bibinfo{title}{An economic model of friendship:
  {H}omophily, minorities, and segregation}}.
\newblock {\emph{\JournalTitle{Econometrica}}} \textbf{\bibinfo{volume}{77}},
  \bibinfo{pages}{1003--1045} (\bibinfo{year}{2009}).

\bibitem{mcdonald2011s}
\bibinfo{author}{McDonald, S.}
\newblock \bibinfo{journal}{\bibinfo{title}{What's in the ``old boys'' network?
  accessing social capital in gendered and racialized networks}}.
\newblock {\emph{\JournalTitle{Social networks}}}
  \textbf{\bibinfo{volume}{33}}, \bibinfo{pages}{317--330}
  (\bibinfo{year}{2011}).

\bibitem{jackson2021inequality}
\bibinfo{author}{Jackson, M.~O.}
\newblock \bibinfo{journal}{\bibinfo{title}{Inequality's economic and social
  roots: {T}he role of social networks and homophily}}.
\newblock {\emph{\JournalTitle{Available at SSRN 3795626}}}
  (\bibinfo{year}{2021}).

\bibitem{levy2019echo}
\bibinfo{author}{Levy, G.} \& \bibinfo{author}{Razin, R.}
\newblock \bibinfo{journal}{\bibinfo{title}{Echo chambers and their effects on
  economic and political outcomes}}.
\newblock {\emph{\JournalTitle{Annual Review of Economics}}}
  \textbf{\bibinfo{volume}{11}}, \bibinfo{pages}{303--328}
  (\bibinfo{year}{2019}).

\bibitem{coffman2018gender}
\bibinfo{author}{Coffman, K.~B.}, \bibinfo{author}{Exley, C.~L.} \&
  \bibinfo{author}{Niederle, M.}
\newblock \emph{\bibinfo{title}{When gender discrimination is not about
  gender}} (\bibinfo{publisher}{Harvard Business School Boston},
  \bibinfo{year}{2018}).

\bibitem{jackson2019human}
\bibinfo{author}{Jackson, M.~O.}
\newblock \emph{\bibinfo{title}{The human network: {How} we're connected and
  why it matters}} (\bibinfo{publisher}{Atlantic Books}, \bibinfo{year}{2019}).

\bibitem{avin2015homophily}
\bibinfo{author}{Avin, C.} \emph{et~al.}
\newblock \bibinfo{title}{Homophily and the glass ceiling effect in social
  networks}.
\newblock In \emph{\bibinfo{booktitle}{Proceedings of the 2015 conference on
  innovations in theoretical computer science}}, \bibinfo{pages}{41--50}
  (\bibinfo{year}{2015}).

\bibitem{karimi2018homophily}
\bibinfo{author}{Karimi, F.}, \bibinfo{author}{G{\'e}nois, M.},
  \bibinfo{author}{Wagner, C.}, \bibinfo{author}{Singer, P.} \&
  \bibinfo{author}{Strohmaier, M.}
\newblock \bibinfo{journal}{\bibinfo{title}{Homophily influences ranking of
  minorities in social networks}}.
\newblock {\emph{\JournalTitle{Scientific reports}}}
  \textbf{\bibinfo{volume}{8}}, \bibinfo{pages}{1--12} (\bibinfo{year}{2018}).

\bibitem{espin2022inequality}
\bibinfo{author}{Esp{\'\i}n-Noboa, L.}, \bibinfo{author}{Wagner, C.},
  \bibinfo{author}{Strohmaier, M.} \& \bibinfo{author}{Karimi, F.}
\newblock \bibinfo{journal}{\bibinfo{title}{Inequality and inequity in
  network-based ranking and recommendation algorithms}}.
\newblock {\emph{\JournalTitle{Scientific reports}}}
  \textbf{\bibinfo{volume}{12}}, \bibinfo{pages}{1--14} (\bibinfo{year}{2022}).

\bibitem{ferrara2022link}
\bibinfo{author}{Ferrara, A.}, \bibinfo{author}{Esp{\'\i}n-Noboa, L.},
  \bibinfo{author}{Karimi, F.} \& \bibinfo{author}{Wagner, C.}
\newblock \bibinfo{title}{Link recommendations: {Their} impact on network
  structure and minorities}.
\newblock In \emph{\bibinfo{booktitle}{14th ACM Web Science Conference 2022}},
  WebSci '22, \bibinfo{pages}{228–238},
  \doiprefix\url{10.1145/3501247.3531583} (\bibinfo{publisher}{Association for
  Computing Machinery}, \bibinfo{address}{New York, NY, USA},
  \bibinfo{year}{2022}).

\bibitem{neuhauser2023improving}
\bibinfo{author}{Neuh{\"a}user, L.}, \bibinfo{author}{Karimi, F.},
  \bibinfo{author}{Bachmann, J.}, \bibinfo{author}{Strohmaier, M.} \&
  \bibinfo{author}{Schaub, M.~T.}
\newblock \bibinfo{journal}{\bibinfo{title}{Improving the visibility of
  minorities through network growth interventions}}.
\newblock {\emph{\JournalTitle{Communications Physics}}}
  \textbf{\bibinfo{volume}{6}}, \bibinfo{pages}{108} (\bibinfo{year}{2023}).

\bibitem{lindquist2015entrepreneurial}
\bibinfo{author}{Lindquist, M.~J.}, \bibinfo{author}{Sol, J.} \&
  \bibinfo{author}{Van~Praag, M.}
\newblock \bibinfo{journal}{\bibinfo{title}{Why do entrepreneurial parents have
  entrepreneurial children?}}
\newblock {\emph{\JournalTitle{Journal of Labor Economics}}}
  \textbf{\bibinfo{volume}{33}}, \bibinfo{pages}{269--296}
  (\bibinfo{year}{2015}).

\bibitem{kossinets2009origins}
\bibinfo{author}{Kossinets, G.} \& \bibinfo{author}{Watts, D.~J.}
\newblock \bibinfo{journal}{\bibinfo{title}{Origins of homophily in an evolving
  social network}}.
\newblock {\emph{\JournalTitle{American journal of sociology}}}
  \textbf{\bibinfo{volume}{115}}, \bibinfo{pages}{405--450}
  (\bibinfo{year}{2009}).

\bibitem{barabasi1999emergence}
\bibinfo{author}{Barab{\'a}si, A.-L.} \& \bibinfo{author}{Albert, R.}
\newblock \bibinfo{journal}{\bibinfo{title}{Emergence of scaling in random
  networks}}.
\newblock {\emph{\JournalTitle{Science}}} \textbf{\bibinfo{volume}{286}},
  \bibinfo{pages}{509--512} (\bibinfo{year}{1999}).

\bibitem{Horowitz2019}
\bibinfo{author}{Horowitz, J.~M.}
\newblock \bibinfo{title}{Most {Americans} say the legacy of slavery still
  affects {Black} people in the {U.S.} today} (\bibinfo{year}{2019}).
\newblock \bibinfo{note}{Publication date: 2019-06-17. Accessed online:
  2023-10-03}.

\bibitem{codd1998prisoners}
\bibinfo{author}{Codd, H.}
\newblock \bibinfo{journal}{\bibinfo{title}{Prisoners' families: {T}he
  ``forgotten victims''}}.
\newblock {\emph{\JournalTitle{Prob. J.}}} \textbf{\bibinfo{volume}{45}},
  \bibinfo{pages}{148} (\bibinfo{year}{1998}).

\bibitem{garip2021network}
\bibinfo{author}{Garip, F.} \& \bibinfo{author}{Molina, M.~D.}
\newblock \bibinfo{title}{Network amplification}.
\newblock In \emph{\bibinfo{booktitle}{Research Handbook on Analytical
  Sociology}}, \bibinfo{pages}{308--320} (\bibinfo{publisher}{Edward Elgar
  Publishing}, \bibinfo{year}{2021}).

\bibitem{chiang2011network}
\bibinfo{author}{Chiang, Y.-S.} \& \bibinfo{author}{Takahashi, N.}
\newblock \bibinfo{journal}{\bibinfo{title}{Network homophily and the evolution
  of the pay-it-forward reciprocity}}.
\newblock {\emph{\JournalTitle{PloS one}}} \textbf{\bibinfo{volume}{6}},
  \bibinfo{pages}{e29188} (\bibinfo{year}{2011}).

\bibitem{laniado2016gender}
\bibinfo{author}{Laniado, D.}, \bibinfo{author}{Volkovich, Y.},
  \bibinfo{author}{Kappler, K.} \& \bibinfo{author}{Kaltenbrunner, A.}
\newblock \bibinfo{journal}{\bibinfo{title}{Gender homophily in online dyadic
  and triadic relationships}}.
\newblock {\emph{\JournalTitle{EPJ Data Science}}}
  \textbf{\bibinfo{volume}{5}}, \bibinfo{pages}{19} (\bibinfo{year}{2016}).

\bibitem{grund2015ethnic}
\bibinfo{author}{Grund, T.~U.} \& \bibinfo{author}{Densley, J.~A.}
\newblock \bibinfo{journal}{\bibinfo{title}{Ethnic homophily and triad closure:
  {M}apping internal gang structure using exponential random graph models}}.
\newblock {\emph{\JournalTitle{Journal of Contemporary Criminal Justice}}}
  \textbf{\bibinfo{volume}{31}}, \bibinfo{pages}{354--370}
  (\bibinfo{year}{2015}).

\bibitem{centola2015social}
\bibinfo{author}{Centola, D.}
\newblock \bibinfo{journal}{\bibinfo{title}{The social origins of networks and
  diffusion}}.
\newblock {\emph{\JournalTitle{American Journal of Sociology}}}
  \textbf{\bibinfo{volume}{120}}, \bibinfo{pages}{1295--1338}
  (\bibinfo{year}{2015}).

\bibitem{zhao2021network}
\bibinfo{author}{Zhao, L.} \& \bibinfo{author}{Garip, F.}
\newblock \bibinfo{journal}{\bibinfo{title}{Network diffusion under homophily
  and consolidation as a mechanism for social inequality}}.
\newblock {\emph{\JournalTitle{Sociological Methods \& Research}}}
  \textbf{\bibinfo{volume}{50}}, \bibinfo{pages}{1150--1185}
  (\bibinfo{year}{2021}).

\bibitem{bian2018evidence}
\bibinfo{author}{Bian, L.}, \bibinfo{author}{Leslie, S.-J.} \&
  \bibinfo{author}{Cimpian, A.}
\newblock \bibinfo{journal}{\bibinfo{title}{Evidence of bias against girls and
  women in contexts that emphasize intellectual ability.}}
\newblock {\emph{\JournalTitle{American Psychologist}}}
  \textbf{\bibinfo{volume}{73}}, \bibinfo{pages}{1139} (\bibinfo{year}{2018}).

\bibitem{nickerson1998confirmation}
\bibinfo{author}{Nickerson, R.~S.}
\newblock \bibinfo{journal}{\bibinfo{title}{Confirmation bias: {A} ubiquitous
  phenomenon in many guises}}.
\newblock {\emph{\JournalTitle{Review of general psychology}}}
  \textbf{\bibinfo{volume}{2}}, \bibinfo{pages}{175--220}
  (\bibinfo{year}{1998}).

\bibitem{schulz2022network}
\bibinfo{author}{Schulz, J.}, \bibinfo{author}{Mayerhoffer, D.~M.} \&
  \bibinfo{author}{Gebhard, A.}
\newblock \bibinfo{journal}{\bibinfo{title}{A network-based explanation of
  inequality perceptions}}.
\newblock {\emph{\JournalTitle{Social Networks}}}
  \textbf{\bibinfo{volume}{70}}, \bibinfo{pages}{306--324}
  (\bibinfo{year}{2022}).

\bibitem{alipourfard2020friendship}
\bibinfo{author}{Alipourfard, N.}, \bibinfo{author}{Nettasinghe, B.},
  \bibinfo{author}{Abeliuk, A.}, \bibinfo{author}{Krishnamurthy, V.} \&
  \bibinfo{author}{Lerman, K.}
\newblock \bibinfo{journal}{\bibinfo{title}{Friendship paradox biases
  perceptions in directed networks}}.
\newblock {\emph{\JournalTitle{Nature communications}}}
  \textbf{\bibinfo{volume}{11}}, \bibinfo{pages}{707} (\bibinfo{year}{2020}).

\bibitem{grandi2023identifying}
\bibinfo{author}{Grandi, U.}, \bibinfo{author}{Kanesh, L.},
  \bibinfo{author}{Lisowski, G.}, \bibinfo{author}{Sridharan, R.} \&
  \bibinfo{author}{Turrini, P.}
\newblock \bibinfo{title}{Identifying and eliminating majority illusion in
  social networks}.
\newblock In \emph{\bibinfo{booktitle}{Proceedings of the AAAI Conference on
  Artificial Intelligence}}, vol.~\bibinfo{volume}{37},
  \bibinfo{pages}{5062--5069} (\bibinfo{year}{2023}).

\bibitem{lee2019homophily}
\bibinfo{author}{Lee, E.} \emph{et~al.}
\newblock \bibinfo{journal}{\bibinfo{title}{Homophily and minority-group size
  explain perception biases in social networks}}.
\newblock {\emph{\JournalTitle{Nature human behaviour}}}
  \textbf{\bibinfo{volume}{3}}, \bibinfo{pages}{1078--1087}
  (\bibinfo{year}{2019}).

\bibitem{pronin2007perception}
\bibinfo{author}{Pronin, E.}
\newblock \bibinfo{journal}{\bibinfo{title}{Perception and misperception of
  bias in human judgment}}.
\newblock {\emph{\JournalTitle{Trends in cognitive sciences}}}
  \textbf{\bibinfo{volume}{11}}, \bibinfo{pages}{37--43}
  (\bibinfo{year}{2007}).

\bibitem{lieberman1988first}
\bibinfo{author}{Lieberman, M.~B.} \& \bibinfo{author}{Montgomery, D.~B.}
\newblock \bibinfo{journal}{\bibinfo{title}{First-mover advantages}}.
\newblock {\emph{\JournalTitle{Strategic management journal}}}
  \textbf{\bibinfo{volume}{9}}, \bibinfo{pages}{41--58} (\bibinfo{year}{1988}).

\bibitem{schuck1974sexism}
\bibinfo{author}{Schuck, V.}
\newblock \bibinfo{journal}{\bibinfo{title}{Sexism and scholarship: {A} brief
  overview of women, academia, and the disciplines}}.
\newblock {\emph{\JournalTitle{Social Science Quarterly}}}
  \bibinfo{pages}{563--585} (\bibinfo{year}{1974}).

\bibitem{menges1983barriers}
\bibinfo{author}{Menges, R.~J.} \& \bibinfo{author}{Exum, W.~H.}
\newblock \bibinfo{journal}{\bibinfo{title}{Barriers to the progress of women
  and minority faculty}}.
\newblock {\emph{\JournalTitle{The Journal of Higher Education}}}
  \textbf{\bibinfo{volume}{54}}, \bibinfo{pages}{123--144}
  (\bibinfo{year}{1983}).

\bibitem{winegarden1972barriers}
\bibinfo{author}{Winegarden, C.}
\newblock \bibinfo{journal}{\bibinfo{title}{Barriers to black employment in
  white-collar jobs: {A} quantitative approach}}.
\newblock {\emph{\JournalTitle{The Review of Black Political Economy}}}
  \textbf{\bibinfo{volume}{2}}, \bibinfo{pages}{13--24} (\bibinfo{year}{1972}).

\bibitem{wilson2013men}
\bibinfo{author}{Wilson, G.} \& \bibinfo{author}{Maume, D.}
\newblock \bibinfo{journal}{\bibinfo{title}{Men's race-based mobility into
  management: {Analyses} at the blue collar and white collar job levels}}.
\newblock {\emph{\JournalTitle{Research in Social Stratification and
  Mobility}}} \textbf{\bibinfo{volume}{33}}, \bibinfo{pages}{1--12}
  (\bibinfo{year}{2013}).

\bibitem{zick2008ethnic}
\bibinfo{author}{Zick, A.}, \bibinfo{author}{Pettigrew, T.~F.} \&
  \bibinfo{author}{Wagner, U.}
\newblock \bibinfo{title}{Ethnic prejudice and discrimination in {Europe}}
  (\bibinfo{year}{2008}).

\bibitem{zbarauskaite2015minority}
\bibinfo{author}{Zbarauskait{\.e}, A.}, \bibinfo{author}{Grigutyt{\.e}, N.} \&
  \bibinfo{author}{Gailien{\.e}, D.}
\newblock \bibinfo{journal}{\bibinfo{title}{Minority ethnic identity and
  discrimination experience in a context of social transformations}}.
\newblock {\emph{\JournalTitle{Procedia-Social and Behavioral Sciences}}}
  \textbf{\bibinfo{volume}{165}}, \bibinfo{pages}{121--130}
  (\bibinfo{year}{2015}).

\bibitem{shepherd2021inequality}
\bibinfo{author}{Shepherd, H.} \& \bibinfo{author}{Garip, F.}
\newblock \bibinfo{journal}{\bibinfo{title}{On inequality}}.
\newblock {\emph{\JournalTitle{Personal Networks: {C}lassic Readings and New
  Directions in Ego-centric Analysis}}} \bibinfo{pages}{630--650}
  (\bibinfo{year}{2021}).

\bibitem{eckles2016estimating}
\bibinfo{author}{Eckles, D.}, \bibinfo{author}{Kizilcec, R.~F.} \&
  \bibinfo{author}{Bakshy, E.}
\newblock \bibinfo{journal}{\bibinfo{title}{Estimating peer effects in networks
  with peer encouragement designs}}.
\newblock {\emph{\JournalTitle{Proceedings of the National Academy of
  Sciences}}} \textbf{\bibinfo{volume}{113}}, \bibinfo{pages}{7316--7322}
  (\bibinfo{year}{2016}).

\bibitem{centola2021change}
\bibinfo{author}{Centola, D.}
\newblock \emph{\bibinfo{title}{Change: {How} to make big things happen}}
  (\bibinfo{publisher}{Hachette UK}, \bibinfo{year}{2021}).

\bibitem{bapna2015your}
\bibinfo{author}{Bapna, R.} \& \bibinfo{author}{Umyarov, A.}
\newblock \bibinfo{journal}{\bibinfo{title}{Do your online friends make you
  pay? a randomized field experiment on peer influence in online social
  networks}}.
\newblock {\emph{\JournalTitle{Management science}}}
  \textbf{\bibinfo{volume}{61}}, \bibinfo{pages}{1902--1920}
  (\bibinfo{year}{2015}).

\bibitem{hoxby2000peer}
\bibinfo{author}{Hoxby, C.~M.}
\newblock \bibinfo{title}{Peer effects in the classroom: {L}earning from gender
  and race variation} (\bibinfo{year}{2000}).

\bibitem{bond201261}
\bibinfo{author}{Bond, R.~M.} \emph{et~al.}
\newblock \bibinfo{journal}{\bibinfo{title}{A 61-million-person experiment in
  social influence and political mobilization}}.
\newblock {\emph{\JournalTitle{Nature}}} \textbf{\bibinfo{volume}{489}},
  \bibinfo{pages}{295--298} (\bibinfo{year}{2012}).

\bibitem{christakis2008collective}
\bibinfo{author}{Christakis, N.~A.} \& \bibinfo{author}{Fowler, J.~H.}
\newblock \bibinfo{journal}{\bibinfo{title}{The collective dynamics of smoking
  in a large social network}}.
\newblock {\emph{\JournalTitle{New England journal of medicine}}}
  \textbf{\bibinfo{volume}{358}}, \bibinfo{pages}{2249--2258}
  (\bibinfo{year}{2008}).

\bibitem{christakis2007spread}
\bibinfo{author}{Christakis, N.~A.} \& \bibinfo{author}{Fowler, J.~H.}
\newblock \bibinfo{journal}{\bibinfo{title}{The spread of obesity in a large
  social network over 32 years}}.
\newblock {\emph{\JournalTitle{New England journal of medicine}}}
  \textbf{\bibinfo{volume}{357}}, \bibinfo{pages}{370--379}
  (\bibinfo{year}{2007}).

\bibitem{pinheiro2014origin}
\bibinfo{author}{Pinheiro, F.~L.}, \bibinfo{author}{Santos, M.~D.},
  \bibinfo{author}{Santos, F.~C.} \& \bibinfo{author}{Pacheco, J.~M.}
\newblock \bibinfo{journal}{\bibinfo{title}{Origin of peer influence in social
  networks}}.
\newblock {\emph{\JournalTitle{Physical review letters}}}
  \textbf{\bibinfo{volume}{112}}, \bibinfo{pages}{098702}
  (\bibinfo{year}{2014}).

\bibitem{heider1946attitudes}
\bibinfo{author}{Heider, F.}
\newblock \bibinfo{journal}{\bibinfo{title}{Attitudes and cognitive
  organization}}.
\newblock {\emph{\JournalTitle{The Journal of psychology}}}
  \textbf{\bibinfo{volume}{21}}, \bibinfo{pages}{107--112}
  (\bibinfo{year}{1946}).

\bibitem{rezapour2024structural}
\bibinfo{author}{Rezapour, R.}, \bibinfo{author}{Dinh, L.},
  \bibinfo{author}{Jiang, L.} \& \bibinfo{author}{Diesner, J.}
\newblock \bibinfo{journal}{\bibinfo{title}{Structural balance in real-world
  social networks: {I}ncorporating direction and transitivity in measuring
  partial balance}}.
\newblock {\emph{\JournalTitle{Social network analysis and mining}}}
  \textbf{\bibinfo{volume}{14}}, \bibinfo{pages}{168} (\bibinfo{year}{2024}).

\bibitem{szell2010multirelational}
\bibinfo{author}{Szell, M.}, \bibinfo{author}{Lambiotte, R.} \&
  \bibinfo{author}{Thurner, S.}
\newblock \bibinfo{journal}{\bibinfo{title}{Multirelational organization of
  large-scale social networks in an online world}}.
\newblock {\emph{\JournalTitle{Proceedings of the National Academy of
  Sciences}}} \textbf{\bibinfo{volume}{107}}, \bibinfo{pages}{13636--13641}
  (\bibinfo{year}{2010}).

\bibitem{dunbar2010many}
\bibinfo{author}{Dunbar, R.}
\newblock \emph{\bibinfo{title}{How many friends does one person need?
  Dunbar’s number and other evolutionary quirks}}
  (\bibinfo{publisher}{Harvard University Press}, \bibinfo{year}{2010}).

\bibitem{beugnot2020gender}
\bibinfo{author}{Beugnot, J.} \& \bibinfo{author}{Peterl{\'e}, E.}
\newblock \bibinfo{journal}{\bibinfo{title}{Gender bias in job referrals: {An}
  experimental test}}.
\newblock {\emph{\JournalTitle{Journal of Economic Psychology}}}
  \textbf{\bibinfo{volume}{76}}, \bibinfo{pages}{102209}
  (\bibinfo{year}{2020}).

\bibitem{borondo2014each}
\bibinfo{author}{Borondo, J.}, \bibinfo{author}{Borondo, F.},
  \bibinfo{author}{Rodriguez-Sickert, C.} \& \bibinfo{author}{Hidalgo, C.~A.}
\newblock \bibinfo{journal}{\bibinfo{title}{To each according to its degree:
  {The} meritocracy and topocracy of embedded markets}}.
\newblock {\emph{\JournalTitle{Scientific reports}}}
  \textbf{\bibinfo{volume}{4}}, \bibinfo{pages}{1--7} (\bibinfo{year}{2014}).

\bibitem{dwork2024equilibria}
\bibinfo{author}{Dwork, C.}, \bibinfo{author}{Hays, C.},
  \bibinfo{author}{Kleinberg, J.} \& \bibinfo{author}{Raghavan, M.}
\newblock \bibinfo{title}{Equilibria, efficiency, and inequality in network
  formation for hiring and opportunity}.
\newblock In \emph{\bibinfo{booktitle}{Proceedings of the 25th ACM Conference
  on Economics and Computation}}, \bibinfo{pages}{347--371}
  (\bibinfo{year}{2024}).

\bibitem{zhang2024network}
\bibinfo{author}{Zhang, Y.}, \bibinfo{author}{Mukhopadhyay, R.} \&
  \bibinfo{author}{Chaintreau, A.}
\newblock \bibinfo{journal}{\bibinfo{title}{Network fairness ambivalence:
  {When} does social network capital mitigate or amplify unfairness?}}
\newblock {\emph{\JournalTitle{Proceedings of the ACM on Measurement and
  Analysis of Computing Systems}}} \textbf{\bibinfo{volume}{8}},
  \bibinfo{pages}{1--28} (\bibinfo{year}{2024}).

\bibitem{zhang2022fairness}
\bibinfo{author}{Zhang, W.}, \bibinfo{author}{Weiss, J.~C.},
  \bibinfo{author}{Zhou, S.} \& \bibinfo{author}{Walsh, T.}
\newblock \bibinfo{journal}{\bibinfo{title}{Fairness amidst non-iid graph data:
  {A} literature review}}.
\newblock {\emph{\JournalTitle{arXiv preprint arXiv:2202.07170}}}
  (\bibinfo{year}{2022}).

\bibitem{choudhary2022survey}
\bibinfo{author}{Choudhary, M.}, \bibinfo{author}{Laclau, C.} \&
  \bibinfo{author}{Largeron, C.}
\newblock \bibinfo{journal}{\bibinfo{title}{A survey on fairness for machine
  learning on graphs}}.
\newblock {\emph{\JournalTitle{arXiv preprint arXiv:2205.05396}}}
  (\bibinfo{year}{2022}).

\bibitem{mehrotra2022revisiting}
\bibinfo{author}{Mehrotra, A.}, \bibinfo{author}{Sachs, J.} \&
  \bibinfo{author}{Celis, L.~E.}
\newblock \bibinfo{journal}{\bibinfo{title}{Revisiting group fairness metrics:
  {The} effect of networks}}.
\newblock {\emph{\JournalTitle{Proceedings of the ACM on Human-Computer
  Interaction}}} \textbf{\bibinfo{volume}{6}}, \bibinfo{pages}{1--29}
  (\bibinfo{year}{2022}).

\bibitem{liu2023group}
\bibinfo{author}{Liu, D.}, \bibinfo{author}{Do, V.}, \bibinfo{author}{Usunier,
  N.} \& \bibinfo{author}{Nickel, M.}
\newblock \bibinfo{title}{Group fairness without demographics using social
  networks}.
\newblock In \emph{\bibinfo{booktitle}{Proceedings of the 2023 ACM Conference
  on Fairness, Accountability, and Transparency}}, \bibinfo{pages}{1432--1449}
  (\bibinfo{year}{2023}).

\bibitem{barnes2025edge}
\bibinfo{author}{Barnes, K.} \emph{et~al.}
\newblock \bibinfo{journal}{\bibinfo{title}{Edge interventions can mitigate
  demographic and prestige disparities in the computer science coauthorship
  network}}.
\newblock {\emph{\JournalTitle{arXiv preprint arXiv:2506.04435}}}
  (\bibinfo{year}{2025}).

\bibitem{beilinson2020fairness}
\bibinfo{author}{Beilinson, H.}
\newblock \emph{\bibinfo{title}{Fairness and Information Access Clustering in
  Social Networks}}.
\newblock \bibinfo{type}{Bachelor's thesis}, \bibinfo{school}{Haverford
  College} (\bibinfo{year}{2020}).
\newblock \bibinfo{note}{Accessed online: 2026-03-30}.

\bibitem{wang2022information}
\bibinfo{author}{Wang, X.}, \bibinfo{author}{Varol, O.} \&
  \bibinfo{author}{Eliassi-Rad, T.}
\newblock \bibinfo{journal}{\bibinfo{title}{Information access equality on
  generative models of complex networks}}.
\newblock {\emph{\JournalTitle{Applied Network Science}}}
  \textbf{\bibinfo{volume}{7}}, \bibinfo{pages}{1--20} (\bibinfo{year}{2022}).

\bibitem{perozzi2014deepwalk}
\bibinfo{author}{Perozzi, B.}, \bibinfo{author}{Al-Rfou, R.} \&
  \bibinfo{author}{Skiena, S.}
\newblock \bibinfo{title}{{DeepWalk}: {Online} learning of social
  representations}.
\newblock In \emph{\bibinfo{booktitle}{Proceedings of the 20th ACM SIGKDD
  international conference on Knowledge discovery and data mining}},
  \bibinfo{pages}{701--710} (\bibinfo{year}{2014}).

\bibitem{dong2022fairness}
\bibinfo{author}{Dong, Y.}, \bibinfo{author}{Ma, J.}, \bibinfo{author}{Chen,
  C.} \& \bibinfo{author}{Li, J.}
\newblock \bibinfo{journal}{\bibinfo{title}{Fairness in graph mining: {A}
  survey}}.
\newblock {\emph{\JournalTitle{arXiv preprint arXiv:2204.09888}}}
  (\bibinfo{year}{2022}).

\bibitem{dong2022edits}
\bibinfo{author}{Dong, Y.}, \bibinfo{author}{Liu, N.},
  \bibinfo{author}{Jalaian, B.} \& \bibinfo{author}{Li, J.}
\newblock \bibinfo{title}{Edits: {Modeling} and mitigating data bias for graph
  neural networks}.
\newblock In \emph{\bibinfo{booktitle}{Proceedings of the ACM web conference
  2022}}, \bibinfo{pages}{1259--1269} (\bibinfo{year}{2022}).

\bibitem{khajehnejad2022crosswalk}
\bibinfo{author}{Khajehnejad, A.} \emph{et~al.}
\newblock \bibinfo{title}{{CrossWalk}: {Fairness-enhanced} node representation
  learning}.
\newblock In \emph{\bibinfo{booktitle}{Proceedings of the AAAI Conference on
  Artificial Intelligence}}, vol.~\bibinfo{volume}{36},
  \bibinfo{pages}{11963--11970} (\bibinfo{year}{2022}).

\bibitem{palowitch2019monet}
\bibinfo{author}{Palowitch, J.} \& \bibinfo{author}{Perozzi, B.}
\newblock \bibinfo{journal}{\bibinfo{title}{{MONET}: {Debiasing} graph
  embeddings via the metadata-orthogonal training unit}}.
\newblock {\emph{\JournalTitle{arXiv preprint arXiv:1909.11793}}}
  (\bibinfo{year}{2019}).

\bibitem{li2021dyadic}
\bibinfo{author}{Li, P.}, \bibinfo{author}{Wang, Y.}, \bibinfo{author}{Zhao,
  H.}, \bibinfo{author}{Hong, P.} \& \bibinfo{author}{Liu, H.}
\newblock \bibinfo{title}{On dyadic fairness: Exploring and mitigating bias in
  graph connections}.
\newblock In \emph{\bibinfo{booktitle}{International Conference on Learning
  Representations}} (\bibinfo{year}{2021}).

\bibitem{yang2022obtaining}
\bibinfo{author}{Yang, M.} \emph{et~al.}
\newblock \bibinfo{journal}{\bibinfo{title}{Obtaining dyadic fairness by
  optimal transport}}.
\newblock {\emph{\JournalTitle{arXiv preprint arXiv:2202.04520}}}
  (\bibinfo{year}{2022}).

\bibitem{laclau2021all}
\bibinfo{author}{Laclau, C.}, \bibinfo{author}{Redko, I.},
  \bibinfo{author}{Choudhary, M.} \& \bibinfo{author}{Largeron, C.}
\newblock \bibinfo{title}{All of the fairness for edge prediction with optimal
  transport}.
\newblock In \emph{\bibinfo{booktitle}{International Conference on Artificial
  Intelligence and Statistics}}, \bibinfo{pages}{1774--1782}
  (\bibinfo{organization}{PMLR}, \bibinfo{year}{2021}).

\bibitem{rahman2019fairwalk}
\bibinfo{author}{Rahman, T.}, \bibinfo{author}{Surma, B.},
  \bibinfo{author}{Backes, M.} \& \bibinfo{author}{Zhang, Y.}
\newblock \bibinfo{title}{Fairwalk: {T}owards fair graph embedding}.
\newblock In \emph{\bibinfo{booktitle}{Proceedings of the Twenty-Eighth
  International Joint Conference on Artificial Intelligence, {IJCAI-19}}},
  \bibinfo{pages}{3289--3295}, \doiprefix\url{10.24963/ijcai.2019/456}
  (\bibinfo{publisher}{International Joint Conferences on Artificial
  Intelligence Organization}, \bibinfo{year}{2019}).

\bibitem{hardt2016equality}
\bibinfo{author}{Hardt, M.}, \bibinfo{author}{Price, E.} \&
  \bibinfo{author}{Srebro, N.}
\newblock \bibinfo{journal}{\bibinfo{title}{Equality of opportunity in
  supervised learning}}.
\newblock {\emph{\JournalTitle{Advances in neural information processing
  systems}}} \textbf{\bibinfo{volume}{29}} (\bibinfo{year}{2016}).

\bibitem{stoica2018algorithmic}
\bibinfo{author}{Stoica, A.-A.}, \bibinfo{author}{Riederer, C.} \&
  \bibinfo{author}{Chaintreau, A.}
\newblock \bibinfo{title}{Algorithmic glass ceiling in social networks: {The}
  effects of social recommendations on network diversity}.
\newblock In \emph{\bibinfo{booktitle}{Proceedings of the 2018 World Wide Web
  Conference}}, \bibinfo{pages}{923--932} (\bibinfo{year}{2018}).

\bibitem{fabbri2020effect}
\bibinfo{author}{Fabbri, F.}, \bibinfo{author}{Bonchi, F.},
  \bibinfo{author}{Boratto, L.} \& \bibinfo{author}{Castillo, C.}
\newblock \bibinfo{title}{The effect of homophily on disparate visibility of
  minorities in people recommender systems}.
\newblock In \emph{\bibinfo{booktitle}{Proceedings of the International AAAI
  Conference on Web and Social Media}}, vol.~\bibinfo{volume}{14},
  \bibinfo{pages}{165--175} (\bibinfo{year}{2020}).

\bibitem{stoica2020seeding}
\bibinfo{author}{Stoica, A.-A.}, \bibinfo{author}{Han, J.~X.} \&
  \bibinfo{author}{Chaintreau, A.}
\newblock \bibinfo{title}{Seeding network influence in biased networks and the
  benefits of diversity}.
\newblock In \emph{\bibinfo{booktitle}{Proceedings of The Web Conference
  2020}}, \bibinfo{pages}{2089--2098} (\bibinfo{year}{2020}).

\bibitem{espin2021explaining}
\bibinfo{author}{Esp{\'\i}n-Noboa, L.}, \bibinfo{author}{Karimi, F.},
  \bibinfo{author}{Ribeiro, B.}, \bibinfo{author}{Lerman, K.} \&
  \bibinfo{author}{Wagner, C.}
\newblock \bibinfo{journal}{\bibinfo{title}{Explaining classification
  performance and bias via network structure and sampling technique}}.
\newblock {\emph{\JournalTitle{Applied Network Science}}}
  \textbf{\bibinfo{volume}{6}}, \bibinfo{pages}{1--25} (\bibinfo{year}{2021}).

\bibitem{saxena2024fairsna}
\bibinfo{author}{Saxena, A.}, \bibinfo{author}{Fletcher, G.} \&
  \bibinfo{author}{Pechenizkiy, M.}
\newblock \bibinfo{journal}{\bibinfo{title}{{FairSNA}: {Algorithmic} fairness
  in social network analysis}}.
\newblock {\emph{\JournalTitle{ACM Computing Surveys}}}
  \textbf{\bibinfo{volume}{56}}, \bibinfo{pages}{1--45} (\bibinfo{year}{2024}).

\bibitem{farnadi2018fairness}
\bibinfo{author}{Farnadi, G.}, \bibinfo{author}{Babaki, B.} \&
  \bibinfo{author}{Getoor, L.}
\newblock \bibinfo{title}{Fairness in relational domains}.
\newblock In \emph{\bibinfo{booktitle}{Proceedings of the 2018 AAAI/ACM
  Conference on AI, Ethics, and Society}}, \bibinfo{pages}{108--114}
  (\bibinfo{year}{2018}).

\bibitem{farnadi2019declarative}
\bibinfo{author}{Farnadi, G.}, \bibinfo{author}{Babaki, B.} \&
  \bibinfo{author}{Getoor, L.}
\newblock \bibinfo{journal}{\bibinfo{title}{A declarative approach to fairness
  in relational domains}}.
\newblock {\emph{\JournalTitle{Bulletin of the IEEE Computer Society Technical
  Committee on Data Engineering}}}  (\bibinfo{year}{2019}).

\bibitem{yang2024your}
\bibinfo{author}{Yang, W.} \emph{et~al.}
\newblock \bibinfo{title}{Your neighbor matters: Towards fair decisions under
  networked interference}.
\newblock In \emph{\bibinfo{booktitle}{Proceedings of the 30th ACM SIGKDD
  Conference on Knowledge Discovery and Data Mining}},
  \bibinfo{pages}{3829--3840} (\bibinfo{year}{2024}).

\bibitem{sium2024individual}
\bibinfo{author}{Sium, Y.}, \bibinfo{author}{Li, Q.} \&
  \bibinfo{author}{Varshney, K.~R.}
\newblock \bibinfo{title}{Individual fairness in graphs using local and global
  structural information}.
\newblock In \emph{\bibinfo{booktitle}{Proceedings of the AAAI/ACM Conference
  on AI, Ethics, and Society}}, vol.~\bibinfo{volume}{7},
  \bibinfo{pages}{1379--1389} (\bibinfo{year}{2024}).

\bibitem{fish2022s}
\bibinfo{author}{Fish, B.} \& \bibinfo{author}{Stark, L.}
\newblock \bibinfo{title}{It’s not fairness, and it’s not fair: {The}
  failure of distributional equality and the promise of relational equality in
  complete-information hiring games}.
\newblock In \emph{\bibinfo{booktitle}{Equity and Access in Algorithms,
  Mechanisms, and Optimization}}, EAAMO '22,
  \doiprefix\url{10.1145/3551624.3555296} (\bibinfo{publisher}{Association for
  Computing Machinery}, \bibinfo{address}{New York, NY, USA},
  \bibinfo{year}{2022}).

\bibitem{birhane2021algorithmic}
\bibinfo{author}{Birhane, A.}
\newblock \bibinfo{journal}{\bibinfo{title}{Algorithmic injustice: {A}
  relational ethics approach}}.
\newblock {\emph{\JournalTitle{Patterns}}} \textbf{\bibinfo{volume}{2}},
  \bibinfo{pages}{100205} (\bibinfo{year}{2021}).

\bibitem{bengtson2023relational}
\bibinfo{author}{Bengtson, A.} \& \bibinfo{author}{Nielsen, L.}
\newblock \bibinfo{journal}{\bibinfo{title}{Relational justice: {Egalitarian}
  and sufficientarian}}.
\newblock {\emph{\JournalTitle{Journal of Applied Philosophy}}}
  \textbf{\bibinfo{volume}{40}}, \bibinfo{pages}{900--918}
  (\bibinfo{year}{2023}).

\bibitem{arnaiz2025structural}
\bibinfo{author}{Arnaiz-Rodriguez, A.}, \bibinfo{author}{Rex, G.~C.} \&
  \bibinfo{author}{Oliver, N.}
\newblock \bibinfo{title}{Structural group unfairness: Measurement and
  mitigation by means of the effective resistance}.
\newblock In \emph{\bibinfo{booktitle}{Proceedings of the International AAAI
  Conference on Web and Social Media}}, vol.~\bibinfo{volume}{19},
  \bibinfo{pages}{83--106} (\bibinfo{year}{2025}).

\bibitem{balepur2024intervening}
\bibinfo{author}{Balepur, N.} \& \bibinfo{author}{Sundaram, H.}
\newblock \bibinfo{title}{Intervening to increase community trust for fair
  network outcomes}.
\newblock In \emph{\bibinfo{booktitle}{The 2024 ACM Conference on Fairness,
  Accountability, and Transparency}}, \bibinfo{pages}{1827--1837}
  (\bibinfo{year}{2024}).

\bibitem{dwork2012fairness}
\bibinfo{author}{Dwork, C.}, \bibinfo{author}{Hardt, M.},
  \bibinfo{author}{Pitassi, T.}, \bibinfo{author}{Reingold, O.} \&
  \bibinfo{author}{Zemel, R.}
\newblock \bibinfo{title}{Fairness through awareness}.
\newblock In \emph{\bibinfo{booktitle}{Proceedings of the 3rd innovations in
  theoretical computer science conference}}, \bibinfo{pages}{214--226}
  (\bibinfo{year}{2012}).

\bibitem{feldman2015certifying}
\bibinfo{author}{Feldman, M.}, \bibinfo{author}{Friedler, S.~A.},
  \bibinfo{author}{Moeller, J.}, \bibinfo{author}{Scheidegger, C.} \&
  \bibinfo{author}{Venkatasubramanian, S.}
\newblock \bibinfo{title}{Certifying and removing disparate impact}.
\newblock In \emph{\bibinfo{booktitle}{Proceedings of the 21st ACM SIGKDD
  International Conference on Knowledge Discovery and Data Mining}},
  \bibinfo{pages}{259--268} (\bibinfo{year}{2015}).

\bibitem{zemel2013learning}
\bibinfo{author}{Zemel, R.}, \bibinfo{author}{Wu, Y.},
  \bibinfo{author}{Swersky, K.}, \bibinfo{author}{Pitassi, T.} \&
  \bibinfo{author}{Dwork, C.}
\newblock \bibinfo{title}{Learning fair representations}.
\newblock In \emph{\bibinfo{booktitle}{International conference on machine
  learning}}, \bibinfo{pages}{325--333} (\bibinfo{organization}{PMLR},
  \bibinfo{year}{2013}).

\bibitem{cohen1989currency}
\bibinfo{author}{Cohen, G.~A.}
\newblock \bibinfo{journal}{\bibinfo{title}{On the currency of egalitarian
  justice}}.
\newblock {\emph{\JournalTitle{Ethics}}} \textbf{\bibinfo{volume}{99}},
  \bibinfo{pages}{906--944} (\bibinfo{year}{1989}).

\bibitem{rawls2009theory}
\bibinfo{author}{Rawls, J.}
\newblock \emph{\bibinfo{title}{A theory of justice: {Revised} edition}}
  (\bibinfo{publisher}{Harvard university press}, \bibinfo{year}{2009}).

\bibitem{koggel2022feminist}
\bibinfo{author}{Koggel, C.~M.}, \bibinfo{author}{Harbin, A.} \&
  \bibinfo{author}{Llewellyn, J.~J.}
\newblock \bibinfo{title}{Feminist relational theory} (\bibinfo{year}{2022}).

\bibitem{Gunkel2025}
\bibinfo{author}{Gunkel, D.~J.} \& \bibinfo{author}{Coeckelbergh, M.}
\newblock \bibinfo{title}{A relational approach to moral standing: {Reframing}
  ethical boundaries in the age of artificial intelligence}.
\newblock In \bibinfo{editor}{Vandenberghe, F.} \& \bibinfo{editor}{Papilloud,
  C.} (eds.) \emph{\bibinfo{booktitle}{New Directions in Relational Sociology,
  Volume Two: Relations All the Way Down}}, \bibinfo{pages}{285--302},
  \doiprefix\url{10.1007/978-3-032-02413-8_12} (\bibinfo{publisher}{Springer
  Nature Switzerland}, \bibinfo{address}{Cham}, \bibinfo{year}{2025}).

\bibitem{wong2020democratizing}
\bibinfo{author}{Wong, P.-H.}
\newblock \bibinfo{journal}{\bibinfo{title}{Democratizing algorithmic
  fairness}}.
\newblock {\emph{\JournalTitle{Philosophy \& Technology}}}
  \textbf{\bibinfo{volume}{33}}, \bibinfo{pages}{225--244}
  (\bibinfo{year}{2020}).

\bibitem{nath2020relational}
\bibinfo{author}{Nath, R.}
\newblock \bibinfo{journal}{\bibinfo{title}{Relational egalitarianism}}.
\newblock {\emph{\JournalTitle{Philosophy Compass}}}
  \textbf{\bibinfo{volume}{15}}, \bibinfo{pages}{e12686}
  (\bibinfo{year}{2020}).

\bibitem{anderson1999point}
\bibinfo{author}{Anderson, E.~S.}
\newblock \bibinfo{journal}{\bibinfo{title}{What is the point of equality?}}
\newblock {\emph{\JournalTitle{Ethics}}} \textbf{\bibinfo{volume}{109}},
  \bibinfo{pages}{287--337} (\bibinfo{year}{1999}).

\bibitem{adler2022prioritarianism}
\bibinfo{author}{Adler, M.~D.} \& \bibinfo{author}{Norheim, O.~F.}
\newblock \emph{\bibinfo{title}{Prioritarianism in practice}}
  (\bibinfo{publisher}{Cambridge University Press}, \bibinfo{year}{2022}).

\bibitem{konow2001fair}
\bibinfo{author}{Konow, J.}
\newblock \bibinfo{journal}{\bibinfo{title}{Fair and square: {T}he four sides
  of distributive justice}}.
\newblock {\emph{\JournalTitle{Journal of Economic Behavior \& Organization}}}
  \textbf{\bibinfo{volume}{46}}, \bibinfo{pages}{137--164}
  (\bibinfo{year}{2001}).

\bibitem{lind1992procedural}
\bibinfo{author}{Lind, E.~A.} \& \bibinfo{author}{Earley, P.~C.}
\newblock \bibinfo{journal}{\bibinfo{title}{Procedural justice and culture}}.
\newblock {\emph{\JournalTitle{International journal of Psychology}}}
  \textbf{\bibinfo{volume}{27}}, \bibinfo{pages}{227--242}
  (\bibinfo{year}{1992}).

\bibitem{leventhal1980should}
\bibinfo{author}{Leventhal, G.~S.}
\newblock \bibinfo{title}{What should be done with equity theory? new
  approaches to the study of fairness in social relationships}.
\newblock In \emph{\bibinfo{booktitle}{Social exchange: Advances in theory and
  research}}, \bibinfo{pages}{27--55} (\bibinfo{publisher}{Springer},
  \bibinfo{year}{1980}).

\bibitem{suresh2021framework}
\bibinfo{author}{Suresh, H.} \& \bibinfo{author}{Guttag, J.}
\newblock \bibinfo{title}{A framework for understanding sources of harm
  throughout the machine learning life cycle}.
\newblock In \emph{\bibinfo{booktitle}{Proceedings of the 1st ACM Conference on
  Equity and Access in Algorithms, Mechanisms, and Optimization}},
  \bibinfo{pages}{1--9} (\bibinfo{year}{2021}).

\bibitem{grgic2018beyond}
\bibinfo{author}{Grgi{\'c}-Hla{\v{c}}a, N.}, \bibinfo{author}{Zafar, M.~B.},
  \bibinfo{author}{Gummadi, K.~P.} \& \bibinfo{author}{Weller, A.}
\newblock \bibinfo{title}{Beyond distributive fairness in algorithmic decision
  making: Feature selection for procedurally fair learning}.
\newblock In \emph{\bibinfo{booktitle}{Proceedings of the AAAI conference on
  artificial intelligence}}, vol.~\bibinfo{volume}{32} (\bibinfo{year}{2018}).

\bibitem{ustun2019actionable}
\bibinfo{author}{Ustun, B.}, \bibinfo{author}{Spangher, A.} \&
  \bibinfo{author}{Liu, Y.}
\newblock \bibinfo{title}{Actionable recourse in linear classification}.
\newblock In \emph{\bibinfo{booktitle}{Proceedings of the conference on
  fairness, accountability, and transparency}}, \bibinfo{pages}{10--19}
  (\bibinfo{year}{2019}).

\bibitem{wagner2017sampling}
\bibinfo{author}{Wagner, C.}, \bibinfo{author}{Singer, P.},
  \bibinfo{author}{Karimi, F.}, \bibinfo{author}{Pfeffer, J.} \&
  \bibinfo{author}{Strohmaier, M.}
\newblock \bibinfo{title}{Sampling from social networks with attributes}.
\newblock In \emph{\bibinfo{booktitle}{Proceedings of the 26th international
  conference on world wide web}}, \bibinfo{pages}{1181--1190}
  (\bibinfo{year}{2017}).

\bibitem{mansoury2020feedback}
\bibinfo{author}{Mansoury, M.}, \bibinfo{author}{Abdollahpouri, H.},
  \bibinfo{author}{Pechenizkiy, M.}, \bibinfo{author}{Mobasher, B.} \&
  \bibinfo{author}{Burke, R.}
\newblock \bibinfo{title}{Feedback loop and bias amplification in recommender
  systems}.
\newblock In \emph{\bibinfo{booktitle}{Proceedings of the 29th ACM
  international conference on information \& knowledge management}},
  \bibinfo{pages}{2145--2148} (\bibinfo{year}{2020}).

\bibitem{colquitt2001justice}
\bibinfo{author}{Colquitt, J.~A.}, \bibinfo{author}{Conlon, D.~E.},
  \bibinfo{author}{Wesson, M.~J.}, \bibinfo{author}{Porter, C.~O.} \&
  \bibinfo{author}{Ng, K.~Y.}
\newblock \bibinfo{journal}{\bibinfo{title}{Justice at the millennium: a
  meta-analytic review of 25 years of organizational justice research.}}
\newblock {\emph{\JournalTitle{Journal of applied psychology}}}
  \textbf{\bibinfo{volume}{86}}, \bibinfo{pages}{425} (\bibinfo{year}{2001}).

\bibitem{tyler1992relational}
\bibinfo{author}{Tyler, T.~R.} \& \bibinfo{author}{Lind, E.~A.}
\newblock \bibinfo{title}{A relational model of authority in groups}.
\newblock In \emph{\bibinfo{booktitle}{Advances in experimental social
  psychology}}, vol.~\bibinfo{volume}{25}, \bibinfo{pages}{115--191}
  (\bibinfo{publisher}{Elsevier}, \bibinfo{year}{1992}).

\bibitem{d2020fairness}
\bibinfo{author}{D'Amour, A.} \emph{et~al.}
\newblock \bibinfo{title}{Fairness is not static: {Deeper} understanding of
  long term fairness via simulation studies}.
\newblock In \emph{\bibinfo{booktitle}{Proceedings of the 2020 Conference on
  Fairness, Accountability, and Transparency}}, \bibinfo{pages}{525--534}
  (\bibinfo{year}{2020}).

\bibitem{si2022enabling}
\bibinfo{author}{Si~Salem, T.}, \bibinfo{author}{Iosifidis, G.} \&
  \bibinfo{author}{Neglia, G.}
\newblock \bibinfo{journal}{\bibinfo{title}{Enabling long-term fairness in
  dynamic resource allocation}}.
\newblock {\emph{\JournalTitle{Proceedings of the ACM on Measurement and
  Analysis of Computing Systems}}} \textbf{\bibinfo{volume}{6}},
  \bibinfo{pages}{1--36} (\bibinfo{year}{2022}).

\bibitem{mepham2000framework}
\bibinfo{author}{Mepham, B.}
\newblock \bibinfo{journal}{\bibinfo{title}{A framework for the ethical
  analysis of novel foods: {The} ethical matrix}}.
\newblock {\emph{\JournalTitle{Journal of agricultural and environmental
  ethics}}} \textbf{\bibinfo{volume}{12}}, \bibinfo{pages}{165--176}
  (\bibinfo{year}{2000}).

\bibitem{oneil2020near}
\bibinfo{author}{O’Neil, C.} \& \bibinfo{author}{Gunn, H.}
\newblock \bibinfo{journal}{\bibinfo{title}{Near-term artificial intelligence
  and the ethical matrix}}.
\newblock {\emph{\JournalTitle{Ethics of Artificial Intelligence}}}
  \bibinfo{pages}{235--69} (\bibinfo{year}{2020}).

\bibitem{kaya2025mapping}
\bibinfo{author}{Kaya, M.} \& \bibinfo{author}{Bogers, T.}
\newblock \bibinfo{journal}{\bibinfo{title}{Mapping stakeholder needs to
  multi-sided fairness in candidate recommendation for algorithmic hiring}}.
\newblock {\emph{\JournalTitle{arXiv preprint arXiv:2508.00908}}}
  (\bibinfo{year}{2025}).

\bibitem{bastarrica2018affirmative}
\bibinfo{author}{Bastarrica, M.~C.}, \bibinfo{author}{Hitschfeld, N.},
  \bibinfo{author}{Samary, M.~M.} \& \bibinfo{author}{Simmonds, J.}
\newblock \bibinfo{title}{Affirmative action for attracting women to {STEM} in
  {Chile}}.
\newblock In \emph{\bibinfo{booktitle}{Proceedings of the 1st International
  Workshop on Gender Equality in Software Engineering}},
  \bibinfo{pages}{45--48} (\bibinfo{year}{2018}).

\bibitem{gomes2019class}
\bibinfo{author}{Gomes~Jr, L.}
\newblock \bibinfo{title}{In-class social networks and academic performance:
  {How} good connections can improve grades}.
\newblock In \emph{\bibinfo{booktitle}{Simp{\'o}sio Brasileiro de Banco de
  Dados (SBBD)}}, \bibinfo{pages}{25--36} (\bibinfo{organization}{SBC},
  \bibinfo{year}{2019}).

\bibitem{chetty2022social2}
\bibinfo{author}{Chetty, R.} \emph{et~al.}
\newblock \bibinfo{journal}{\bibinfo{title}{Social capital {II}: {D}eterminants
  of economic connectedness}}.
\newblock {\emph{\JournalTitle{Nature}}} \textbf{\bibinfo{volume}{608}},
  \bibinfo{pages}{122--134}, \doiprefix\url{10.1038/s41586-022-04997-3}
  (\bibinfo{year}{2022}).

\bibitem{erdHos1960evolution}
\bibinfo{author}{Erd{\H{o}}s, P.} \& \bibinfo{author}{R{\'e}nyi, A.}
\newblock \bibinfo{journal}{\bibinfo{title}{On the evolution of random
  graphs}}.
\newblock {\emph{\JournalTitle{Publ. Math. Inst. Hungar. Acad. Sci}}}
  \textbf{\bibinfo{volume}{5}}, \bibinfo{pages}{4} (\bibinfo{year}{1960}).

\bibitem{bower2022random}
\bibinfo{author}{Bower, A.}, \bibinfo{author}{Lum, K.},
  \bibinfo{author}{Lazovich, T.}, \bibinfo{author}{Yee, K.} \&
  \bibinfo{author}{Belli, L.}
\newblock \bibinfo{title}{Random isn’t always fair: {Candidate} set imbalance
  \& exposure inequality in recommender systems}.
\newblock In \emph{\bibinfo{booktitle}{FAccTRec Workshop at the ACM Conference
  on Recommender Systems (RecSys)}} (\bibinfo{year}{2022}).
\newblock \bibinfo{note}{Workshop paper}.

\bibitem{tsioutsiouliklis2022link}
\bibinfo{author}{Tsioutsiouliklis, S.}, \bibinfo{author}{Pitoura, E.},
  \bibinfo{author}{Semertzidis, K.} \& \bibinfo{author}{Tsaparas, P.}
\newblock \bibinfo{title}{Link recommendations for {PageRank} fairness}.
\newblock In \emph{\bibinfo{booktitle}{Proceedings of the ACM Web Conference
  2022}}, \bibinfo{pages}{3541--3551} (\bibinfo{year}{2022}).

\bibitem{current2022fairegm}
\bibinfo{author}{Current, S.}, \bibinfo{author}{He, Y.},
  \bibinfo{author}{Gurukar, S.} \& \bibinfo{author}{Parthasarathy, S.}
\newblock \bibinfo{title}{Fairegm: {Fair} link prediction and recommendation
  via emulated graph modification}.
\newblock In \emph{\bibinfo{booktitle}{Proceedings of the 2nd ACM Conference on
  Equity and Access in Algorithms, Mechanisms, and Optimization}}, EAAMO '22,
  \doiprefix\url{10.1145/3551624.3555287} (\bibinfo{publisher}{Association for
  Computing Machinery}, \bibinfo{address}{New York, NY, USA},
  \bibinfo{year}{2022}).

\bibitem{becker2023improving}
\bibinfo{author}{Becker, R.}, \bibinfo{author}{D'Angelo, G.} \&
  \bibinfo{author}{Ghobadi, S.}
\newblock \bibinfo{title}{Improving fairness in information exposure by adding
  links}.
\newblock In \emph{\bibinfo{booktitle}{Proceedings of the Thirty-Seventh AAAI
  Conference on Artificial Intelligence and Thirty-Fifth Conference on
  Innovative Applications of Artificial Intelligence and Thirteenth Symposium
  on Educational Advances in Artificial Intelligence}},
  AAAI'23/IAAI'23/EAAI'23, \doiprefix\url{10.1609/aaai.v37i12.26652}
  (\bibinfo{publisher}{AAAI Press}, \bibinfo{year}{2023}).

\bibitem{reymert2021bibliometrics}
\bibinfo{author}{Reymert, I.}
\newblock \bibinfo{journal}{\bibinfo{title}{Bibliometrics in academic
  recruitment: {A} screening tool rather than a game changer}}.
\newblock {\emph{\JournalTitle{Minerva}}} \textbf{\bibinfo{volume}{59}},
  \bibinfo{pages}{53--78} (\bibinfo{year}{2021}).

\bibitem{laurano2015true}
\bibinfo{author}{Laurano, M.}
\newblock \bibinfo{title}{The true cost of a bad hire} (\bibinfo{year}{2015}).

\bibitem{acharya2020rational}
\bibinfo{author}{Acharya, S.} \& \bibinfo{author}{Wee, S.~L.}
\newblock \bibinfo{journal}{\bibinfo{title}{Rational inattention in hiring
  decisions}}.
\newblock {\emph{\JournalTitle{American Economic Journal: Macroeconomics}}}
  \textbf{\bibinfo{volume}{12}}, \bibinfo{pages}{1--40} (\bibinfo{year}{2020}).

\bibitem{conroy2021rethinking}
\bibinfo{author}{Conroy, G.}
\newblock \bibinfo{title}{Rethinking research assessment: 7 sources of bias to
  watch out for at your institution} (\bibinfo{year}{2021}).

\bibitem{bolick2008takes}
\bibinfo{author}{Bolick, T.}
\newblock \bibinfo{journal}{\bibinfo{title}{``it takes a community'': {Social}
  capital, autism spectrum disorders, and the real world}}.
\newblock {\emph{\JournalTitle{Topics in Language Disorders}}}
  \textbf{\bibinfo{volume}{28}}, \bibinfo{pages}{375--387}
  (\bibinfo{year}{2008}).

\bibitem{crespi2016autism}
\bibinfo{author}{Crespi, B.~J.}
\newblock \bibinfo{journal}{\bibinfo{title}{Autism as a disorder of high
  intelligence}}.
\newblock {\emph{\JournalTitle{Frontiers in neuroscience}}}
  \textbf{\bibinfo{volume}{10}}, \bibinfo{pages}{206417}
  (\bibinfo{year}{2016}).

\bibitem{herrera2023quantifying}
\bibinfo{author}{Herrera-Guzm{\'a}n, Y.}, \bibinfo{author}{Gates, A.~J.},
  \bibinfo{author}{Candia, C.} \& \bibinfo{author}{Barab{\'a}si, A.-L.}
\newblock \bibinfo{journal}{\bibinfo{title}{Quantifying hierarchy and prestige
  in {US} ballet academies as social predictors of career success}}.
\newblock {\emph{\JournalTitle{Scientific Reports}}}
  \textbf{\bibinfo{volume}{13}}, \bibinfo{pages}{18594} (\bibinfo{year}{2023}).

\bibitem{robertson1999corruption}
\bibinfo{author}{Robertson-Snape, F.}
\newblock \bibinfo{journal}{\bibinfo{title}{Corruption, collusion and nepotism
  in {Indonesia}}}.
\newblock {\emph{\JournalTitle{Third world quarterly}}}
  \textbf{\bibinfo{volume}{20}}, \bibinfo{pages}{589--602}
  (\bibinfo{year}{1999}).

\bibitem{inzlicht2018effort}
\bibinfo{author}{Inzlicht, M.}, \bibinfo{author}{Shenhav, A.} \&
  \bibinfo{author}{Olivola, C.~Y.}
\newblock \bibinfo{journal}{\bibinfo{title}{The effort paradox: {Effort} is
  both costly and valued}}.
\newblock {\emph{\JournalTitle{Trends in cognitive sciences}}}
  \textbf{\bibinfo{volume}{22}}, \bibinfo{pages}{337--349}
  (\bibinfo{year}{2018}).

\bibitem{lin2024effort}
\bibinfo{author}{Lin, H.}, \bibinfo{author}{Westbrook, A.},
  \bibinfo{author}{Fan, F.} \& \bibinfo{author}{Inzlicht, M.}
\newblock \bibinfo{journal}{\bibinfo{title}{An experimental manipulation of the
  value of effort}}.
\newblock {\emph{\JournalTitle{Nature Human Behaviour}}}
  \doiprefix\url{10.1038/s41562-024-01842-7} (\bibinfo{year}{2024}).

\bibitem{cohn2015fair}
\bibinfo{author}{Cohn, A.}, \bibinfo{author}{Fehr, E.} \&
  \bibinfo{author}{Goette, L.}
\newblock \bibinfo{journal}{\bibinfo{title}{Fair wages and effort provision:
  {Combining} evidence from a choice experiment and a field experiment}}.
\newblock {\emph{\JournalTitle{Management Science}}}
  \textbf{\bibinfo{volume}{61}}, \bibinfo{pages}{1777--1794}
  (\bibinfo{year}{2015}).

\bibitem{nicholls1976effort}
\bibinfo{author}{Nicholls, J.~G.}
\newblock \bibinfo{journal}{\bibinfo{title}{Effort is virtuous, but it's better
  to have ability: {Evaluative} responses to perceptions of effort and
  ability}}.
\newblock {\emph{\JournalTitle{Journal of Research in Personality}}}
  \textbf{\bibinfo{volume}{10}}, \bibinfo{pages}{306--315}
  (\bibinfo{year}{1976}).

\bibitem{freyer2022inherited}
\bibinfo{author}{Freyer, T.} \& \bibinfo{author}{G{\"u}nther, L.~R.}
\newblock \bibinfo{title}{Inherited inequality and the dilemma of meritocracy}.
\newblock \bibinfo{type}{Tech. Rep.}, \bibinfo{institution}{ECONtribute
  Discussion Paper} (\bibinfo{year}{2022}).

\bibitem{alexander1985fair}
\bibinfo{author}{Alexander, L.~A.}
\newblock \bibinfo{journal}{\bibinfo{title}{Fair equality of opportunity: {John
  Rawls}’ (best) forgotten principle}}.
\newblock {\emph{\JournalTitle{Philosophy research archives}}}
  \textbf{\bibinfo{volume}{11}}, \bibinfo{pages}{197--208}
  (\bibinfo{year}{1985}).

\bibitem{baker2005national}
\bibinfo{author}{Baker, D.} \& \bibinfo{author}{LeTendre, G.~K.}
\newblock \emph{\bibinfo{title}{National differences, global similarities:
  {World} culture and the future of schooling}} (\bibinfo{publisher}{Stanford
  University Press}, \bibinfo{year}{2005}).

\bibitem{brown2016credentials}
\bibinfo{author}{Brown, P.}, \bibinfo{author}{Power, S.},
  \bibinfo{author}{Tholen, G.} \& \bibinfo{author}{Allouch, A.}
\newblock \bibinfo{journal}{\bibinfo{title}{Credentials, talent and cultural
  capital: {A} comparative study of educational elites in {England} and
  {France}}}.
\newblock {\emph{\JournalTitle{British Journal of Sociology of Education}}}
  \textbf{\bibinfo{volume}{37}}, \bibinfo{pages}{191--211}
  (\bibinfo{year}{2016}).

\bibitem{chua2021economic}
\bibinfo{author}{Chua, V.}
\newblock \bibinfo{journal}{\bibinfo{title}{Economic sociology in singapore:
  {Meritocracy} and the missing embeddedness}}.
\newblock {\emph{\JournalTitle{Economic Sociology: Perspectives and
  Conversations}}} \textbf{\bibinfo{volume}{23}}, \bibinfo{pages}{19--22}
  (\bibinfo{year}{2021}).

\bibitem{messick1979fairness}
\bibinfo{author}{Messick, D.~M.} \& \bibinfo{author}{Sentis, K.~P.}
\newblock \bibinfo{journal}{\bibinfo{title}{Fairness and preference}}.
\newblock {\emph{\JournalTitle{Journal of Experimental Social Psychology}}}
  \textbf{\bibinfo{volume}{15}}, \bibinfo{pages}{418--434}
  (\bibinfo{year}{1979}).

\bibitem{nguyen2024definitions}
\bibinfo{author}{Nguyen, T.}, \bibinfo{author}{Alam, S.}, \bibinfo{author}{Hu,
  C.}, \bibinfo{author}{Albiston, C.} \& \bibinfo{author}{Salehi, N.}
\newblock \bibinfo{journal}{\bibinfo{title}{Definitions of fairness differ
  across socioeconomic groups \& shape perceptions of algorithmic decisions}}.
\newblock {\emph{\JournalTitle{Proceedings of the ACM on Human-Computer
  Interaction}}} \textbf{\bibinfo{volume}{8}}, \bibinfo{pages}{1--31}
  (\bibinfo{year}{2024}).

\bibitem{yurrita2022towards}
\bibinfo{author}{Yurrita, M.}, \bibinfo{author}{Murray-Rust, D.},
  \bibinfo{author}{Balayn, A.} \& \bibinfo{author}{Bozzon, A.}
\newblock \bibinfo{title}{Towards a multi-stakeholder value-based assessment
  framework for algorithmic systems}.
\newblock In \emph{\bibinfo{booktitle}{Proceedings of the 2022 ACM Conference
  on Fairness, Accountability, and Transparency}}, \bibinfo{pages}{535--563}
  (\bibinfo{year}{2022}).

\bibitem{europe2019guidelines}
\bibinfo{author}{{High-Level Expert Group on Artificial Intelligence}}.
\newblock \bibinfo{journal}{\bibinfo{title}{Ethics guidelines for trustworthy
  {AI}}}.
\newblock {\emph{\JournalTitle{{The European Commission}}}}
  (\bibinfo{year}{2019}).
\newblock \bibinfo{note}{Accessed online: 2023-03-13},
  \eprint{https://digital-strategy.ec.europa.eu/en/library/ethics-guidelines-trustworthy-ai}.

\bibitem{bell2023possibility}
\bibinfo{author}{Bell, A.} \emph{et~al.}
\newblock \bibinfo{title}{The possibility of fairness: {Revisiting} the
  impossibility theorem in practice}.
\newblock In \emph{\bibinfo{booktitle}{Proceedings of the 2023 ACM Conference
  on Fairness, Accountability, and Transparency}}, \bibinfo{pages}{400--422}
  (\bibinfo{year}{2023}).

\bibitem{hu2024achieving}
\bibinfo{author}{Hu, Z.}, \bibinfo{author}{Zhang, Z.}, \bibinfo{author}{Feng,
  W.} \& \bibinfo{author}{Liu, Q.}
\newblock \bibinfo{title}{Achieving universal fairness in machine learning: {A}
  multi-objective optimization perspective}.
\newblock In \emph{\bibinfo{booktitle}{International Conference on Knowledge
  Science, Engineering and Management}}, \bibinfo{pages}{164--179}
  (\bibinfo{organization}{Springer}, \bibinfo{year}{2024}).

\bibitem{kleinberg2016inherent}
\bibinfo{author}{Kleinberg, J.}, \bibinfo{author}{Mullainathan, S.} \&
  \bibinfo{author}{Raghavan, M.}
\newblock \bibinfo{journal}{\bibinfo{title}{Inherent trade-offs in the fair
  determination of risk scores}}.
\newblock {\emph{\JournalTitle{arXiv preprint arXiv:1609.05807}}}
  (\bibinfo{year}{2016}).

\bibitem{zehlike2025beyond}
\bibinfo{author}{Zehlike, M.}, \bibinfo{author}{Loosley, A.},
  \bibinfo{author}{Jonsson, H.}, \bibinfo{author}{Wiedemann, E.} \&
  \bibinfo{author}{Hacker, P.}
\newblock \bibinfo{journal}{\bibinfo{title}{Beyond incompatibility:
  {Trade-offs} between mutually exclusive fairness criteria in machine learning
  and law}}.
\newblock {\emph{\JournalTitle{Artificial Intelligence}}}
  \textbf{\bibinfo{volume}{340}}, \bibinfo{pages}{104280}
  (\bibinfo{year}{2025}).

\bibitem{corradi2025admission}
\bibinfo{author}{Corradi, B.}, \bibinfo{author}{Espinosa, D.},
  \bibinfo{author}{Rodr{\'\i}guez, C.} \& \bibinfo{author}{Espinoza, {\'O}.}
\newblock \bibinfo{journal}{\bibinfo{title}{Is admission enough? university
  persistence of students admitted through affirmative action policies in
  chile}}.
\newblock {\emph{\JournalTitle{Higher Education Policy}}}
  \bibinfo{pages}{1--21} (\bibinfo{year}{2025}).

\bibitem{mishra2020social}
\bibinfo{author}{Mishra, S.}
\newblock \bibinfo{journal}{\bibinfo{title}{Social networks, social capital,
  social support and academic success in higher education: {A} systematic
  review with a special focus on ‘underrepresented’ students}}.
\newblock {\emph{\JournalTitle{Educational Research Review}}}
  \textbf{\bibinfo{volume}{29}}, \bibinfo{pages}{100307}
  (\bibinfo{year}{2020}).

\bibitem{ferrazzi2014never}
\bibinfo{author}{Ferrazzi, K.} \& \bibinfo{author}{Raz, T.}
\newblock \emph{\bibinfo{title}{Never Eat Alone, Expanded and Updated: {A}nd
  other secrets to success, one relationship at a time}}
  (\bibinfo{publisher}{Currency}, \bibinfo{year}{2014}).

\bibitem{small2019role}
\bibinfo{author}{Small, M.~L.} \& \bibinfo{author}{Adler, L.}
\newblock \bibinfo{journal}{\bibinfo{title}{The role of space in the formation
  of social ties}}.
\newblock {\emph{\JournalTitle{Annual Review of Sociology}}}
  \textbf{\bibinfo{volume}{45}}, \bibinfo{pages}{111--132}
  (\bibinfo{year}{2019}).

\bibitem{helliwell2006well}
\bibinfo{author}{Helliwell, J.~F.}
\newblock \bibinfo{journal}{\bibinfo{title}{Well-being, social capital and
  public policy: what's new?}}
\newblock {\emph{\JournalTitle{The economic journal}}}
  \textbf{\bibinfo{volume}{116}}, \bibinfo{pages}{C34--C45}
  (\bibinfo{year}{2006}).

\bibitem{gilani2020creating}
\bibinfo{author}{Gilani, D.}
\newblock \bibinfo{journal}{\bibinfo{title}{Creating connections: {T}he role of
  universities in enhancing graduates’ social capital and challenging
  nepotism}}.
\newblock {\emph{\JournalTitle{Perspectives: {P}olicy and Practice in Higher
  Education}}} \textbf{\bibinfo{volume}{24}}, \bibinfo{pages}{14--18}
  (\bibinfo{year}{2020}).

\bibitem{bachmann2026cumulative}
\bibinfo{author}{Bachmann, J.}, \bibinfo{author}{Esp{\'\i}n-Noboa, L.},
  \bibinfo{author}{I{\~n}iguez, G.} \& \bibinfo{author}{Karimi, F.}
\newblock \bibinfo{journal}{\bibinfo{title}{Cumulative advantage of brokerage
  in physics}}.
\newblock {\emph{\JournalTitle{Quantitative Science Studies}}}
  \textbf{\bibinfo{volume}{7}}, \bibinfo{pages}{680--694}
  (\bibinfo{year}{2026}).

\bibitem{greenwald1995implicit}
\bibinfo{author}{Greenwald, A.~G.} \& \bibinfo{author}{Banaji, M.~R.}
\newblock \bibinfo{journal}{\bibinfo{title}{Implicit social cognition:
  {A}ttitudes, self-esteem, and stereotypes.}}
\newblock {\emph{\JournalTitle{Psychological review}}}
  \textbf{\bibinfo{volume}{102}}, \bibinfo{pages}{4} (\bibinfo{year}{1995}).

\bibitem{greco2016multiple}
\bibinfo{author}{Greco, S.}, \bibinfo{author}{Figueira, J.} \&
  \bibinfo{author}{Ehrgott, M.}
\newblock \emph{\bibinfo{title}{Multiple criteria decision analysis}},
  vol.~\bibinfo{volume}{37} (\bibinfo{publisher}{Springer},
  \bibinfo{year}{2016}).

\bibitem{wu2022multi}
\bibinfo{author}{Wu, H.}, \bibinfo{author}{Ma, C.}, \bibinfo{author}{Mitra,
  B.}, \bibinfo{author}{Diaz, F.} \& \bibinfo{author}{Liu, X.}
\newblock \bibinfo{journal}{\bibinfo{title}{A multi-objective optimization
  framework for multi-stakeholder fairness-aware recommendation}}.
\newblock {\emph{\JournalTitle{ACM Transactions on Information Systems}}}
  \textbf{\bibinfo{volume}{41}}, \bibinfo{pages}{1--29} (\bibinfo{year}{2022}).

\bibitem{goldner2025multidimensional}
\bibinfo{author}{Goldner, K.} \& \bibinfo{author}{Lundy, T.}
\newblock \bibinfo{title}{Multidimensional bayesian utility maximization:
  {Tight} approximations to welfare}.
\newblock In \emph{\bibinfo{booktitle}{Advances in Neural Information
  Processing Systems (NeurIPS)}}, \doiprefix\url{10.48550/arXiv.2402.12340}
  (\bibinfo{year}{2025}).
\newblock \bibinfo{note}{Poster}, \eprint{2402.12340}.

\bibitem{samson2018multi}
\bibinfo{author}{Samson, D.}, \bibinfo{author}{Foley, P.},
  \bibinfo{author}{Gan, H.~S.} \& \bibinfo{author}{Gloet, M.}
\newblock \bibinfo{journal}{\bibinfo{title}{Multi-stakeholder decision
  theory}}.
\newblock {\emph{\JournalTitle{Annals of Operations Research}}}
  \textbf{\bibinfo{volume}{268}}, \bibinfo{pages}{357--386}
  (\bibinfo{year}{2018}).

\bibitem{hilbe2014cooperation}
\bibinfo{author}{Hilbe, C.}, \bibinfo{author}{Wu, B.},
  \bibinfo{author}{Traulsen, A.} \& \bibinfo{author}{Nowak, M.~A.}
\newblock \bibinfo{journal}{\bibinfo{title}{Cooperation and control in
  multiplayer social dilemmas}}.
\newblock {\emph{\JournalTitle{Proceedings of the National Academy of
  Sciences}}} \textbf{\bibinfo{volume}{111}}, \bibinfo{pages}{16425--16430}
  (\bibinfo{year}{2014}).

\bibitem{salahshour2025perceptual}
\bibinfo{author}{Salahshour, M.}
\newblock \bibinfo{journal}{\bibinfo{title}{Perceptual rationality: {An}
  evolutionary game theory of perceptually rational decision-making}}.
\newblock {\emph{\JournalTitle{Royal Society Open Science}}}
  \textbf{\bibinfo{volume}{12}} (\bibinfo{year}{2025}).

\bibitem{loor2017refocusing}
\bibinfo{author}{Loor, M.}, \bibinfo{author}{Tapia-Rosero, A.} \&
  \bibinfo{author}{De~Tr{\'e}, G.}
\newblock \bibinfo{title}{Refocusing attention on unobserved attributes to
  reach consensus in decision making problems involving a heterogeneous group
  of experts}.
\newblock In \emph{\bibinfo{booktitle}{Proceedings of the Conference of the
  European Society for Fuzzy Logic and Technology}}, \bibinfo{pages}{405--416}
  (\bibinfo{organization}{Springer}, \bibinfo{year}{2017}).

\bibitem{tapia2016fusion}
\bibinfo{author}{Tapia-Rosero, A.}, \bibinfo{author}{Bronselaer, A.},
  \bibinfo{author}{De~Mol, R.} \& \bibinfo{author}{De~Tr{\'e}, G.}
\newblock \bibinfo{journal}{\bibinfo{title}{Fusion of preferences from
  different perspectives in a decision-making context}}.
\newblock {\emph{\JournalTitle{Information fusion}}}
  \textbf{\bibinfo{volume}{29}}, \bibinfo{pages}{120--131}
  (\bibinfo{year}{2016}).

\bibitem{aplak2013fuzzy}
\bibinfo{author}{Aplak, H.~S.} \& \bibinfo{author}{T{\"u}rkbey, O.}
\newblock \bibinfo{journal}{\bibinfo{title}{Fuzzy logic based game theory
  applications in multi-criteria decision making process}}.
\newblock {\emph{\JournalTitle{Journal of Intelligent \& Fuzzy Systems}}}
  \textbf{\bibinfo{volume}{25}}, \bibinfo{pages}{359--371}
  (\bibinfo{year}{2013}).

\bibitem{das2021method}
\bibinfo{author}{Das, A.} \& \bibinfo{author}{Geisler, W.~S.}
\newblock \bibinfo{journal}{\bibinfo{title}{A method to integrate and classify
  normal distributions}}.
\newblock {\emph{\JournalTitle{Journal of Vision}}}
  \textbf{\bibinfo{volume}{21}}, \bibinfo{pages}{1--1} (\bibinfo{year}{2021}).

\bibitem{radicchi2008universality}
\bibinfo{author}{Radicchi, F.}, \bibinfo{author}{Fortunato, S.} \&
  \bibinfo{author}{Castellano, C.}
\newblock \bibinfo{journal}{\bibinfo{title}{Universality of citation
  distributions: {Toward} an objective measure of scientific impact}}.
\newblock {\emph{\JournalTitle{Proceedings of the National Academy of
  Sciences}}} \textbf{\bibinfo{volume}{105}}, \bibinfo{pages}{17268--17272}
  (\bibinfo{year}{2008}).

\bibitem{xie2020predicting}
\bibinfo{author}{Xie, Z.}
\newblock \bibinfo{journal}{\bibinfo{title}{Predicting publication productivity
  for researchers: {A} piecewise {Poisson} model}}.
\newblock {\emph{\JournalTitle{Journal of Informetrics}}}
  \textbf{\bibinfo{volume}{14}}, \bibinfo{pages}{101065}
  (\bibinfo{year}{2020}).

\bibitem{spearman2010survey}
\bibinfo{author}{Spearman, C.~M.}, \bibinfo{author}{Quigley, M.~J.},
  \bibinfo{author}{Quigley, M.~R.} \& \bibinfo{author}{Wilberger, J.~E.}
\newblock \bibinfo{journal}{\bibinfo{title}{Survey of the h index for all of
  academic neurosurgery: {A}nother power-law phenomenon?}}
\newblock {\emph{\JournalTitle{Journal of neurosurgery}}}
  \textbf{\bibinfo{volume}{113}}, \bibinfo{pages}{929--933}
  (\bibinfo{year}{2010}).

\end{thebibliography}
\end{document}